\documentclass[11pt]{article}
\pdfoutput=1
\usepackage{jheppub} 
\usepackage{enumerate}
\usepackage{amsmath,amssymb,epsfig,cancel}
\usepackage{color}
\usepackage{subfig}
\usepackage{graphicx}
\usepackage{caption}
\usepackage{braket}
\usepackage[english]{babel}

\title{The non-Integrability of Strings in Massive Type IIA  and their Holographic  duals}

\author{Carlos N\'u\~nez$^{1}$,}
\author{Jos\'e Manuel Pen\'{\i}n$^{2}$,}
\author{Dibakar Roychowdhury$^{1}$,}
\author{and Jeroen van Gorsel$^{1}$}

\affiliation{$^1$ Department of Physics, Swansea University, Swansea SA2 8PP, United Kingdom}
\affiliation{$^2$  Departamento de F\'{\i}sica de Part\'{\i}culas Universidade de Santiago de Compostela
and
Instituto Galego de F\'{\i}sica de Altas Enerx\'{\i}as (IGFAE) E-15782 Santiago de Compostela, Spain}

\emailAdd{ c.nunez@swansea.ac.uk, jmanpen@gmail.com, dibakarphys@gmail.com, jeroen.van.gorsel@gmail.com} 

\abstract{In this work we study various aspects of six-dimensional ${\cal N}=(1,0)$ SCFTs. We consider the construction of their string duals in Massive IIA
and discuss  some observables in given examples.  We study the dynamics of string solitons wrapping and rotating on the Massive IIA background  and show that the associated Hamiltonian system  is both non-integrable and chaotic, implying the non-integrability of the dual CFT. Our procedure
is analytic, using well developed mathematical techniques, and numerical, by the explicit calculation of power spectra, Lyapunov coefficients and Poincar\'e sections.
\\[10pt]
 } 

\subheader{}

\begin{document}
\def\Tr{{\textrm{Tr}}}




\maketitle

\newpage

\section{Introduction}


Quantum field  theories in six dimensions were objects of curiosity before the  1990's. Reasoning based on perturbation theory
around a Gaussian fixed point (which implied the existence of a continuum Lagrangian) suggested that no such theories existed by themselves, needing some UV completion.
But, observations based on the possibility of encountering a strongly coupled  fixed point and having at the same time  anomaly cancellations,
gave credibility to the  existence of these theories, making them objects of interest \cite{Seiberg:1996qx}. 
These ideas  were supported
by the construction of  Hanany-Witten brane set-ups
\cite{Hanany:1996ie} for  six-dimensional field theories. Indeed,
the papers \cite{Hanany:1997sa}, found the first of those realisations. 

These ideas were subsequently developed and in the past few years, the different versions of the $(2,0)$ six-dimensional SCFT were used as an effective way to organise and understand various features of lower dimensional CFTs---see for example \cite{Gaiotto:2009we}. The Maldacena conjecture  \cite{Maldacena:1997re} gave another important piece of evidence to ascertain the existence of these theories. 

For six-dimensional CFTs with minimal ${\cal N}=(1,0)$ SUSY, the same holographic ideas were used by the authors of \cite{Apruzzi:2013yva}-\cite{Apruzzi:2016rny}. These authors constructed backgrounds in Type II supergravity that realise the $SO(2,6)$ (with an $AdS_7$ factor) and the necessary $SU(2)$ R-symmetry. These backgrounds are dual to the SCFTs  realised at low energies, on the six-dimensional intersection among NS5-D6-D8-branes.
The  six-dimensional ${\cal N}=(1,0)$ SUSY CFTs provides us with a well defined holographic pair (quiver CFT-background) on which different ideas and methods developed in the last twenty years can be tested. The extension of these ideas to the case in which orientifold planes are present have been carefully discussed in \cite{Apruzzi:2017nck}.

One of these developments is the possibility of finding integrability of the field theory. Classical integrability, first
formalised by Liouville, is a frame to solve Hamiltonian problems via quadratures.  Indeed, if the system has the same number of conserved quantities as coordinates in phase space, one could move to action-angle variables ($I_k$, $\theta_k$) and solve $\theta_k= I_k t$. This was further developed later under the name of the classical inverse scattering method. Of course, integrability does not equal solvability. Integrability refers to the property of systems to exhibit regular orbits in phase space (in contrast to chaotic ones). When a system displays classical integrability, there are various methods to find the exact solution to the system. These properties extend, with some limitations to the quantum case. 

In contrast, in systems that do not display integrability one cannot in general, write a closed analytic solution. Some non-integrable systems, present the feature
 of a strong sensitivity to the initial conditions. This notion is formalised by the (largest) Lyapunov exponent, a quantity that characterises the rate of separation of two trajectories that start being arbitrarily close together in phase space.
Another quantity characterising chaotic systems (or the transition from integrable to chaotic by a growing non-integrable perturbation) are the Poincar\'e sections
\cite{Ott:2002book}. For integrable systems, the phase space is foliated  by the so-called KAM tori \cite{Ott:2002book}, but these start to disappear as the non-integrable perturbation grows in influence.
 
Showing the integrability of a system is usually quite a difficult task. In some situations, it may be easier to prove  that it is non-integrable. There are different methods developed to this end. We will use those in the papers
\cite{Basu:2011di},\cite{Basu:2011fw}. The idea is to find a  string soliton (in the holographic language, this soliton captures the dynamics of a long, spinning, heavy operator in the dual CFT) and show that the dynamics of such an object is non-integrable in the sense defined by Liouville. Various techniques developed by mathematicians studying this problem are explained and used in this paper. This can be complemented numerically, by studying the chaos indicators (Poincar\'e sections, Lyapunov exponent) of the Hamiltonian system in question.

In the rest of this work, we deal with six-dimensional ${\cal N}=(1,0)$ SUSY CFTs. We discuss the constructions of the dual backgrounds and study some of their properties and observables. After that, we focus  on their (non-) integrability properties. The plan of the paper is the following: in Section \ref{scftsxx}, we will discuss generalities of the CFTs of interest in this work. 
We shall briefly review the formalism developed in the papers  \cite{Apruzzi:2013yva}-
 \cite{Cremonesi:2015bld}.  We give new examples of pairs (CFT-backgrounds) and 
 present some developments that might be useful in future studies of these systems.
In Section \ref{stringdynamics}, we discuss the (non-) integrability of these CFTs. By focusing on the generic examples developed
in Section \ref{scftsxx}, we analytically show their non-integrable behaviour. In Section \ref{sec:numerics}, we provide a careful numerical analysis of the material in Section \ref{stringdynamics}, showing that indeed, those systems are chaotic.
We  summarise and give an outlook for future developments in Section \ref{concl}. Various interesting appendices complement our technical presentation.

\section{Six-dimensional SCFTs and their holographic description}\label{scftsxx}
In dimension higher than four, when flowing up in energies, a Yang-Mills theory becomes strongly coupled and non-renormalisable. 
Hence,  the field theory needs  a UV-completion. It was suggested in 
\cite{Seiberg:1996qx}, that such a completion is given in terms of a conformal field theory at strong coupling. The existence of these field theories was supported by the computation
of anomalies
in \cite{Danielsson:1997kt}, that showed that the inclusion of vector, hyper and tensor multiplets gives place to anomaly-free field theories.

Such expectations were realised with the brane construction of six-dimensional field theories \cite{Hanany:1997sa}. These constructions are based on intersections of 
NS5-D6-D8-branes. All the branes extend along $\mathbb{R}^{1,5}$. The NS5-branes are located at fixed $x_{6i}$ positions. The D6-branes extend on the $x_6$ intervals (between pairs of NS5-branes) and are point-like objects in the $[x_7,x_8,x_9]$ directions. Finally, the D8-branes extend on $[x_7,x_8,x_9]$ being localised in $x_6$. The system preserves eight supercharges, which is associated with the chiral ${\cal N}=(1,0)$ super-algebra. The isometries of the brane set-up are $SO(1,5)\times SO(3)$. The $SO(3)\sim SU(2)$ is the R-symmetry algebra of the $\mathcal{N}=(1,0)$ conformal algebra.

The anomaly cancellation implies that for every gauge group the number of flavour fields doubles the number of gauge fields $N_f= 2 N_c$. In the brane set-up  $N_c$ is the number of D6-branes in a given interval $[x_{6,i}, x_{6,i+1}]$ between two NS five-branes. On the other hand, the number $N_f= N_{D6,i+1}+ N_{D6,i-1}+ N_{D8}$ counts the  D6-branes in the two adjacent intervals and the number of D8-branes in the interval (these are hypermultiplets in the six-dimensional low energy field theory realised on the D6-branes).

Unlike all other Hanany-Witten brane set-ups realising field theories in lower dimensions, the tensor multiplets living on the NS5-branes play an important role in the cancellation of anomalies and provide self-dual two-forms, which give place (in the case in which the NS5-branes become coincident) to tensionless strings.  In fact, the positions of the NS5-branes are not fixed like in lower dimensional set-ups, but they are represented by a real scalar field $\Phi_i$, which gets a VEV. This scalar field couples to the gauge field strength on the D6-branes, leading to a term in the Lagrangian
$ L\sim (\Phi_{i+1}-\Phi_i)F_{mn}^2+ ...$. When the five-branes become coincident, the effective gauge coupling $\frac{1}{g_{6}^2}= \langle \Phi_{i+1}-\Phi_{i} \rangle$ diverges. In this limit, the field theory flows to  the conjectured CFT  
\cite{Seiberg:1996qx}. Some pieces of evidence support this proposal: the number of supercharges preserved by the brane set-up, the isometries---associated with the unique ${\cal N}=(1,0)$ superconformal algebra, the massless string solitons corresponding with D2-branes that extend in $[x_0,x_1,x_6]$ and end on the five-branes. A detailed explanation of the story summarised above can be found in \cite{Fazzi:2017trx}.

Important evidence for the existence of the ${\cal N}=(1,0)$ SCFTs, comes from holography. Indeed, the paper \cite{Apruzzi:2013yva} started a very fertile line of research, searching for supersymmetric $AdS_7$ solutions in Type II supergravities.

The metric was proposed to be of the form,
\begin{equation}
ds^2= f_1(z) ds^2_{AdS_7}+f_2(z) dz^2+f_3(z) (d\chi^2+\sin^2\chi d\xi^2),\nonumber
\end{equation}
with Neveu-Schwarz and Ramond-Ramond fields preserving the isometries $SO(2,6)\times SO(3)$ and eight supercharges.

In the particular case of Massive type IIA supergravity a system of 
BPS equations was written and a family of solutions found. The paper
\cite{Gaiotto:2014lca}, pointed out the type of field theories these Massive IIA solutions are holographically dual to. Various efforts to solve the BPS equations and interpret the solutions followed
\cite{Passias:2015gya}. These  lead to the formulation by Cremonesi and Tomasiello \cite{Cremonesi:2015bld}, where a precision test was put forward, calculating the $a$-central charge, both in the CFT and in holography. Other interesting developments deal with flows away from, and compactifications of the six-dimensional SCFTs, see for example \cite{Apruzzi:2016rny}. In the following, we summarise the Massive IIA backgrounds as written by Cremonesi and Tomasiello. This is the language in which we shall present the different findings of this paper.
\subsection{Cremonesi-Tomasiello formulation of the holographic duals to ${\cal N}=(1,0)$ SCFTs}\label{cremotoma}
After various manipulations, Cremonesi and Tomasiello \cite{Cremonesi:2015bld}  wrote the Massive IIA backgrounds dual to the six-dimensional conformal field theories as,
\begin{eqnarray}
& & ds^2=f_1(z) ds^2_{AdS_7}+f_2(z) dz^2+f_3(z) d \Omega^2(\chi, \xi),\nonumber\\
& & B_2=f_4(z)  \mathrm{Vol}_{S^2},\;\;\; F_2= f_5(z)  \mathrm{Vol}_{S^2}, \;\;\; e^{\phi}=f_6(z).\label{backgroundads7xm3}
\end{eqnarray}
We have defined  $d \Omega^2(\chi, \xi)=d \chi^2 + \sin^2 \chi\;d \xi^2$ and $\mathrm{Vol}_{S^2}=\sin\chi\;d\chi\wedge d\xi$. 
The functions $f_i(z)$ are written in terms
of another function $\alpha(z)$ and its derivatives,
\begin{eqnarray}
& & f_1(z)= 8 \sqrt{2} \pi  \sqrt{-\frac{\alpha}{{\alpha''}}},\;\;\; f_2(z)= \sqrt{2} \pi \sqrt{-\frac{{\alpha''}}{{\alpha}}},\;\; 
f_3(z)=\sqrt{2} \pi \sqrt{-\frac{{\alpha''}}{\alpha}}\left( \frac{\alpha^2}{{\alpha'}^2-2 \alpha {\alpha''}}\right),\nonumber\\
& &  f_4(z)=\pi \left(-z +\frac{\alpha {\alpha'}}{{{\alpha'}}^2-2 \alpha {\alpha''}}\right),\;\;\; f_5(z)=\left( \frac{{\alpha''}}{162 \pi^2}+ \frac{\pi F_0 \alpha {\alpha'}}{ {\alpha'}^2-2 \alpha {\alpha''}}  \right),\nonumber\\
& &f_6(z)=2^{\frac{5}{4}} \pi^{\frac{5}{2}}3^4 \frac{(-\alpha/ {\alpha''})^{\frac{3}{4}}}{\sqrt{{\alpha'}^2-2 \alpha {\alpha''}}}.
\label{functionsf}
\end{eqnarray}
The different geometries specified by $\alpha(z)$ are supersymmetric solutions of the Massive IIA equations of motion (with mass parameter $F_0$), if $\alpha(z)$ solves the differential equation
\begin{equation}
{\alpha'''}=-162 \pi^3 F_0.
\label{alphathird}
\end{equation}
Given a six-dimensional ${\cal N}=(1,0)$ super-conformal field theory encoded in a quiver diagram, Cremonesi and Tomasiello \cite{Cremonesi:2015bld} gave a recipe to find the precise solution to eq. (\ref{alphathird}), such  that when replaced in eq. (\ref{backgroundads7xm3})
gives the holographic dual to the SCFT. We discuss now some interesting examples.
\subsubsection{Examples of SCFTs and their holographic  type IIA backgrounds}\label{examplesquivers}
Given a quiver diagram encoding the dynamics of
a six-dimensional ${\cal N}=(1,0)$ SCFT,
 we revise here the prescription of \cite{Cremonesi:2015bld} to construct the function $\alpha(z)$.
 
It is clear that solutions to the eq. (\ref{alphathird}) are of the form
\begin{equation}
\alpha(z)= c_0 + c_1 z + c_2 z^2-27\pi^3 F_0 z^3.\nonumber
\end{equation}
In general the solutions we search for, are continuous, differentiable and piecewise defined  in the interval $(i, i+1)$ with $i$ an integer. 
The first region is $(0,1)$, here the function $\alpha(z)$ must satisfy that $\alpha(z=0)=0$. In the last region, the interval $(P, P+1)$, the other boundary condition is
$\alpha(z=P+1)=0$. In general, this allows for a solution of the form,

$$
\alpha(z) =-81\pi^2 \left\{
        \begin{array}{ll}
a_1 z +\frac{a_2}{2} z^2 + \frac{a_3}{6} z^3  & \quad 0 \leq z\leq 1 \\
b_0 +b_1 (z-1) + \frac{b_2}{2}(z-1)^2 + \frac{b_3}{6}(z-1)^3 & \quad 1\leq z\leq 2\\
c_0 + c_1(z-2) + \frac{c_2}{2} (z-2)^2 +\frac{c_3}{6}(z-2)^3 & \quad 2\leq z\leq 3\\
.... & \quad i\leq z\leq i+1\\
p_0   + p_1(z-P) + \frac{p_2}{2} (z-P)^2 +\frac{p_3}{6}(z-P)^3 & \quad P\leq z\leq P+1        
        \end{array}
    \right.
$$
where the constants $(a_2,a_3)$, $(b_2,b_3)$,..., $(p_2,p_3)$ are determined by inspecting the function $R(z)$ that describes the ranks of the gauge groups. In fact, this implies that $a_2=0$ and $a_3$ is the slope corresponding to the first node in the quiver. The other coefficients are determined by imposing continuity of the functions
$\alpha(z)$, $\alpha'(z)$, and that $\alpha(z=P+1)=0$. The procedure to be followed is better understood by inspecting some examples.

\begin{figure}[htb]
\begin{center}
\includegraphics[width=5in,angle=0]{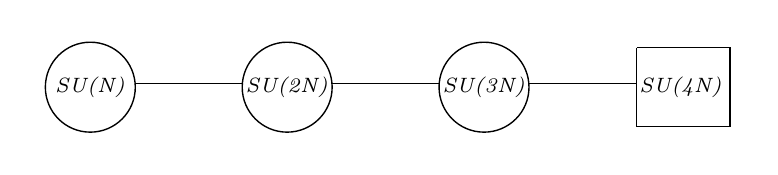}
\caption{The quiver encoding the dynamics of our first example CFT.}\label{figure1xx}
\end{center}
\end{figure}

Let us first consider the quiver depicted in Figure \ref{figure1xx}, with three gauge groups $SU(N)\times SU(2N)\times SU(3N)$ ending with a flavour group $SU(4N)$. Notice that each node satisfies $N_f=2N_c$. 
The function $R(z)$ describing the ranks of this first quiver is
$$
R_1(z) =N \left\{
        \begin{array}{ll}
z  & \quad 0 \leq z\leq 3 \\
12-3z & \quad 3\leq z\leq 4,      
        \end{array}
    \right.
$$
indicating the presence of gauge groups $SU(N)$ at $z=1$, $SU(2N)$ at $z=2$ and $SU(3N)$ at $z=3$. In this sense, the  $z$-direction of the supergravity background encodes the field theory information.  The slope of the first three nodes is $s=N$, which translates to $a_3=b_3=c_3=N$.  Similarly $p_3=-3N$. The  change in slope $\Delta s=-4N$ indicates  the presence of the $SU(4N)$ flavour group.  On the other hand, in the first interval $0\leq z\leq 1$, there is no gauge group, hence $a_2=0$, while the gauge group in the second interval is $SU(N)$, indicating that $b_2=1$. Similarly $c_2=2$ and $p_2=3$, reflecting the presence of the $SU(2N)$ and $SU(3N)$ gauge groups. With this, we can write,
$$
\alpha(z) =-81\pi^2 N \left\{
        \begin{array}{ll}
a_1 z + \frac{1}{6} z^3  & \quad 0 \leq z\leq 1 \\
b_0 +b_1 (z-1) + \frac{1}{2}(z-1)^2 + \frac{1}{6}(z-1)^3 & \quad 1\leq z\leq 2\\
c_0 + c_1(z-2) + \frac{2}{2} (z-2)^2 +\frac{1}{6}(z-2)^3 & \quad 2\leq z\leq 3\\
p_0   + p_1(z-3) + \frac{3}{2} (z-3)^2 -\frac{1}{2}(z-3)^3 & \quad 3\leq z\leq 4.        
        \end{array}
    \right.
$$
where the remaining constants are determined by imposing continuity of $\alpha, \alpha'$ and that $\alpha(z=4)=0$. This gives,
\begin{equation}
a_1=-\frac{5}{2},\;\; b_1=-2,\;\; c_1=-\frac{1}{2},\;\; p_1=2; \;\;\; b_0=-\frac{7}{3}, \;\; c_0=-\frac{11}{3},\;\; p_0= -3.
\end{equation}
The function $\alpha(z)$ describing the background in eq. (\ref{backgroundads7xm3}), dual  to the quiver CFT in Figure \ref{figure1xx} reads
\begin{eqnarray}
& &\alpha_1(z) =-81\pi^2 N \left\{
        \begin{array}{ll}
-\frac{5}{2} z + \frac{1}{6} z^3  & \quad 0 \leq z\leq 1 \\
-\frac{7}{3} -2 (z-1) + \frac{1}{2}(z-1)^2 + \frac{1}{6}(z-1)^3 & \quad 1\leq z\leq 2\\
-\frac{11}{3}-\frac{1}{2}(z-2) + \frac{2}{2} (z-2)^2 +\frac{1}{6}(z-2)^3 & \quad 2\leq z\leq 3\\
-3   + 2(z-3) + \frac{3}{2} (z-3)^2 -\frac{1}{2}(z-3)^3 & \quad 3\leq z\leq 4.        
\end{array}
 \right.\label{quiver4final}
\end{eqnarray}
We have worked with a quiver with three colour nodes and one flavour node. Strictly speaking, the supergravity description
is valid if the number of colour nodes is taken to be large \cite{Cremonesi:2015bld}. Our example in eq. (\ref{quiver4final})   illustrates the procedure. 
{In order to have a better holographic description of the CFT, we should  work with a quiver with $SU(N)\times SU(2N)\times SU(3N)\times SU(4N)\times ....\times SU(PN)$ 
closed by an $SU(PN+N)$-flavour group (and taking $P$ to be large). In that case, we write the function,
\begin{eqnarray}
& &-\frac{\alpha_1(z) }{81\pi^2 N }=\left\{
 \begin{array}{ll}
a_1 z + \frac{1}{6} z^3  & \quad 0 \leq z\leq 1 \\  
(k a_1 +\frac{k^3}{6}) \!\!+\!\!(a_1\!+\!\frac{k^2}{2})\! (z-k)\! +\! \frac{k}{2}\!(z-k)^2 \!+\! \frac{1}{6}\!(z-k)^3\!\! & \quad k\!\leq\! z\leq \!(k+1),\! \nonumber\\
(P a_1 +\frac{P^3}{6}) + (a_1+\frac{P^2}{2})(z-P) +\frac{P}{2} (z-P)^2   -\frac{P}{6}(z-P)^3 & \quad P\leq z\leq P+1,        
 \end{array}
 \right.
\end{eqnarray}
where
\begin{equation}
k=1,....,\!P-\!1,\;\;\;\;\;-6 a_1= P^2+2P. \label{quiver4Pfinal}
\end{equation} }
The case of a quiver with increasing ranks, not closed by the flavour group, is described holographically by the function $\alpha_1(z)=-81\pi^2 N (a_1 z+\frac{z^3}{6})$, being $a_1$ a free parameter.

\begin{figure}[h!]
    \centering
    {{\includegraphics[width=4.5cm]{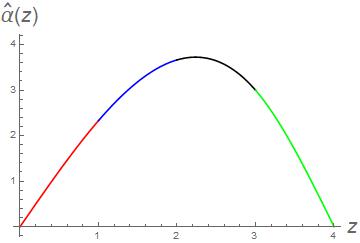} }}%
    \qquad
    {{\includegraphics[width=4.5cm]{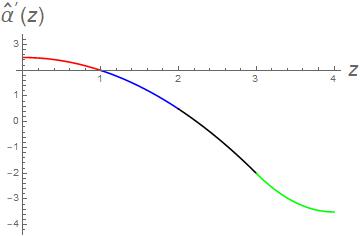} }}%
    \qquad
    {{\includegraphics[width=4.5cm]{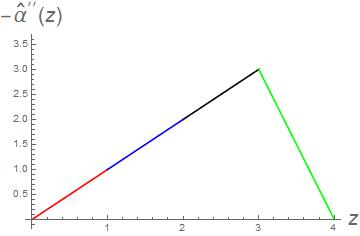} }}

\caption{The function $\hat{\alpha}(z) \equiv \frac{\alpha(z)}{81 \pi^2 N}$ and its derivatives, that describe the CFT associated with the quiver in Figure \ref{figure1xx}.}
\label{alphaderivatives1}

\end{figure}

\begin{figure}%
    \centering
    {{\includegraphics[width=4.5cm]{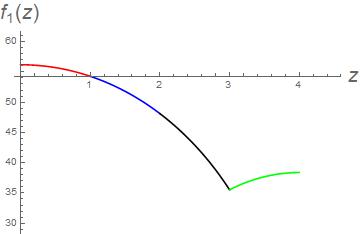} }}%
    \qquad
    {{\includegraphics[width=4.5cm]{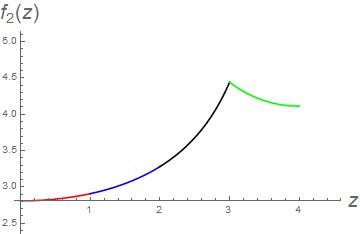} }}%
    \qquad
    {{\includegraphics[width=4.5cm]{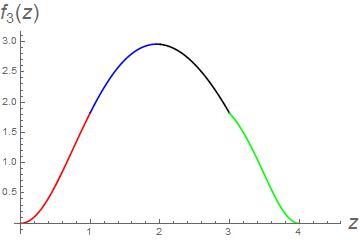} }}
    \qquad
    {{\includegraphics[width=4.5cm]{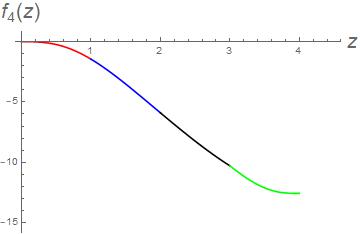} }}
    \qquad
    {{\includegraphics[width=4.5cm]{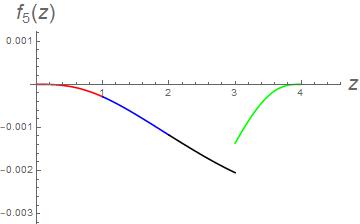} }}
    \qquad
    {{\includegraphics[width=4.5cm]{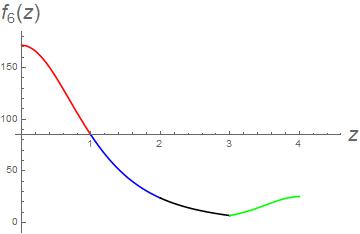} }}

\caption{From top-left to bottom-right, the functions $f_1(z),...., f_6(z)$, that describe the CFT associated with the quiver in Figure \ref{figure1xx}.}
\label{fsq1}
\end{figure}

\begin{figure}%
    \centering
    {{\includegraphics[width=9.5cm]{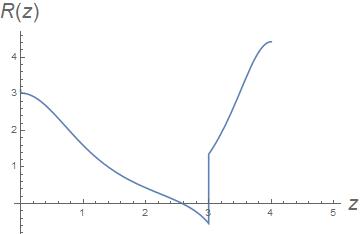} }}%
 
\caption{The Ricci scalar associated to the quiver in Figure \ref{figure1xx}.}
\label{R1}
\end{figure}

It is instructive to plot the function $-\frac{\alpha_1(z)}{81\pi^2}$ and its derivatives for the background defined by eq. (\ref{quiver4final}), see Figure \ref{alphaderivatives1}. We also plot the fields defining the background and the Ricci scalar, see Figures \ref{fsq1} and \ref{R1}. 
None of these functions are divergent for the  $\alpha(z)$ in eq. (\ref{quiver4final}).  We shall use this background in the coming sections to study the dynamics of  a string configuration that rotates and winds on it.

    


\begin{figure}[htb]
\begin{center}
\includegraphics[width=5in,angle=0]{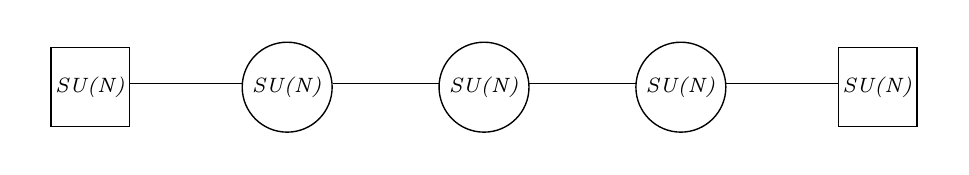}
\caption{The quiver encoding the dynamics of our second example CFT.}\label{fig-quiver2}
\end{center}
\end{figure}

As a second example,  we can work out the function $\alpha(z)$ for the quiver in Figure \ref{fig-quiver2}. 
This quiver starts with an $SU(N)$-flavour node followed by three nodes $SU(N)$-colour, and  it is closed by
a final $SU(N)$-flavour node. The function describing the ranks  is,
$$
R_2(z) =N \left\{
        \begin{array}{ll}
z  & \quad 0 \leq z\leq 1 \\
1 & \quad 1\leq z\leq 3 \\
4-z & \quad 3\leq z\leq 4,      
        \end{array}
    \right.
$$
and the function that determines the holographic description $\alpha_2(z)$ is,
\begin{eqnarray}
& &\alpha_2(z) =-81\pi^2 N \left\{
        \begin{array}{ll}
-\frac{3}{2} z + \frac{1}{6} z^3  & \quad 0 \leq z\leq 1 \\
-\frac{4}{3} - (z-1) + \frac{1}{2}(z-1)^2  & \quad 1\leq z\leq 2\\
-\frac{11}{6} + \frac{1}{2} (z-2)^2  & \quad 2\leq z\leq 3\\
-\frac{4}{3}   + (z-3) + \frac{1}{2} (z-3)^2 -\frac{1}{6}(z-3)^3 & \quad 3\leq z\leq 4.        
        \end{array}
    \right.\label{quiver2final}
\end{eqnarray}
The holographic description of this CFT is trustable when the number of nodes is large. We take the above quiver to be long enough for the illustrative purposes we aim at. 

\begin{figure}[htb]
\begin{center}
\includegraphics[width=6in,angle=0]{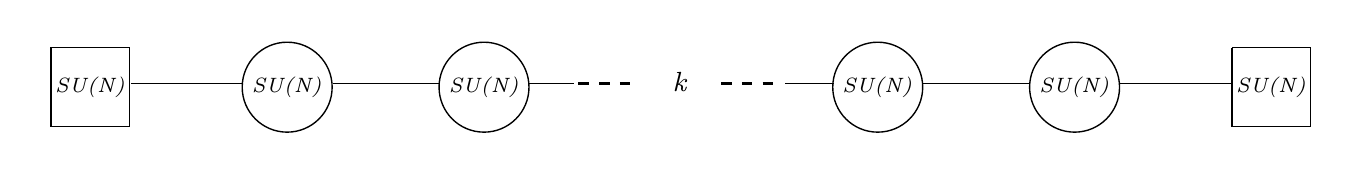}
\caption{The quiver encoding the dynamics of the  third example CFT. The long tail of colour $SU(N)$ ends with a flavour group.}\label{fig-quiver3}
\end{center}
\end{figure}

Finally, we shall consider an {\it endless} quiver. The quiver starts with an $SU(N)$-flavour group and is continued by an infinite tail of  $SU(N)$-colour groups. As a consequence, the $z$-coordinate is unbounded. There is one integration constant that remains undetermined. The  function  describing the ranks is
$$
R_3(z) =N \left\{
        \begin{array}{ll}
z  & \quad 0 \leq z\leq 1 \\
1 & \quad 1\leq z\leq \infty,      
        \end{array}
    \right.
$$
The function $\alpha_3(z)$ reads,
\begin{eqnarray}
& &\alpha_3(z) =-81\pi^2 N \left\{
        \begin{array}{ll}
a_1 z + \frac{1}{6} z^3  & \quad 0 \leq z\leq 1 \\
(a_1+\frac{1}{6}) +(a_1+\frac{1}{2}) (z-1) + \frac{1}{2}(z-1)^2  & \quad 1\leq z\leq 2\\
(2 a_1+\frac{7}{6}) +(a_1+\frac{3}{2}) (z-2) + \frac{1}{2}(z-2)^2  & \quad 2\leq z\leq 3\\
(3 a_1+\frac{19}{6}) +(a_1+\frac{5}{2}) (z-3) + \frac{1}{2}(z-3)^2  & \quad 3\leq z\leq 4\\
(4 a_1+\frac{37}{6}) +(a_1+\frac{7}{2}) (z-4) + \frac{1}{2}(z-4)^2  & \quad 4\leq z\leq 5\\
....\\
(P a_1+\frac{3P^2-3P+1}{6}) +(a_1+\frac{2P-1}{2}) (z-P) + \frac{1}{2}(z-P)^2  & \quad P\leq z\leq P+1\\
....       .      
        \end{array}
    \right.\label{quiver3final}
\end{eqnarray}
Ending the quiver with a flavour $SU(P+1)$ node, reflects in a cap-off  of the geometry at $z=P+1$. This is achieved by adding a term $-\frac{1}{6}(z-P)^3$ to the last line and choosing $a_1=-\frac{P}{2}$. This would correspond to the quiver  in Figure \ref{fig-quiver3}.  The various background functions associated with the holographic description of the quivers in eqs. (\ref{quiver2final})-(\ref{quiver3final}) are  displayed in Appendix \ref{detailsquivers}.
Let us now analyse some observables characterising these CFTs.

\subsection{Page charges and central charge}
In this short section, we will discuss the Page and central charges characterising the backgrounds described in Section \ref{examplesquivers}.
Most of this material, in different notation and for slightly different examples, was discussed in  
\cite{Apruzzi:2013yva}-\cite{Apruzzi:2016rny}. 

In the backgrounds of eq. (\ref{backgroundads7xm3}), we have that $B_2\wedge B_2=0$. This gives for  the Page charges,
\begin{eqnarray}
& & Q_{Dp}=\frac{1}{2\kappa_{10}^2 T_{Dp}} \int F_{8-p}- B_2\wedge F_{6-p}, \nonumber\\
& &  Q_{NS5}= \frac{1}{2\kappa_{10}^2 T_{NS5}}\int H_3,\;\;\;\;\;2\kappa_{10}^2 T_{Dp}= (2\pi)^{7-p} g_s\alpha'^{\frac{7-p}{2}},\;\;\;2\kappa_{10}^2 T_{NS5}= (2\pi)^{2} g_s\alpha'.\nonumber
\end{eqnarray}
As in the rest of this paper, we set $\alpha'=g_s=1$. We start computing the charge of NS-five-branes. Using that $H_3=dB_2$ and integrating on a three-manifold $\Sigma_3=[z,\chi,\xi]$ we have,
\begin{eqnarray}
& & Q_{NS5}=\frac{1}{4\pi^2}\int H_3= \frac{1}{4\pi^2}
\int_{\Omega_2} B_2(z=z_*)- B_2(z=0)= -(P+1).
\end{eqnarray}
We have used the expression for $B_2$ in eq. (\ref{backgroundads7xm3}), the explicit expression for the function $f_4(z)$ in eq. (\ref{functionsf})
and the fact that $\alpha(0)=\alpha(z_*=P+1)=0$ in our backgrounds.

The Page charge of D6-branes is,
\begin{eqnarray}
& & Q_{D6}=\frac{1}{2\pi}\int_{\Omega_2} F_2-F_0 B_2=\frac{1}{81\pi^2}(\alpha'' +162\pi^3 F_0 z)=\frac{1}{81\pi^2}(\alpha'' -\alpha''' z).
\end{eqnarray}
We have used the expression for $F_2,B_2$ in eqs. (\ref{backgroundads7xm3})-(\ref{functionsf}) and the differential equation (\ref{alphathird}) that guarantees BPS solutions.
For the two generic quivers described around eqs. (\ref{quiver4Pfinal}) and (\ref{quiver3final}) we find
\begin{eqnarray}
& &Q_{D6,1} = -N \left\{
        \begin{array}{ll}
0 & \quad 0 \leq z\leq 1 \\
k  & \quad k\leq z\leq (k+1),\;\;\;\; k=1,2,3,4...P-1\\
P & \quad P\leq z\leq P+1\\
\end{array}
    \right.\label{charged61}
\end{eqnarray}
and 
\begin{eqnarray}
& &Q_{D6,2} = -N \left\{
        \begin{array}{ll}
0 & \quad 0 \leq z\leq 1 \\
1 & \quad k\leq z\leq (k+1),\;\;\;\; k=1,2,3,4...P-1\\
1 & \quad P\leq z\leq P+1\\
\end{array}
    \right.\label{charged62}
\end{eqnarray}
The negative sign reflects a choice in orientation.
The  D8-brane charge can be found by studying $F_0$ from the equation (\ref{alphathird}).

It is also interesting to calculate the central charge of these ${\cal N}=(1,0)$ SCFTs. A detailed study is  presented in 
\cite{Cremonesi:2015bld}. Here we do a related calculation as described in \cite{Klebanov:2007ws}. As explained in these papers,  in any background (with dilaton $\Phi$) dual to a $d+1$ dimensional field theory,
\begin{eqnarray}
& & ds^2 =\alpha\;dx_{1,d}^2+ \alpha\beta\;dR^2 + g_{ij} d\theta^i d\theta^j,\;\;\;\; \Phi(R,\theta^i),\nonumber
\end{eqnarray}
we calculate the quantities,
\begin{eqnarray}
& & V_{int}=\int d\theta_i \sqrt{e^{-4\Phi} \det[g_{int}] \alpha^d}, \;\;\;\; H= V_{int}^2,\nonumber
\end{eqnarray}
and together with the Newton constant $G_{N}=8\pi^6$ (in the units we choose here), we calculate a monotonic quantity
\begin{equation}
 c=\frac{d^d}{G_N} \frac{\beta^{d/2} H^{(\frac{2d+1}{2} )}}{(H')^d }.\label{centralcharge}
 \end{equation}
Using the expressions in eqs. (\ref{backgroundads7xm3})-(\ref{functionsf}) and Poincar\'e coordinates to parametrise the $AdS_7$ space, we find
\begin{eqnarray}
& & \alpha=f_1(z) R^2,\;\;\; \beta=\frac{1}{R^4}, \;\;\; d=5,\;\; ds_{int}^2=f_2 dz^2+ f_3 d\Omega_2,\nonumber\\
& & V_{int}= {\cal N} R^5,\;\;\; {\cal N}= - 2(\frac{2}{3})^{8} \int_{0}^{z_*} \alpha(z)\alpha''(z) dz,\nonumber\\
& & c= \frac{1}{16G_N}(\frac{2}{3})^{8} \int_{0}^{z_*} -\alpha(z)\alpha''(z) dz.\label{central}
\end{eqnarray}
Notice that eq. (\ref{central}) is reminiscent of eq. (4.10) in the paper \cite{Cremonesi:2015bld}. In fact, $\alpha''(z)$ is proportional to the rank-function $R(z)$.

We calculate the integral $ \int_{0}^{z_*} -\alpha(z)\alpha''(z) dz$ for the two generic quivers described around eqs. (\ref{quiver4Pfinal}) and (\ref{quiver3final}). For the background described around eq. (\ref{quiver4Pfinal}), we find 
\begin{eqnarray}
& &\int_0^{z_*}\frac{\alpha_1 \alpha_1''}{ (81 \pi^2 N )^2}=\!\left\{
        \begin{array}{ll}
\frac{10a_1 +1}{30}& \quad 0 \leq z\leq 1 \\
\frac{1}{30}\left(1+ \!5k +\!10k^2\! +\!10k^3 \!+\!5 k^4\! +\!10 a_{\!1}\!(1+3 k+3 k^2) \right) & \quad k\leq z\leq (k+1),\; \nonumber\\
 \frac{P^4 +P^3}{12} +\frac{a_1 P^2}{2} +\frac{P^2}{30} +\frac{a_1 P}{6}& \quad P\leq z\leq P+1.\\
\end{array}
    \right.\label{aa''1}
\end{eqnarray}
We sum over all the intervals, and use that $k=1,....,(P-1)$ and  $-6 a_1=P^2+2P$. In the holographic limit  of large $P$ (when the background above are trustable duals to the corresponding quivers) we obtain,
\begin{equation}
c_1=\frac{2}{45 \pi^2} N^2 P^5 \left(1+O(1/P)\right).
\label{centralchargequiver1}
\end{equation}

For the background described around eq. (\ref{quiver3final}), we obtain
\begin{eqnarray}
& &\int_0^{z_*}\alpha_3 \alpha_3''= (81 \pi^2 N )^2\left\{
        \begin{array}{ll}
\frac{10a_1 +1}{30}& \quad 0 \leq z\leq 1 \\
\frac{1}{12}\left(1+ 6 k^2+ 6 a_1 + 12 k a_1 \right) & \quad k\leq z\leq (k+1),\;\;\;\;\nonumber\\
 \frac{P^2 }{4} +\frac{a_1 P}{2} -\frac{P}{12} +\frac{a_1 }{6} +\frac{1}{30}& \quad P\leq z\leq P+1.\\
\end{array}
    \right.\label{aa''2}
\end{eqnarray}
Summing over all the intervals $ k=1,...,(P-1)$, we obtain, in the holographic limit,
\begin{equation}
c_3=\frac{1}{6\pi^2} N^2 P^3 \left(1+ O(1/P) \right) .
\label{centralchargequiver2}
\end{equation}
This concludes the analysis of the characteristic charges of the quivers, as calculated from the dual backgrounds.

\subsection{A Formal Elaboration}
In this section we discuss a formal development that might find some applications in further studies of six-dimensional CFTs and their dual backgrounds.

Consider the differential equation (\ref{alphathird}) defining the function $\alpha(z)$. We explained a way of solving it in Section \ref{cremotoma}. Another way may be to extend the function $F_0$ in an even-periodic way, with period $T=2(P+1)$ and attempt a solution in terms of a Fourier series. For example, in the case of the background defined around eq. (\ref{quiver4Pfinal}),
$$
\alpha_1'''=-162 \pi^3 F_0=-81 \pi^2 N \left\{
        \begin{array}{ll}
-P  & \quad -(P+1) \leq z\leq -P \\
1 & \quad -P\leq z\leq P \\
-P & \quad P\leq z\leq P+1.      
        \end{array}
    \right.
$$
We then decompose this even function in terms of a cosine-Fourier series,
\begin{equation}
\alpha_1'''=-162 \pi N (P+1) \sum_{n=1}^{\infty} \left[\frac{\sin(\frac{n\pi P}{P+1})}{n}  \right] \cos \Big( \frac{n\pi}{P+1} z \Big).\label{fourier1}
\end{equation}
This implies that after integrating three times,
\begin{equation}
\alpha_1=\frac{162  N (P+1)^4}{\pi^2} \sum_{n=1}^{\infty} \left[\frac{\sin(\frac{n\pi P}{P+1})}{n^4}  \right] \sin \Big( \frac{n\pi}{P+1} z \Big).\label{fourier2}
\end{equation}
We have chosen the three integration constants to zero. Notice that in this way, we satisfy the boundary conditions $\alpha(z=0)=\alpha(z=P+1)=0$. Evenly-extending $F_0$ guarantees that the function $\alpha$ is odd. 

For the example of the background with $\alpha(z)$ defined around eq. (\ref{quiver3final}), we proceed similarly. In this case we have
$$
\alpha_3'''=-162 \pi^3 F_0=-81 \pi^2 N \left\{
        \begin{array}{ll}
-1  & \quad -(P+1) \leq z\leq -P \\
0 & \quad -P\leq z\leq 1 \\
1 & \quad -1\leq z\leq 1\\
0 & \quad 1\leq z\leq P\\
-1 & \quad P\leq z\leq P+1.      
        \end{array}
    \right.
$$
and by the same reasoning as above we find,
\begin{equation}
\alpha_3=\frac{162  N (P+1)^3}{\pi^2} \sum_{n=1}^{\infty} \left[\frac{\sin(\frac{n\pi }{P+1}) +\sin(\frac{n \pi P}{P+1})   }{n^4}  \right] \sin \Big( \frac{n\pi}{P+1} z \Big).\label{fourier3}
\end{equation}
Some observations are in order. First, if we use the functions $\alpha(z)$ as defined by eqs. (\ref{fourier2}) and (\ref{fourier3}), their plot looks exactly like those in Figures \ref{alphaderivatives1} and \ref{alphaderivatives2}. Second, studying the Fourier series one can 
check that it can be written as a sum of poly-logarithmic functions $Li_n(z)=\sum_{k=1}^{\infty} \frac{z^k}{k^n}$. Third, the third derivative of $\alpha(z)$ is afflicted by the Gibbs phenomenon at the discontinuity points.

More interestingly is the observation that we are writing $\alpha(z)\sim \sum \sin (\omega z)$, a sum of harmonics. Each of the harmonics solves the equation (\ref{alphathird}), defining a background that has $\alpha''=-\omega^2\alpha$ and hence has constant warp factors of the $AdS_7$ and $z$-direction  $f_1(z)\sim f_2(z)\sim 1$. It also has the warp factor of the $S^2(\chi,\xi)$, $f_3(z)\sim \frac{\sin^2(\omega z)}{3-\cos^2(\omega z)}$, again vanishing in the two ends of the space (but depending on the harmonic in some intermediate points too). The background generated by each harmonic is non-singular as can be seen by computing the Ricci scalar and the dilaton, both bounded.
Each of these harmonics contributes to the central charge as 
\begin{equation}
c\sim - \int_0^{P+1} \alpha\alpha''\sim\frac{P+1}{2},\nonumber
\end{equation}
therefore, the coefficient of the Fourier series is relevant for this counting. Indeed, let us study this in detail. For the expansions in eqs. (\ref{fourier2}) and (\ref{fourier3}) we compute $-\alpha(z)\alpha''(z)$ and integrate it in $[0,P+1]$. Using orthogonality relations and the definition in eq. (\ref{centralcharge}), we find
\begin{eqnarray}
& & c_1=\frac{1}{16 G_N}\Big(\frac{2}{3}\Big)^8 \frac{(162)^2 }{2\pi^2} N^2 (P+1)^7 \sum_{n=1}^{\infty} \left[\frac{\sin(\frac{n\pi P}{P+1})}{n^3}  \right]^2,\nonumber\\
& & c_3 =\frac{1}{16 G_N}\Big(\frac{2}{3}\Big)^8 \frac{(162)^2 \pi^2}{2 \pi^2} N^2 (P+1)^5 \sum_{n=1}^{\infty} \left[\frac{\sin(\frac{n\pi }{P+1}) +\sin(\frac{n\pi P}{P+1})}{n^3}  \right]^2.\label{centralxx}
\end{eqnarray}
The sums can be explicitly evaluated in terms of poly-logarithmic functions. We can use that these formulas are good approximations in the limit of very long linear quivers and expanding at first order for $P\to\infty$, we find
\begin{equation}
  c_1\sim\frac{ 2N^2 P^5}{45 \pi^2},\;\;\;\;\; c_3\sim \frac{N^2 P^3}{6 \pi^2},
\end{equation}
in coincidence with the leading order result of eqs. (\ref{centralchargequiver1}) and (\ref{centralchargequiver2}).

We now move into the study of integrability for these quivers. We will use the holographic perspective as described above.
\section{Dynamics of strings on $AdS_7\times M_3$ backgrounds}\label{stringdynamics}
In this section, we study the dynamics of classical strings moving in backgrounds dual to ${\cal N}=(1,0)$ SCFTs. We will apply the formalism
developed in the papers \cite{Basu:2011fw}, \cite{Basu:2011di} to show that a given string soliton is non-integrable in the Liouvillian sense. This in turns translates into the  non-integrability of the SCFT, as discussed in the introduction.

We will study the dynamics derived from the Polyakov action,
\begin{equation}
S_P=\frac{1}{4 \pi \alpha'} \int_{\Sigma} d^2 \sigma ( G_{\mu \rho} h^{\alpha \beta}+ B_{\mu \rho} \epsilon^{\alpha \beta}) \partial_{\alpha} X^{\mu} \partial_{\beta} X^{\rho},\label{polyakovaction}
\end{equation}
supplemented by the Virasoro constraint $T_{ab}=0$, with
\begin{equation}
T_{a b}=  \partial_a X^{\mu} \partial_b X^{\rho} G_{\mu \rho}-\frac{1}{2} h_{a b} h^{c d} \partial_c X^{\mu} \partial_d X^{\rho} G_{\mu \rho}.
\end{equation}
In the equations above, we choose by convention $-h_{\tau\tau}=h_{\sigma\sigma}=1$ and $\epsilon^{\tau\sigma}=1$. 
We consider a string soliton sitting at the centre of the $AdS_7$ space  given by,
\begin{eqnarray}
t=t(\tau),\;\;z=z(\tau),\;\;
\chi=\chi (\tau),\;\;\;\;
\xi= \nu \sigma,\label{solitonxx}
\end{eqnarray}
where the parameter $\nu$ indicates how many times the string winds around the $\xi$-direction. The equations of motion derived from the action in eq. (\ref{polyakovaction}) are equivalent to those derived from the effective Lagrangian (in eq. (\ref{functionsf}), the reader can find the definitions of $f_i$),
\begin{equation}
\mathcal{L}=f_1(z) \dot{t}^2-f_2(z) \dot{z}^2-f_3(z) \dot{\chi}^2+\nu^2 f_3(z) \sin^2 \chi +2 \nu f_4(z) \sin \chi \dot{\chi}.
\end{equation}
These equations of motion read,
\begin{eqnarray}
& & 2 f_1(z) \dot{t}=2 E \nonumber \\
& & 2 f_3(z) \ddot{\chi}= 2 \nu f'_4(z) \dot{z} \sin \chi-2 f_3'(z) \dot{\chi} \dot{z}-2 \nu^2 f_3(z) \sin \chi \cos \chi \nonumber \\
& & 2 f_2(z) \ddot{z}=-\frac{f'_1(z)}{f_1(z)^2}E^2-\dot{z}^2 f'_2(z)+f'_3(z)(\dot{\chi}^2-\nu^2 \sin^2 \chi)-2 \nu \dot{\chi} \sin \chi f'_4(z).\label{eqsEL}
\end{eqnarray} 
The dot indicates a derivative with respect to $\tau$ and the prime, as above, a derivative with respect to $z$. Notice also that the $t$-equation, the first in (\ref{eqsEL}) was used in the 
equation for $z(\tau)$.
These equations need to be supplemented  by the Virasoro constraint. On this configuration it takes the form,
\begin{eqnarray}
& & 2 T_{\tau\tau}=2T_{\sigma\sigma}= -f_1(z)\dot{t}^2  +f_2(z) \dot{z}^2+ f_3(z) \dot{\chi}^2+\nu^2 f_3(z) \sin^2 \chi  =0,\nonumber\\
& & T_{\sigma\tau}=0.\label{virasoro}
\end{eqnarray}
Using the Euler-Lagrange eqs. (\ref{eqsEL}), the reader can check that  $\partial_\sigma T_{\sigma\sigma}=\partial_\tau T_{\sigma\sigma}=0$.
Hence the constraint $T_{\sigma\sigma}=T_{\tau\tau}=0$ can be satisfied by a judicious choice of the integration constant $E$ in the first equation of 
(\ref{eqsEL}).

It is useful to define the conjugate momenta and the effective Hamiltonian,
\begin{eqnarray}
& & \dot{t}=\frac{p_t}{2 f_1(z)}, \;\; \dot{\chi}=-\frac{1}{2 f_3(z)}(p_\chi-2 \nu f_4(z) \sin \chi),\;\;\dot{z}=-\frac{p_z}{2 f_2(z)},\nonumber\\
& & \mathcal{H}=\frac{p_t^2}{4 f_1(z)}-\frac{p_z^2}{4 f_2(z)}-\frac{1}{4 f_3(z)}(p_\chi-2 \nu f_4(z) \sin \chi)^2-\nu^2 f_3(z) \sin^2 \chi.\label{hamiltonian}
\end{eqnarray}
The Hamilton equations are,
\begin{eqnarray}
& & \dot{t}=\frac{p_t}{2 f_1(z)},\;\;
\dot{z}=-\frac{p_z}{2 f_2(z)},\;\;
\dot{\chi}=-\frac{1}{2 f_3(z)}(p_\chi-2 \nu f_4(z) \sin \chi),\;\;
\dot{p}_t=0,\nonumber\\
& & \dot{p}_\chi=2 \nu^2 \left(\frac{f_4(z)^2}{f_3(z)}+f_3(z)\right) \sin \chi \cos \chi- \nu \frac{f_4(z)}{f_3(z)} p_\chi \cos \chi,\nonumber\\
& & 
\dot{p}_z=\frac{p_t^2}{4 f_1(z)^2} f'_1(z)-\frac{p_z^2}{4 f_2(z)^2}f'_2(z)+\nu^2 f'_3(z) \sin^2 \chi \nonumber \\ 
& & -\frac{f'_3(z)}{4 f_3(z)^2} (p_\chi-2 \nu f_4 \sin \chi)^2-\frac{\nu f'_4(z)}{f_3(z)} \sin \chi (p_\chi-2 \nu f_4(z) \sin \chi). \label{hamiltoneqs}
\end{eqnarray}
The reader  can check that these equations are equivalent to the Euler-Lagrange equations (\ref{eqsEL}).

In what follows, we will use the formalism of \cite{Basu:2011fw}-\cite{Giataganas:2014hma} to analytically study the (non) integrability of the string soliton in eq. (\ref{solitonxx}). 
\subsection{Liouvillian Integrability}
The strategy we will use to prove (non-) integrability in the Liouville sense is the one described in \cite{Basu:2011fw} and exploited in various papers \cite{Zayas:2010fs}-\cite{Roychowdhury:2017vdo}. 

We inspect the equations in (\ref{eqsEL}). In our case, we have only two equations, those for $z(\tau)$ and $\chi(\tau)$. We  shall find a simple solution for one of these equations. Then, we study a fluctuation of the remaining equation (evaluated in the  solution found above). We call this the Normal Variational Equation (NVE).  We apply  Kovacic's criterium to this fluctuated NVE equation, in order   to determine the classical Liouvillian integrability (or non-integrability), of the system.

 In fact, for the case of eqs. (\ref{eqsEL}), one can check that the choice $\chi(\tau)=\dot{\chi}(\tau)=\ddot{\chi}(\tau)=0$, reduces the system of equations to,
\begin{equation}
2 f_2(z) \ddot{z}=-\frac{f'_1(z)}{f_1(z)^2}E^2-\dot{z}^2 f'_2(z).\label{llmm}
\end{equation}
Using the explicit expression for $f_1(z),f_2(z)$, this equation reads,
\begin{equation}
2 \sqrt{-\frac{\alpha''}{\alpha}} \ddot{z}= \left(\frac{\alpha \alpha'''-\alpha' \alpha'' }{2 \alpha^2}\right) \sqrt{-\frac{\alpha}{\alpha''}}(\dot{z}^2-\frac{E^2}{16 \pi^2}),
\end{equation}
which, after a convenient choice of integration constants, admits a simple solution
\begin{equation}
z_{sol}=  \frac{E}{4 \pi}\tau.\label{zsol}
\end{equation}
We  now study the equation for the functions $\chi(\tau)=0+\epsilon x(\tau)$ and expand for small values of $\epsilon$. We obtain the NVE,
\begin{eqnarray}
& &\ddot{x}(\tau)+{\cal B}\dot{x}(\tau)+{\cal A}x(\tau)=0,\nonumber\\
& &{\cal B}=\frac{E f_3'(z)}{4\pi f_3(z)}|_{z_{sol}}, \;\;\;\; {\cal A}=(\nu^2-\nu \frac{E f_4'(z)}{4\pi f_3(z)})|_{z_{sol} }.\label{NVE}
\end{eqnarray}
More explicitly, using the expressions in eq. (\ref{functionsf}) the coefficients ${\cal A}$ and ${\cal B}$ are
\begin{eqnarray}
& & {\cal A}=\bigg( \nu^2- \frac{E\nu}{4\pi \sqrt{2}}\frac{1}{\sqrt{-\alpha \alpha''}} \frac{(-3 \alpha'^2 \alpha''+6 \alpha \alpha''^2-2 \alpha \alpha' \alpha''')}{(-\alpha'^2+2 \alpha \alpha'')} \bigg)_{z_{sol}},\nonumber\\
& & 
{\cal B}=\frac{E}{8\pi} \bigg( 3\frac{\alpha'}{\alpha}+ \frac{(\alpha'^2+2 \alpha \alpha'' )}{(\alpha'^2-2 \alpha \alpha'' )} \frac{\alpha'''}{\alpha'' } \bigg)_{z_{sol}}.\label{AB}
\end{eqnarray}
The Liouvillian integrability of the string soliton depends on the function $\alpha(z)$ defining the background. 
Below, we shall  study this integrability.
\subsection{Analytical study of the (non) integrability of the SCFTs}
In this section, we apply the Kovacic algorithm \cite{kovacic} to the eqs. (\ref{NVE})-(\ref{AB}). For a summary of Kovacic's procedure see Appendix \ref{kovalg}. We particularise in the cases studied in Section \ref{scftsxx}, more concretely for the quivers in Figures \ref{figure1xx} and \ref{fig-quiver2}. The whole problem boils down to study the presence |(or not) of Liouvillian solutions to eqs. (\ref{NVE}), given the different functions $\alpha(z)$.

Consider first the case in which the function $\alpha(z)$ is
\begin{equation}
\alpha(z)= -81\pi^2 k \left( \frac{1}{2}z^2-\frac{2 R_0^2}{81\pi^2 k^2} \right).\label{alphamassless}
\end{equation}
This function corresponds to a background in (massless) Type IIA---since $\alpha'''=0$. In this background there are $k$ D6-branes. Once lifted to eleven dimensions  the metric is $AdS_7\times S^4/Z_k$.
Notice that the coordinate range is $|z|\leq \frac{2R_0}{9\pi k}$. The background is singular at the ends of the space. We find that the NVE equation (\ref{NVE}) reads in this case,
\begin{equation}
\ddot{x}(\tau) -\frac{243E^2 k^2 \pi^2  \tau}{16\pi^2 \left(4 R_0^2-81 k^2 \pi^2 (\frac{E\tau}{4\pi})^2\right)} \dot{x}(\tau)+\nu \left[ \nu+\frac{27 k E \pi }{4\pi \sqrt{4 R_0^2-81 k^2 \pi^2 (\frac{E\tau}{4\pi})^2}} \right]x(\tau)=0.\label{nvemassless}
\end{equation}
This equation is hard to solve exactly. We observe that for large values of the parameter $R_0$ (or for very short times), the eq. (\ref{nvemassless}) reduces to an oscillator equation. This indicates that in such a regime of parameters, the string soliton is Liouville integrable and possibly the full CFT is  integrable
 in that limit too. This may be reminiscent of the `islands of integrability' discussed in \cite{Basu:2012ae}. In fact they appear in the regime in which $E\tau\to 0$. Nevertheless, for finite $R_0$  (or $E\tau\sim R_0$) we failed to find a Liouvillian solution. 

Let us perform a more refined analysis---the details of the logic behind the analysis are in Appendix \ref{kovalg}. 

The first step is to write the NVE as a second order differential equation with rational coefficients.
We choose $64 R_0^2=81 k^2 E^2=1 $ to ease the algebra (not  loosing generality). 
The NVE equation reads,
\begin{equation}
\ddot{x} - \frac{3\tau}{1-\tau^2}\dot{x} +\left(1+\frac{3}{\sqrt{1-\tau^2}}\right) x=0.\nonumber
\end{equation}
We change  variables to $\tau=\sqrt{1-v^2}$. The NVE differential equation in this new variable reads,
 \begin{equation}
x''(v) + {\cal C}(v) x'(v)  +{\cal D}(v) x(v)=0,\;\;\; {\cal C}=\frac{1}{\frac{dv}{d\tau}} \left({\cal B}(v) +\frac{d}{dv}(\frac{dv}{d\tau})    \right), \;\;\;
 {\cal D}=\frac{{\cal A}(v)}{(\frac{dv}{d\tau})^2} .
\end{equation}
Where, in this particular case we have
\begin{eqnarray}
& &v=\sqrt{1-\tau^2},\;\;\; \frac{dv}{d\tau}=-\frac{\sqrt{1-v^2}}{v},\;\;\; \frac{d}{dv}(\frac{dv}{d\tau})=\frac{1}{v^2\sqrt{1-v^2}},\nonumber\\
& & {\cal C}= \frac{3v^2-4}{v- v^3},\;\;\;\; {\cal D}=\frac{v^2+3v}{1-v^2}.
\end{eqnarray}
Following the analysis detailed in Appendix \ref{kovalg}, we construct a function $4 V(v)= 4 {\cal D} -{\cal C}^2 -2 {\cal C}'$,
\begin{equation}
4 V=-4+\frac{3}{4(v-1)^2}- \frac{17}{4(v-1)} -\frac{24}{v^2}+\frac{3}{4(v+1)^2} -\frac{31}{4(v+1)}.\label{veffmassless}
\end{equation}
The pole structure of this function
is analysed  according to the criteria in Appendix \ref{kovalg}.  The existence of poles of order-one and the fact that the function $V$ is of order-one at infinity, implies that none of the three possible cases detailed in Appendix \ref{kovalg} can be satisfied. Therefore, the equation has a non-Liouvillian solutions. The string soliton is non-integrable, and also is non-integrable the associated CFT. For a study of the integrability of a membrane equations in the eleven dimensional lift of this solution see Appendix \ref{appendixM2brane}.

%
\subsection{Integrability for the quivers of Figures \ref{figure1xx} and \ref{fig-quiver2}}
Consider now the quiver CFT with holographic dual defined by the function $\alpha_1(z)$ in eq. (\ref{quiver4final}). Finding an exact solution to eqs. (\ref{NVE})-(\ref{AB}) is  challenging. We may attempt to rewrite eq. (\ref{NVE}) by redefining $x(\tau)= e^{-\frac{1}{2}\int {\cal B} d\tau} f(\tau)$, leading to an equation in the Schr\"odinger form,
\begin{equation}
f''(\tau) + V(\tau) f(\tau)=0, \;\;\;\; V(\tau)= {\cal A}-\frac{1}{4}{\cal B}^2-\frac{1}{2}{\cal B}'|_{z_{sol}}.\label{schroedingerform}
\end{equation}
To solve exactly this last equation is a daunting task. Nevertheless, we can simplify matters if we study the problem very close to $z\sim \tau\sim 0$, that is for short times. Indeed, choosing $E=4\pi$ and $\nu=1$ to avoid cluttered expressions we find for a series expansion in $\tau$,
\begin{eqnarray}
 {\cal A}\sim \gamma_1 -\gamma_2 \tau^2,\;\;\;\; {\cal B}\sim \frac{\gamma_4}{\tau}-\gamma_3 \tau.\label{expansiones}
\end{eqnarray}
The explicit expression of the coefficients $\gamma_1,\gamma_2,\gamma_3,\gamma_4$ is not important for this analysis. Using the leading terms in eq. (\ref{expansiones}), the differential eqs. (\ref{NVE}),(\ref{schroedingerform}) admit Liouvillian solutions. Nevertheless, when the  subleading terms are included, both equations present solutions that contain Hermite polynomials,  and hypergeometric functions ${}_1F_1$. This implies the non-Liouvillian character of the solution, indicating non-integrability of the string soliton of eq. (\ref{solitonxx}).  

A more refined study is presented in Appendix \ref{kovalg}. We change variables to have an NVE with rational coefficients. The necessary conditions for this NVE to admit Liouvillian solutions are not satisfied---the details are given below eq. (\ref{calacalb}).
This translates into the non-integrability of the ${\cal N}=(1,0)$ SCFT described by the quiver in Figure \ref{figure1xx}. The same can be concluded about any background whose defining function $\alpha(z)$ starts as $\alpha(z)\sim a_1 z + a_3 z^3$ close to $z=0$. 

The non-Liouvillian integrability can also be studied numerically. In what follows, we provide a detailed numerical analysis of different observables that suggest 
that the system of equations (\ref{eqsEL}),(\ref{hamiltoneqs}) is non-Liouvillian and  chaotic.

\section{Numerical Analysis}\label{sec:numerics}
In this section, we carry out some explicit numerical computations that provide a solid back-up to our findings of  \textit{analytic} non-integrability associated with $ \mathcal{N}=(1,0) $ SCFTs in six dimensions. We study the dynamics of  classical strings on the backgrounds in eqs. (\ref{backgroundads7xm3})-(\ref{alphathird}). We demonstrate that the phase space dynamics of classical strings on these backgrounds is  chaotic (and  hence non-integrable). Let us mention that non-integrability does not necessarily imply chaos. However, as far as the Gauge/Gravity duality is concerned, all the examples encountered, that have been found to be non-integrable were also chaotic in general. 

The evolution of a dynamical system is given by a set of deterministic differential equations that allows us to calculate the state of a system at a time $t$, knowing an earlier state of the system at some initial time $t_0$. A dynamical system is said to be chaotic when it is exponentially sensitive to its initial conditions, making it practically impossible to accurately predict the long term dynamical behaviour. Indeed, when we have two adjacent initial conditions $x_1(t_0)$ and $x_2(t_0)= x_1(t_0) + \epsilon$, we say the system exhibits chaotic dynamics when $|x_1(t) - x_2(t)| \sim e^{\lambda t}$, provided that the trajectory of our system in phase-space remains \emph{bounded}. This boundedness of the trajectories is to rule out the trivial case where the trajectories move off to infinity and only diverge exponentially because they are moving apart \cite{Ott:2002book}. 

In our case we are studying the motion of classical strings that sit at the centre of $AdS_7$ spacetime, while moving and rotating in an internal space of the form $\mathbb{R}\times S^2$. This is described by the system in eqs. (\ref{eqsEL})-(\ref{virasoro}) or their analog (\ref{hamiltoneqs}). The coordinates $z,\chi$ are bounded and the respective momenta along the $p_z$ and $p_\chi$-directions are bounded also due to the conserved Hamiltonian in eq. (\ref{hamiltonian}).

The trajectory of this string embedding in the phase space will therefore  be bounded if the $z$-coordinate itself is bounded. In this case, the Lyapunov exponent (that measures the exponential divergence of initial conditions) indeed provides a good observable to determine whether the dynamics of this classical string embedding is chaotic or not.

The numerical analysis that follows is quite dense. The plan is the following: first, in Section \ref{subsec_NumericalEvolution}, we examine the motion of classical strings over the background solutions  (\ref{quiver4final})  and (\ref{quiver2final}), by numerically evaluating the equations of motion. We calculate the corresponding power spectra and discuss how this is indicative of chaotic dynamics. In Section \ref{subsec_Lyapunov}, we explore the Lyapunov spectrum \cite{Lyapunov}, demonstrating that the dynamical behaviour is indeed chaotic. We end the analysis in Section \ref{subsec_PoincareSect}, discussing the Poincar\'{e} sections of these solutions and their implications on the chaotic dynamics associated to classical string configurations considered in this paper. Appendix \ref{appendixlyapunov} complements the numerical analysis with some rigorous definitions.

\subsection{Numerical Evolution and Power Spectra}\label{subsec_NumericalEvolution}
The equations of motion for the classical strings (\ref{eqsEL})-(\ref{virasoro}) are considerably simplified  with the choice $\ddot{\chi} = \dot{\chi} = \chi =0$ (or $\chi = \pi$), i.e. when the string stays fixed at the north or the south pole of the 2-sphere. With this choice, the remaining equation for the motion along $z$ is,
\begin{equation}\label{eq_chiToZeroLim}
2 f_2 (z) \ddot{z} + \dot{z}^2 f_2'(z) + \frac{f_1'(z)}{f_1(z)^2}E^2 = 0, 
\end{equation}
which is eq. (\ref{llmm}).

\begin{figure}[h!]
{
 \centering
 \subfloat[\small $z(t)$ and $p_z(t)$ in blue and yellow respectively \normalsize]{
   \label{fig_lowE_z}
     \includegraphics[width=0.5\textwidth]{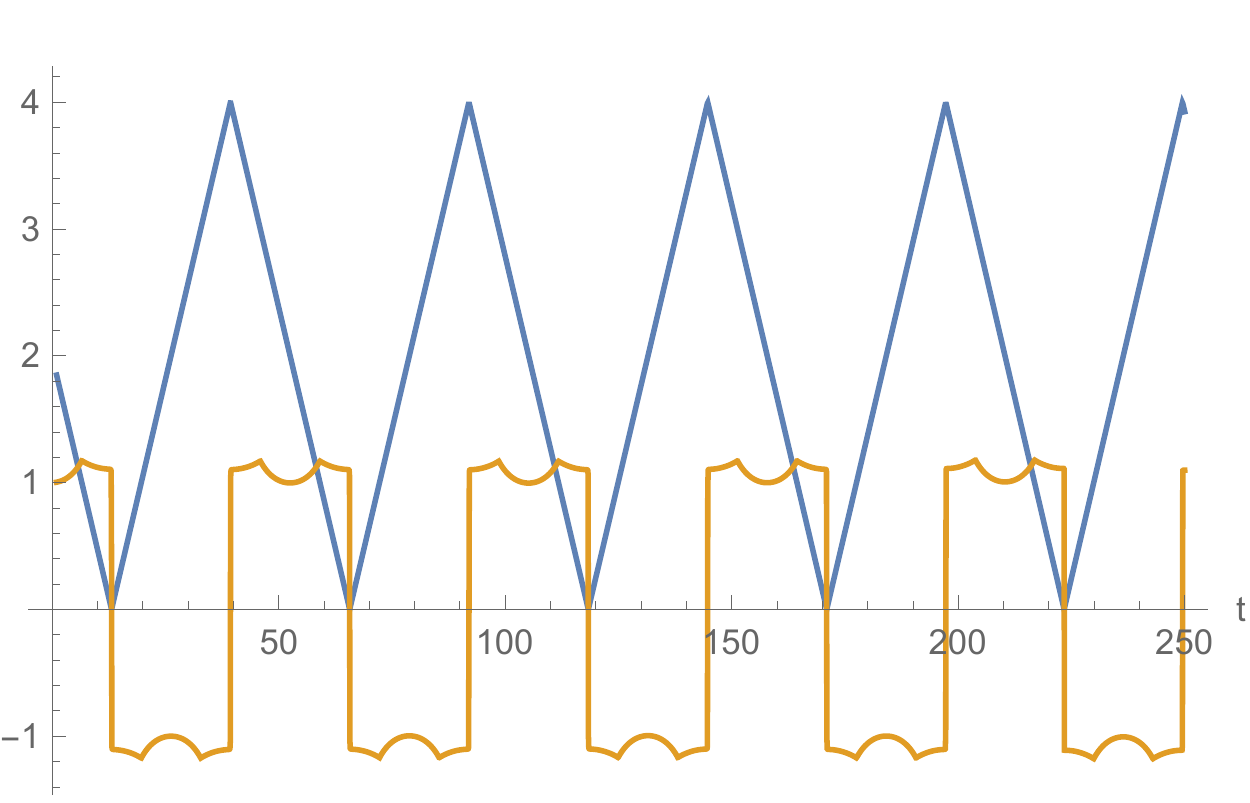}}
 \subfloat[\small $\chi(t)$ 
   \normalsize]{
   \label{fig_lowE_chi}
    \includegraphics[width=0.5\textwidth]{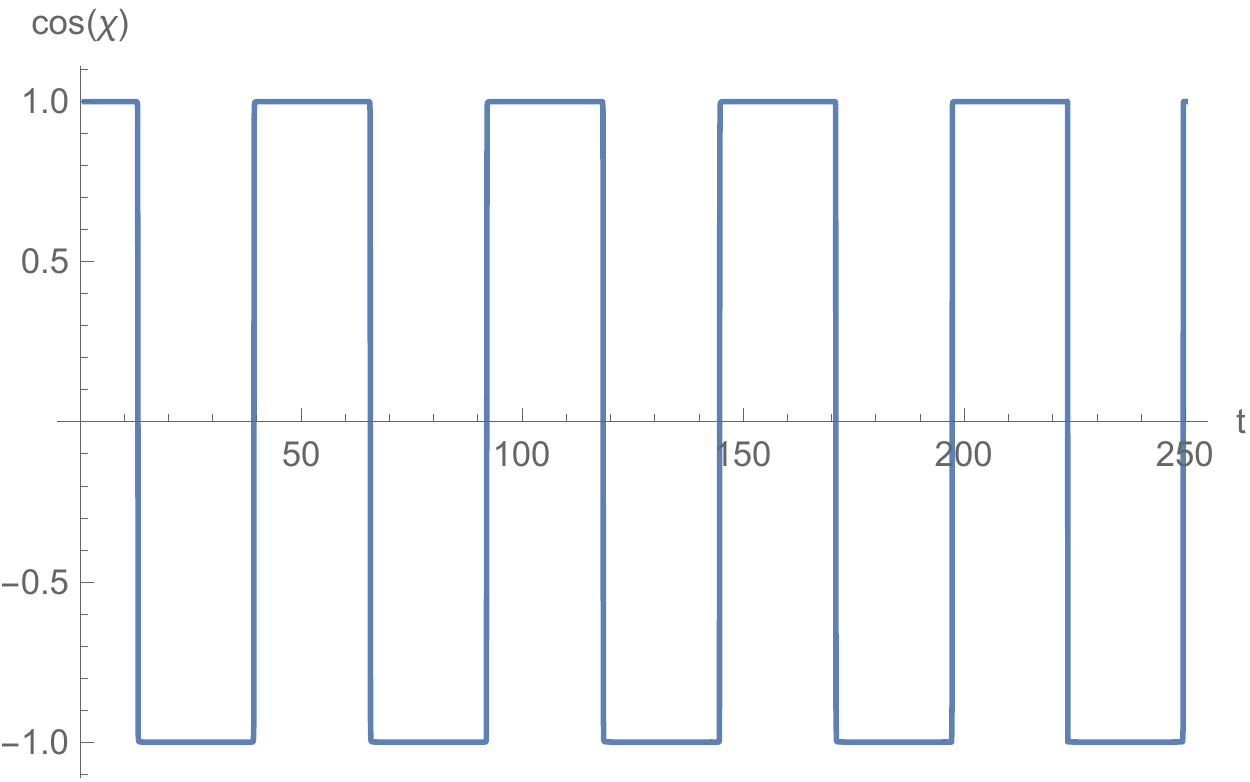}}

\caption{Numerical evolution for a string on the background solution (\ref{quiver2final}) (quiver 3) with initial conditions $\chi(0) = 0.001$, $p_\chi(0) = 0$, $z(0) = 2$ and $p_z(0) = 1$, corresponding to an energy $E \approx 3.83$.}\label{fig_lowE}
}
\end{figure}

Let us study how the classical string dynamics becomes increasingly disorganised as we allow the strings to move further away from the poles of the two-sphere. First, consider a classical string on the background solutions that were already discussed in eqs. (\ref{quiver4final}) and (\ref{quiver2final}). Below, we refer to them  as quiver 1 and quiver 3 respectively. 
In Figure \ref{fig_lowE_z}, we see that when the string stays very close to the poles of the two-sphere, it moves along the $z$-direction until it hits the end of the $z$-domain. Then,  it turns around and moves back along the $z$-direction. On the two-sphere, the string starts out located near the north pole (at, $\chi$=0.01). However, when the string turns around along the $z$-direction it moves almost instantaneously from the north-pole to the south-pole (see, Figure \ref{fig_lowE_chi}).

\begin{figure}[h!]
{
 \centering
 \subfloat[\small $\chi(0) = 0.01$, $E \approx 3.83$, $t_{\mathrm{max}} = 400$ \normalsize]{
   \label{fig_lowE_chi0001_path}
     \includegraphics[width=0.5\textwidth]{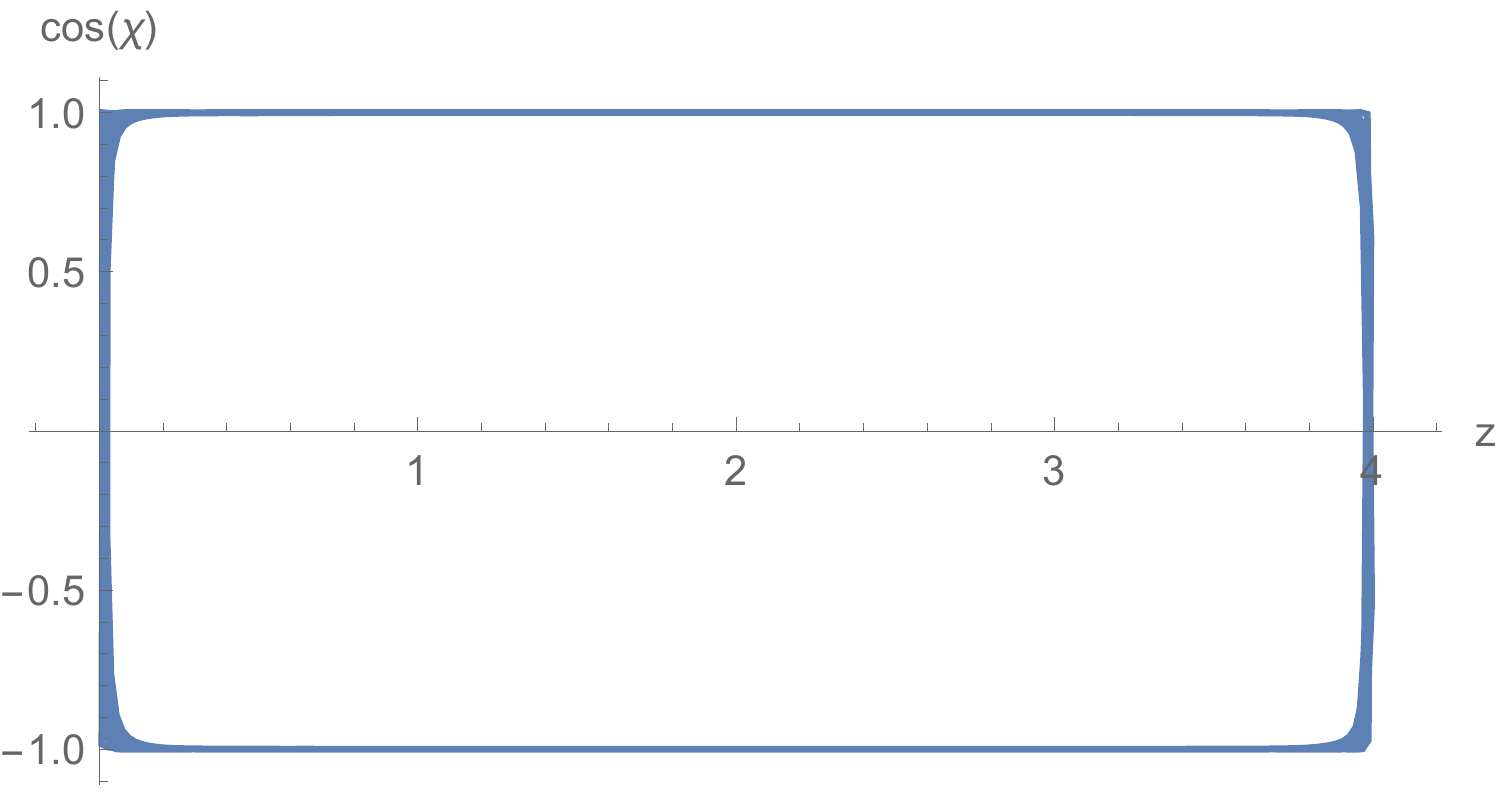}}
 \subfloat[\small $\chi(0) = 0.10$, $E \approx 6.75$, $t_{\mathrm{max}} = 400$ \normalsize]{
   \label{fig_lowE_chi01_path}
    \includegraphics[width=0.5\textwidth]{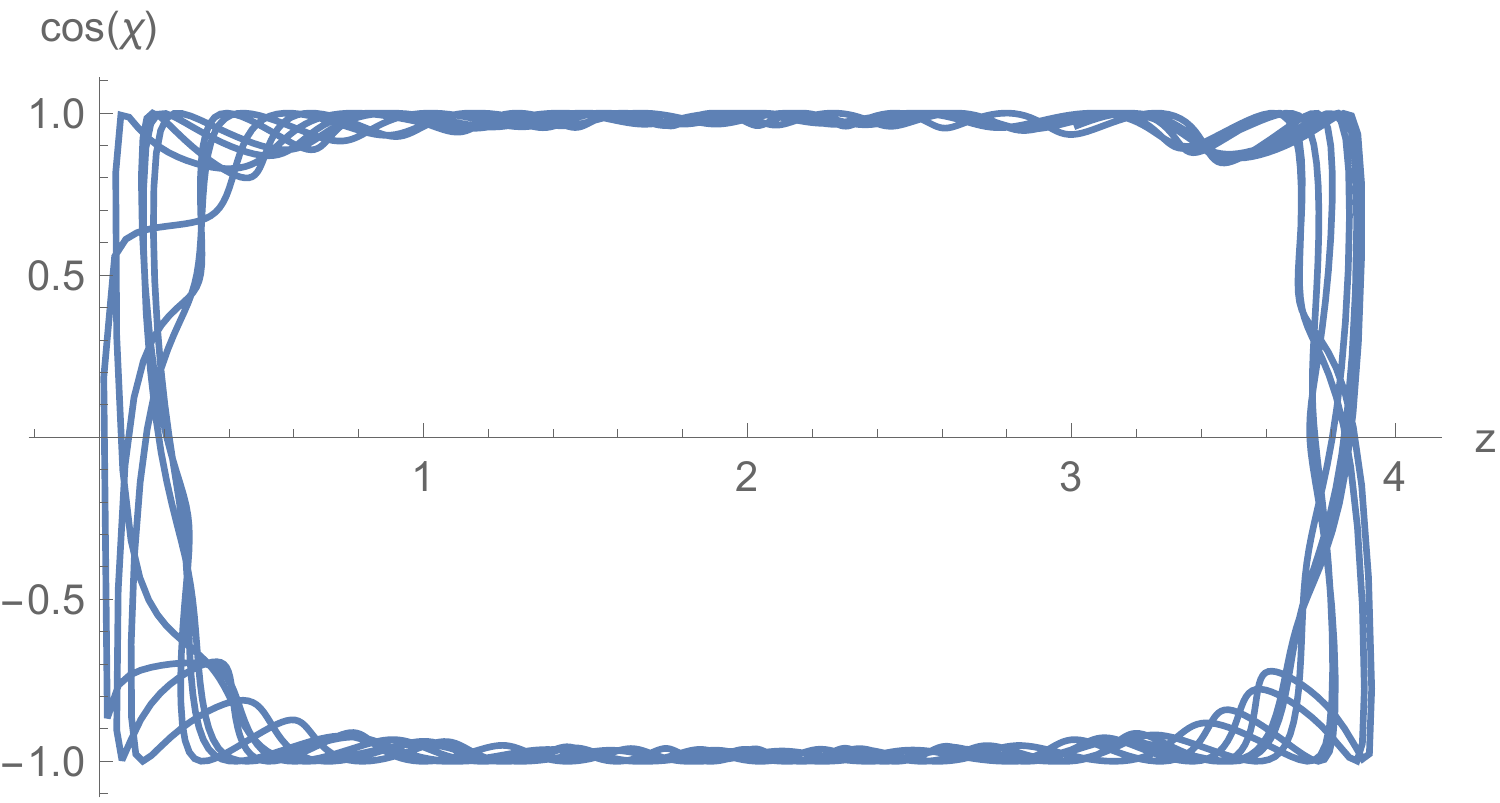}}\\    
 \centering
 \subfloat[\small $\chi(0) = 0.25$, $E \approx 14.31$, $t_{\mathrm{max}} = 400$ \normalsize]{
   \label{fig_lowE_chi025_path}
     \includegraphics[width=0.5\textwidth]{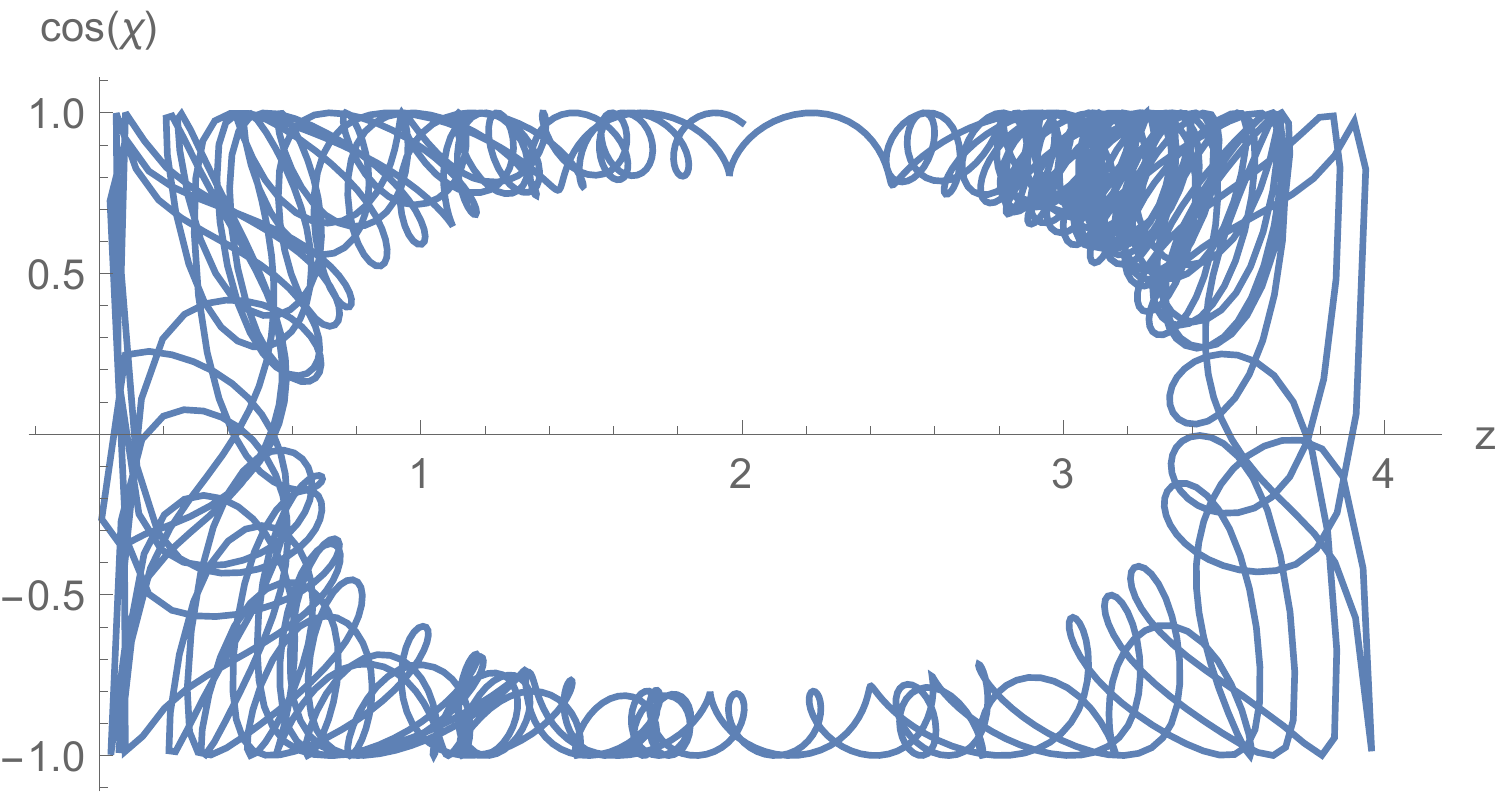}}
 \subfloat[\small $\chi(0) = 0.9$, $E \approx 43.82$, $t_{\mathrm{max}} = 100$ \normalsize]{
   \label{fig_lowE_chi05_path}
    \includegraphics[width=0.5\textwidth]{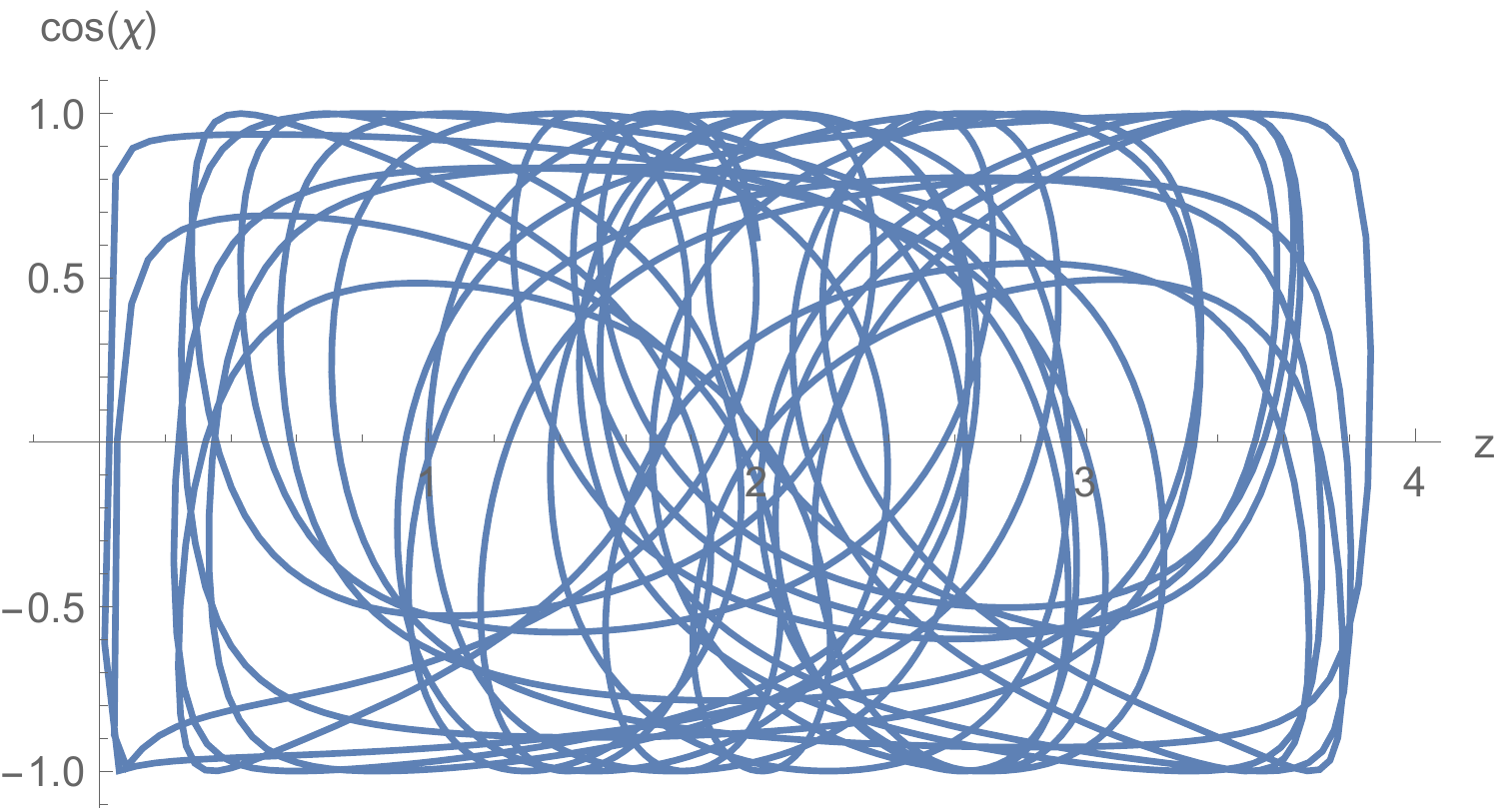}}

\caption{Different trajectories in the $(z, \chi)$-plane for a string-embedding of the form in eq. (\ref{solitonxx}) on the background in (\ref{quiver2final}), we run the evolution up to to $t = t_{\mathrm{max}}$ and only change the initial condition $\chi(0)$.}\label{fig_lowE_path}
}
\end{figure}

Now, we move the initial position of the string away from the poles (that are located at $\chi = 0$ or, $\chi = \pi$) to the middle of the two-sphere, located at $\chi(0) = \pi/2$. We keep all the other initial conditions the same. In Figure \ref{fig_lowE_chi01_path}, we consider an initial $\chi(0) = 0.1$, corresponding to $E \approx 6.75$. We see that the square shaped trajectory that the string traces out in the $(z, \chi)$-plane gets deformed. The dynamics of the system becomes more complicated as the additional $\chi$-dependence weighs in eqs. (\ref{eqsEL}). Similar conclusions could be drawn for the background in eq. (\ref{quiver4final})--quiver 1. Like in the previous example, the trajectories start looking  unstructured as we increase $\chi(0)$.

We now discuss the power spectra \cite{Ott:2002book}. By taking the Fourier transform of the numerical evolutions in Figure \ref{fig_lowE_path}, we can distinguish whether $z(t)$ and $\chi(t)$ are periodic, quasi-periodic or chaotic. When a signal is perfectly periodic with a frequency $\omega$, its Fourier spectrum will show a vertical line at the characteristic frequency of the system. 

\begin{figure}[h!]
{
 \centering
 \subfloat[\small $\chi(0) = 0.01$, $E \approx 3.83$ \normalsize]{
   \label{fig_PowerSpectrum_chi0001}
     \includegraphics[width=0.5\textwidth]{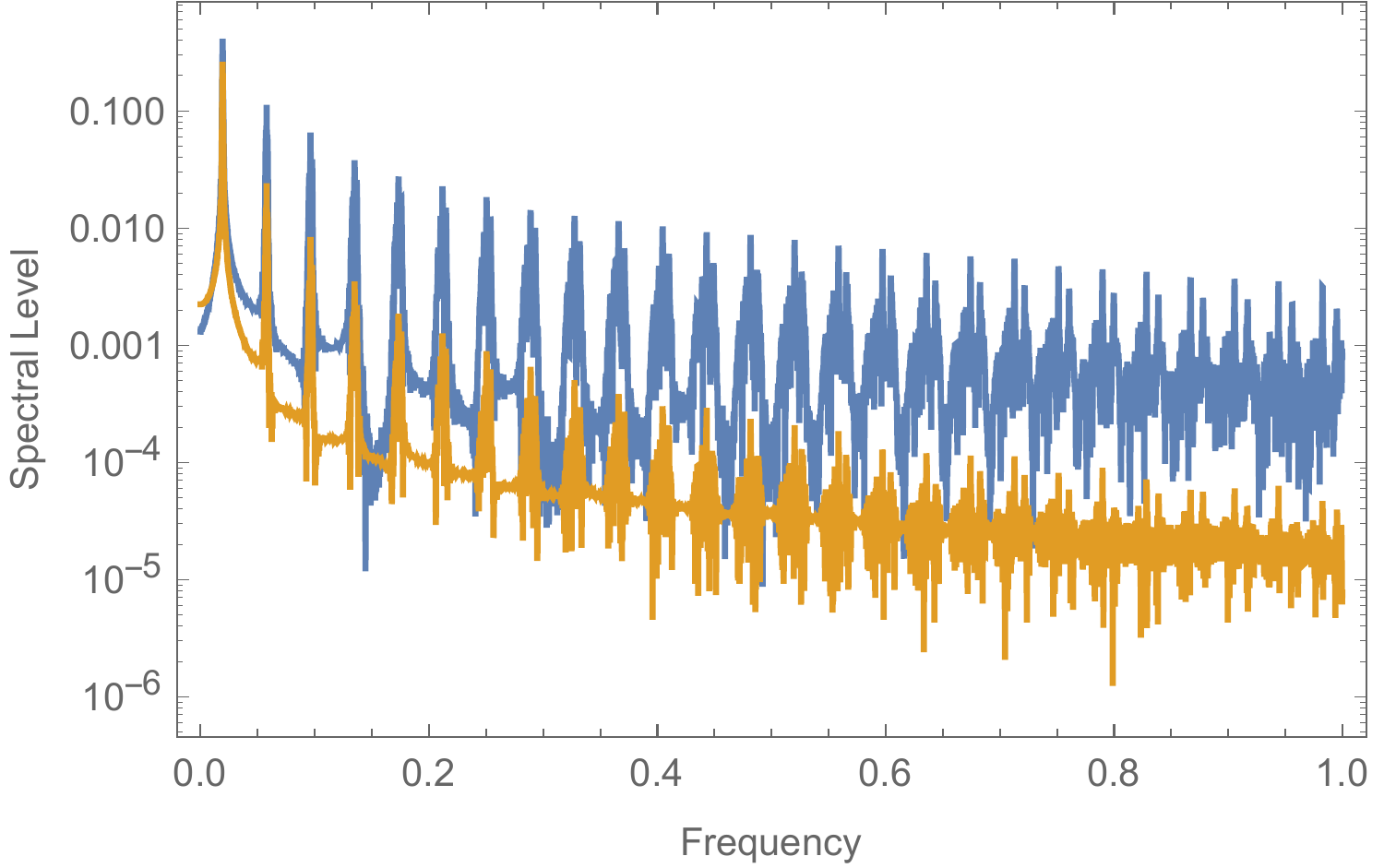}}
 \subfloat[\small $\chi(0) = 0.10$, $E \approx 6.75$ \normalsize]{
   \label{fig_PowerSpectrum_chi01}
    \includegraphics[width=0.5\textwidth]{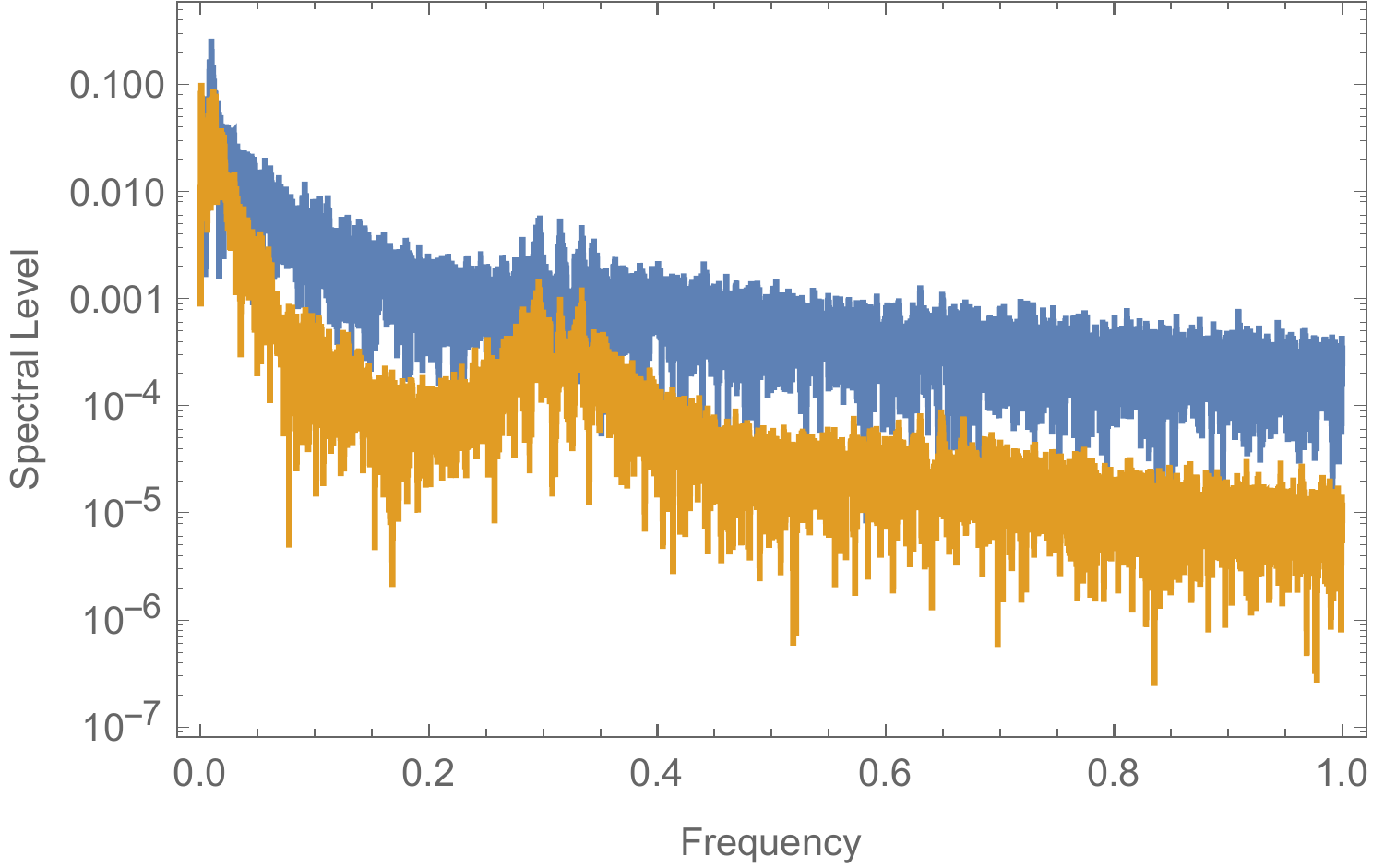}}\\    
 \centering
 \subfloat[\small $\chi(0) = 0.25$, $E \approx 14.31$ \normalsize]{
   \label{PowerSpectrum_chi025}
     \includegraphics[width=0.5\textwidth]{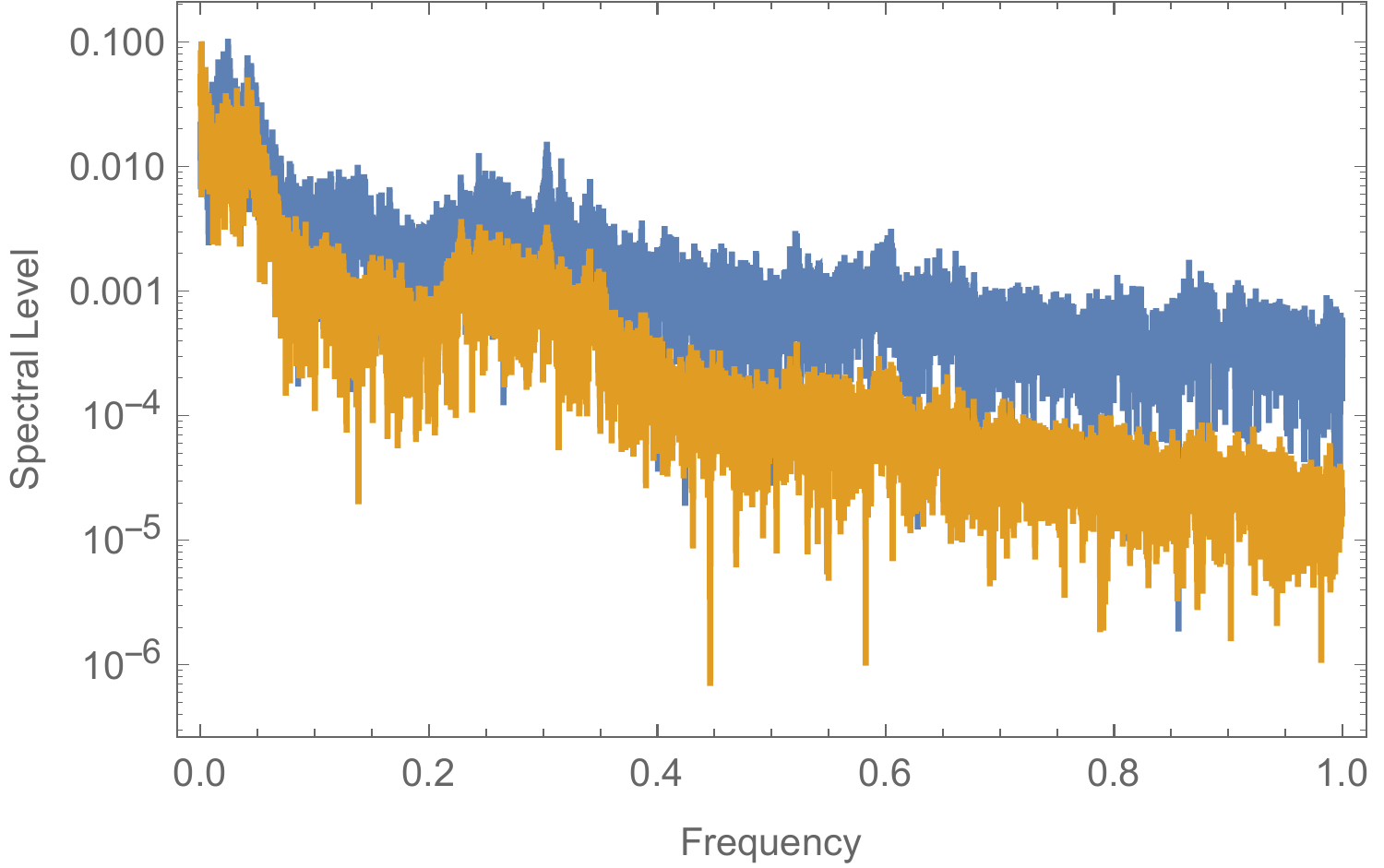}}
 \subfloat[\small $\chi(0) = 0.90$, $E \approx 43.82$ \normalsize]{
   \label{PowerSpectrum_chi05}
    \includegraphics[width=0.5\textwidth]{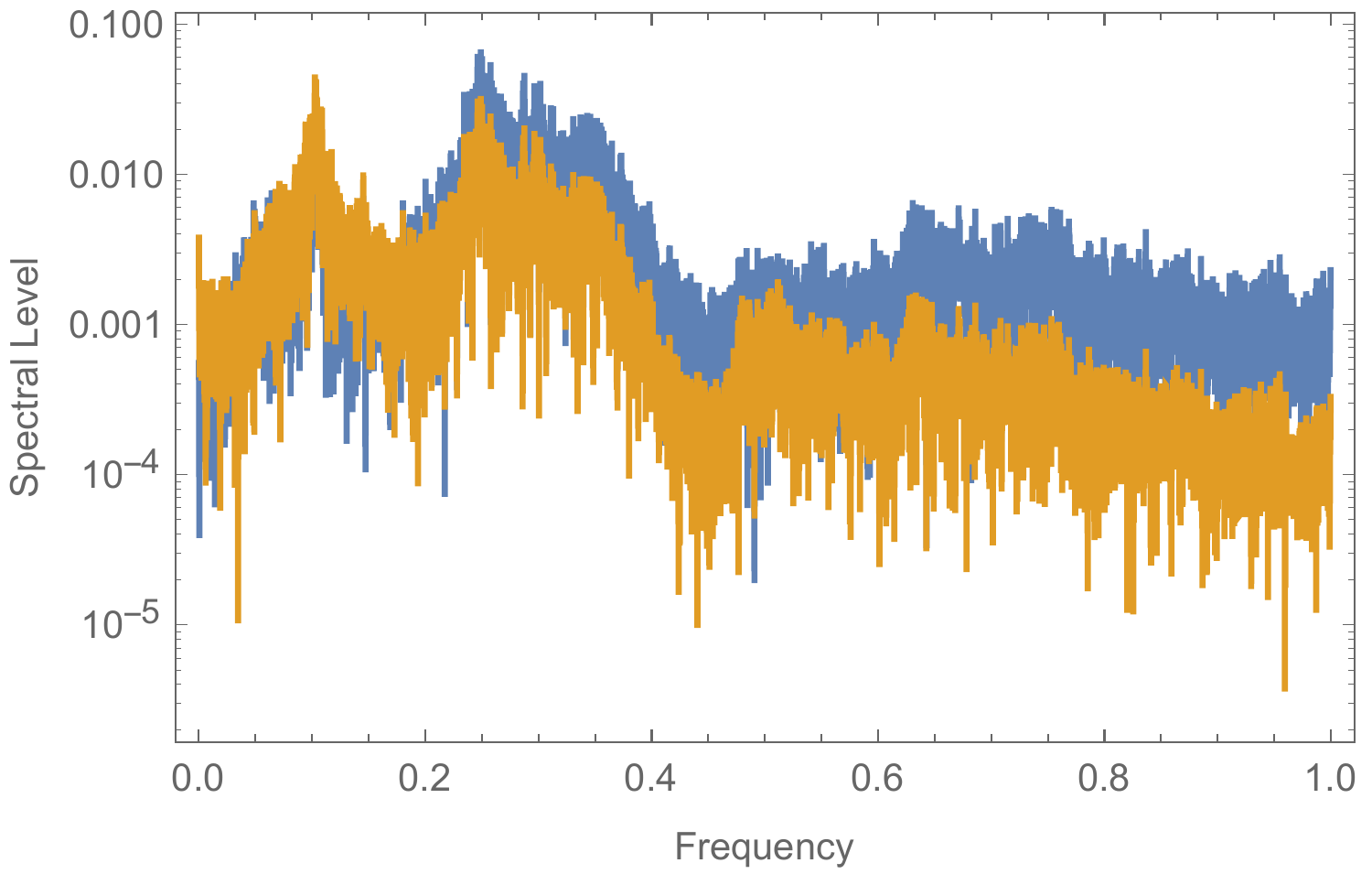}}

\caption{Power spectra for the trajectories in figures \ref{fig_lowE_chi0001_path}-\ref{fig_lowE_chi05_path}. Here the spectra for both $z(t)$ and $\chi(t)$ are shown in yellow and blue respectively. To calculate these spectra we ran a numerical evolution up to $t = 5000$ (roughly 100 oscillations along the $z$-direction), with a resolution of $10$ data-points per time-unit.}
}
\end{figure}

Notice that, for very low values of $\chi(0)$, the string moves almost periodically along the $z$-direction with a jigsaw motion (see Figure \ref{fig_lowE_z}) and along the $\chi$-direction with a square-wave profile (see Figure \ref{fig_lowE_chi}). From Figure \ref{fig_lowE_chi0001_path}, we can see that this motion is {\it not exactly} periodic, as the path of the string in the $(z, \chi)$-plane does not exactly close on itself. We can see a confirmation of this in the corresponding Fourier spectrum--- see Figure \ref{fig_PowerSpectrum_chi0001}. In fact, we clearly see a {\it fundamental frequency} of  value $0.02$ and the corresponding oscillations along the $z$-axis that have a period of roughly $55t$. In addition to this, we see the higher harmonics of the jigsaw and box shaped waveforms. The finite width of these peaks however suggests that this is not a periodic signal but that there is instead some noise present. Such a noisy power spectrum is a typical characteristic of a deterministic chaotic system.

As we increase the value of $\chi(0)$, we first see that the jigsaw and box-shaped waveforms become distorted--- see Figure \ref{fig_lowE_chi01_path}. This is reflected by the corresponding power spectrum, loosing their higher harmonics as shown in Figure \ref{fig_PowerSpectrum_chi01}.  Increasing $\chi(0)$ even further, we see that a broad band of noise around a frequency $0.35$ starts to overpower the spectrum ---see Figures \ref{fig_PowerSpectrum_chi01}-\ref{PowerSpectrum_chi025}. At even higher values of $\chi(0)$, we even loose the initial peak at frequency  $0.02$. The spectrum becomes primarily dominated by noise.

These plots and its analysis have been done for the background  in eq. (\ref{quiver2final})---quiver 3. For the background defined by eq. (\ref{quiver4final}), we see exactly the same qualitative behaviour, for roughly the same values of $\chi(0)$. 

\subsection{The Lyapunov Spectrum}\label{subsec_Lyapunov}
The Lyapunov spectrum is generally introduced as a measure of the dynamical information loss in a chaotic system. This loss of information is what sources  the dynamical Kolmogorov-Sinai (KS) entropy production within a chaotic system. Typically, for dynamical systems with a non-zero Lyapunov, the time evolution associated with two nearby trajectories in the phase space turns out to be highly sensitive to a tiny change in the initial conditions that is eventually amplified exponentially at sufficiently late times. In other words, the existence of a non-zero Lyapunov exponent,  for a point $X=(q,p)$ in the phase space with initial condition $X_0=(q(t=0),p(t=0))$ is,
\begin{eqnarray}
\lambda = \lim_{\tau \rightarrow \infty}\lim_{\Delta X_0 \rightarrow 0}\frac{1}{\tau}\log \frac{\Delta X (X_{0}, \tau)}{\Delta X (X_0 ,0)}
\end{eqnarray}
is intimately related to the degree of randomness associated with the dynamical phase space of a Hamiltonian system.
It is  typically introduced as a quantitative measure of the rate of separation between two infinitesimally close trajectories in the phase space. The function $ \Delta X (X_0 , \tau)$ measures the separation between two  infinitesimally close trajectories (at very late times) as a function of this initial location. Typically, for chaotic systems, one ends up with
\begin{eqnarray}
\Delta X (X_{0}, \tau) \sim \Delta X (X_{0}, 0)e^{\lambda \tau}.
\end{eqnarray}
\begin{figure}[h!]
{
 \centering
 \subfloat[\small  LCE massless: $t=0, z=0.05, \chi=0.05, p_t=0.5, p_z=4.60155, p_\chi=0.01$. Integration: $\tau=0.1, K=500, T=0.009$ \normalsize]{
   \label{LCE massless}
     \includegraphics[width=0.5\textwidth]{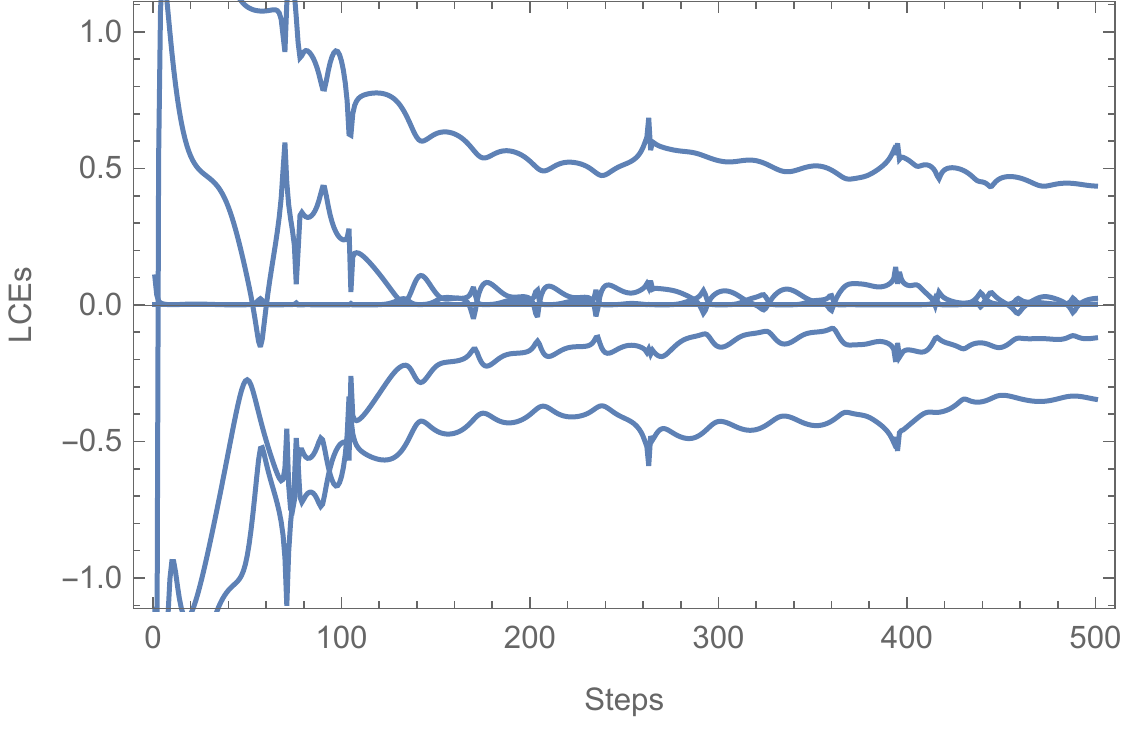}}
 \subfloat[\small LCE quiver 1:  $t=0, z=0.15, \chi=0.15, p_t=15, p_z=0.845631, p_\chi=0.01$. Integration: $\tau=0.1, K=150, T=0.01$ \normalsize]{
   \label{LCE quiver 1}
    \includegraphics[width=0.5\textwidth]{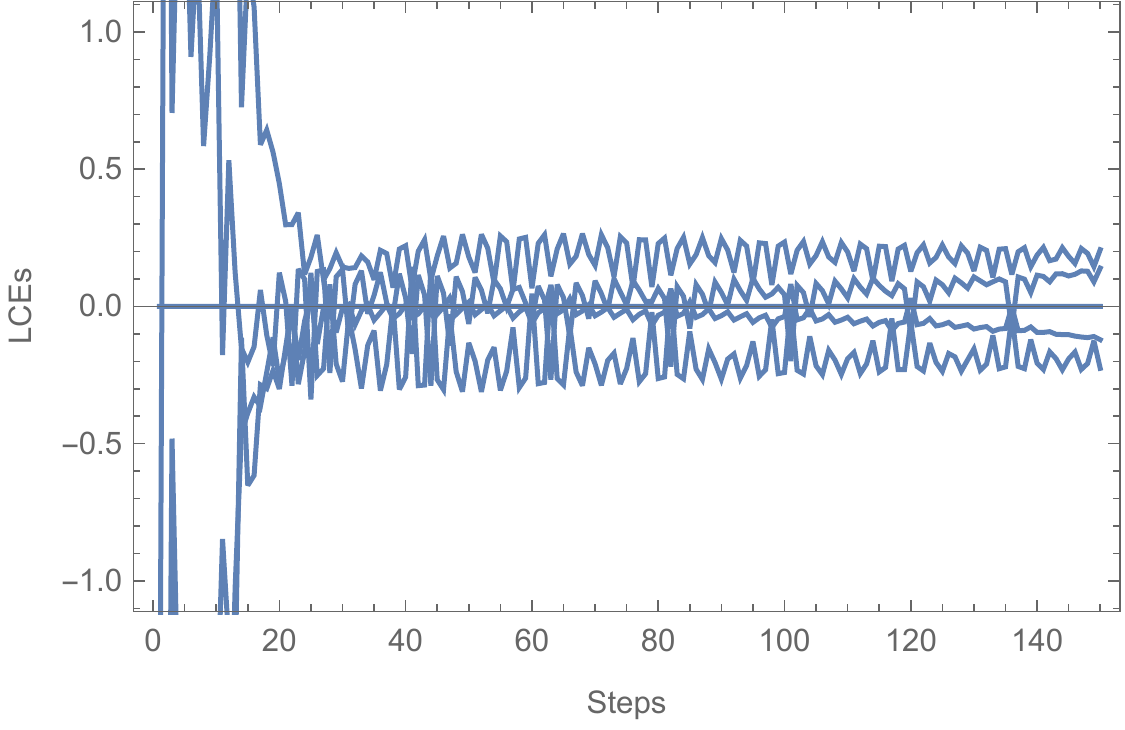}}\\
\subfloat[\small LCE quiver 2  $t=0, z=0.09, \chi=0.09, p_t=15, p_z=0.491105, p_\chi=0.22$. Integration: $\tau=0.5, K=200, T=0.02$ \normalsize]{
   \label{LCE quiver 2}
    \includegraphics[width=0.5\textwidth]{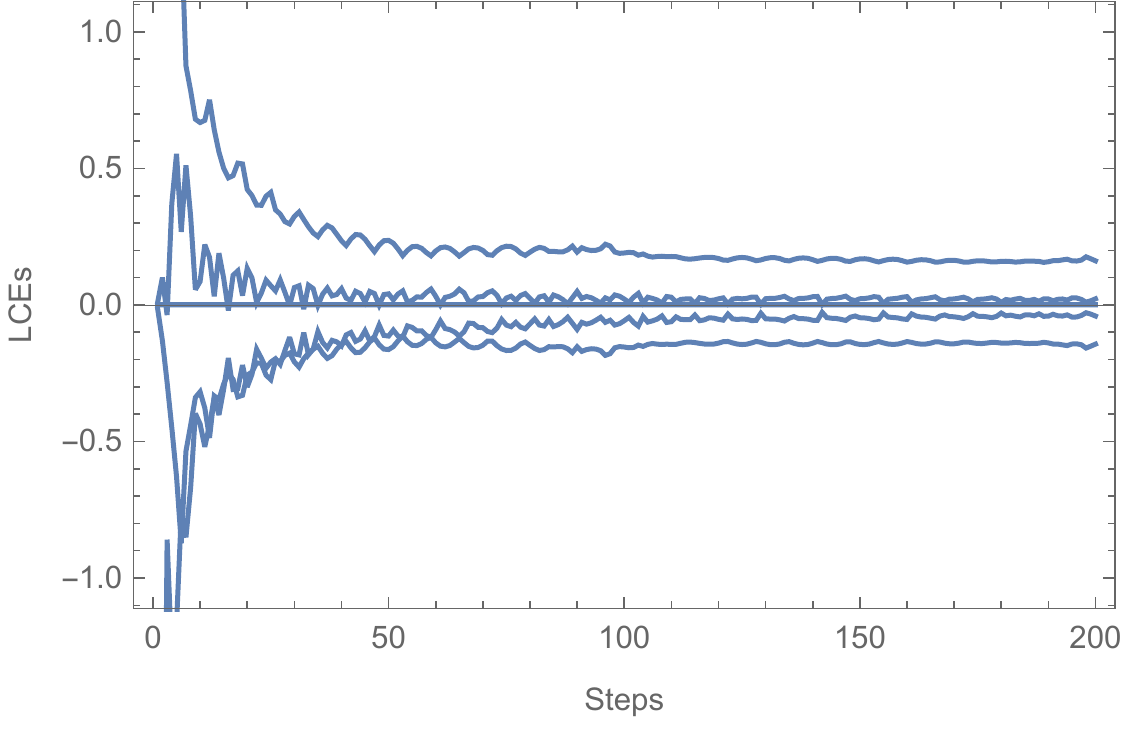}}
 \subfloat[\small LCE quiver 3  $t=0, z=0.09, \chi=0.09, p_t=9, p_z=0.821405, p_\chi=0.22$. Integration: $\tau=0.5, K=200, T=0.02$ \normalsize]{
   \label{LCE quiver 3}
    \includegraphics[width=0.5\textwidth]{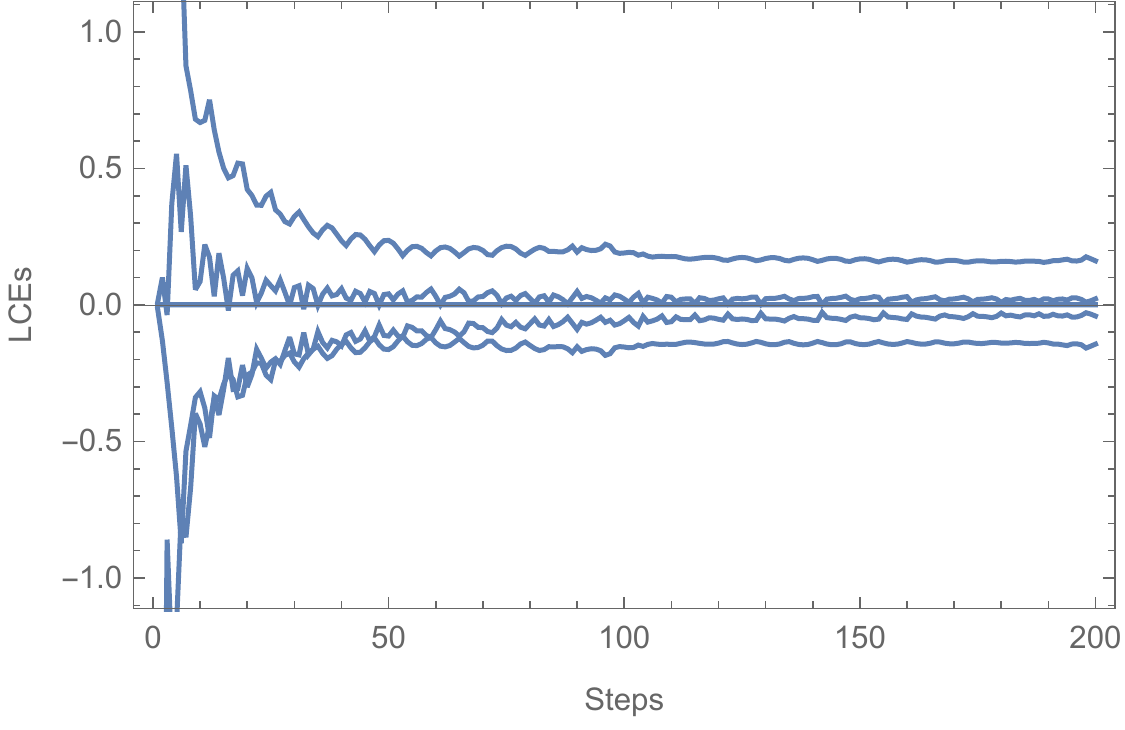}}\\  
\caption{LCEs for different quivers }
}
\end{figure}

Below, we provide a detailed analysis of the computation of the Lyapunov exponents \cite{Lyapunov} corresponding to various background solutions, characterised by a function $ \alpha(z) $.

We will also be interested in the {\it massless background solution} described in  eq. (\ref{alphamassless}) and in the solution corresponding to the quiver that {\it never ends } described by eq. (\ref{quiver3final})---without the `closure'. We denote this last one as quiver 2 in the plots below. We set $ P=10 $  to be the length of the quiver for the purposes of the numerical analysis.

The computation of the Lyapunov exponents is solely based on the prescription of \cite{Lyapunov}\footnote{The details of the numerical techniques and the precise definition of $K$ and $T$ are provided in the Appendix \ref{appendixlyapunov}.}. The initial conditions are fixed to satisfy the Hamiltonian constraint, $ H=0 $. This in turn implies that for a $ 2N $ dimensional phase space,
\begin{eqnarray}
\sum_{i=1}^{2N}\lambda_{i}=0
\end{eqnarray}
where we denote by $ \lambda_{i} $ the $i$-th {\it Lyapunov Characteristic Exponent} (LCE). It  is the measure of the exponential growth associated to the $ i $th direction in the phase space. For some  systems, the sum of all positive Lyapunov exponents measures the KS entropy production during the dynamic evolution in the phase space.


We substitute some appropriate initial conditions into the dynamical equations in order to generate a solution. Choosing  these initial conditions for the phase space variables to satisfy the vanishing of the Hamiltonian, we find the corresponding Lyapunov spectrum for each of the quiver configurations, which clearly have non-zero LCEs---see Figures \ref{LCE massless}-\ref{LCE quiver 3}. Notice that, in our analysis, we are eventually computing four LCEs ($ \lambda_{i} $) that characterise a four-dimensional dynamical phase space. The addition of all them gives a vanishing number as it corresponds to a Hamiltonian system. At this stage, it is worthwhile to mention an interesting limit associated with the massless solution mentioned above. This is the limit in which we set the parameter $ R_{0}\rightarrow \infty $, which as can be seen from eq. (\ref{nvemassless}) yields the NVE,
\begin{equation}
\ddot{x}(\tau)+\nu^2 x(\tau)=0
\end{equation}
 which is therefore trivially integrable.

\begin{figure}[htb!]
\centering
\includegraphics[width=200pt]
{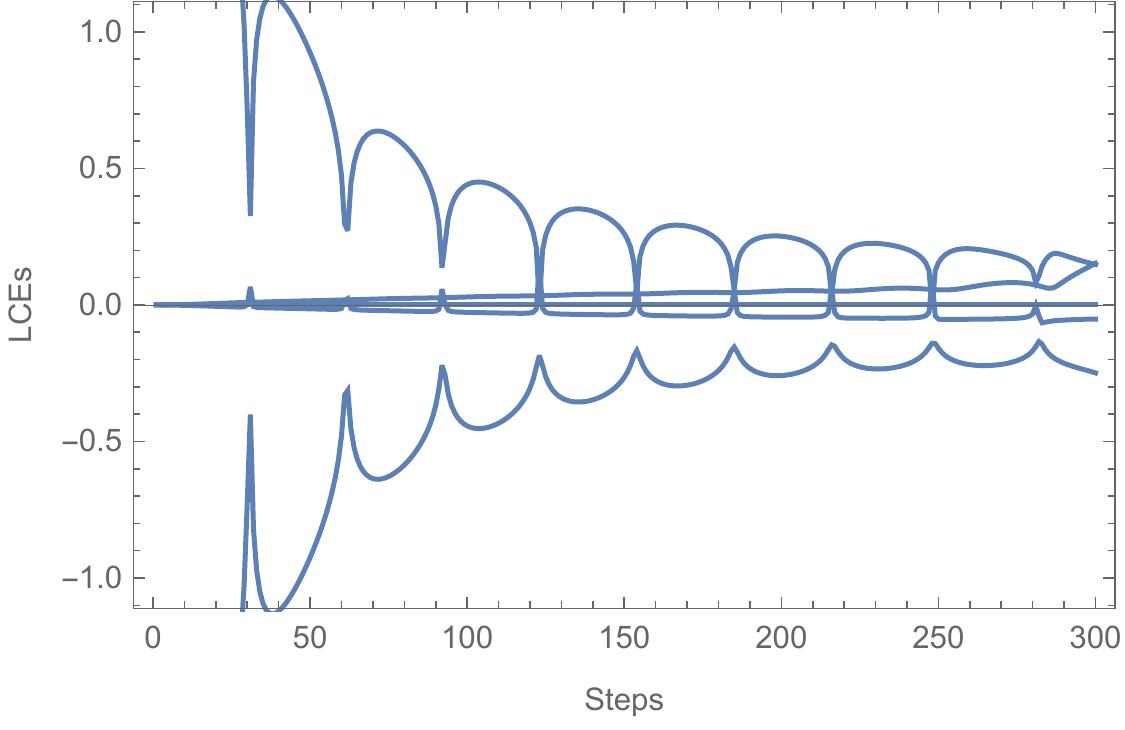}
\caption{LCEs for the massless solution with large $ R_{0} $.}\label{LCE massless2}
\end{figure}
To perform the corresponding (numerical) computation on the LCEs, we choose the  parameters
\begin{equation}
k=1, ~R_0=500,~ \nu=1
\end{equation}
taking as initial conditions for the phase space variables:
\begin{equation}
t=0,~ z=0.05, ~\chi=0.05,~ p_t=100,~ p_z=0.159689,~p_\chi=0.01
\end{equation}
that satisfy the vanishing of the Hamiltonian. From the plot in Figure \ref{LCE massless2}, it is easy to notice that the largest LCE is substantially reduced for large enough times.

\subsection{Poincar\'e Sections and Non-Integrability}\label{subsec_PoincareSect}
An $N$-dimensional integrable system possesses $N$ independent integrals of motion that are in \emph{involution}, namely, the Poisson bracket of any two of these conserved quantities vanishes. As a consequence of this, the corresponding phase space trajectories are confined to the surface of an $N$-dimensional KAM torus \cite{Ott:2002book}. When we change our variables to action-angle variables $(q_i, p_i) \to (\phi_i, J_i)$, such that our Hamiltonian only depends on $J_i$, the corresponding trajectories on this KAM torus are completely specified in terms of $N$ frequencies $(\omega_i)$ that specify the velocities along the different angles on this torus. When there is no set of integers $n^i$ such that $\omega_i n^i = 0$, these trajectories are said to be \emph{quasi-periodic}, they do not close on themselves but fill the surface of a KAM torus. 

As a consequence of this, we can see whether a system is integrable or not, by taking cross-sections of its phase-space trajectories. When we plot for example $(\phi_1, J_1)$ every time $\phi_2=0$ we will see a large number of foliated circular KAM curves associated with the 2-dimensional cross-sections of these foliated KAM tori. Such a cross-section is known as a Poincar\'{e} section \cite{Ott:2002book}. The KAM theorem tells us how these KAM curves will change when we perturb an integrable Hamiltonian with a small deformation $\epsilon H'$, where $\epsilon \ll 1$. The resonant tori - for which these trajectories close on themselves, will be destroyed by this perturbation. However, a large number of these non-resonant KAM tori will survive. As we continue to increase the strength of this perturbation, more and more of these tori are destroyed until the motion becomes seemingly random and we loose all of the KAM curves in our Poincar\'{e} section.\\

\begin{figure}[h!]
{
 \centering
 \subfloat[\small $E = 3$ \normalsize]{
   \label{fig_lowE_chi0001_Poincare}
     \includegraphics[width=0.5\textwidth]{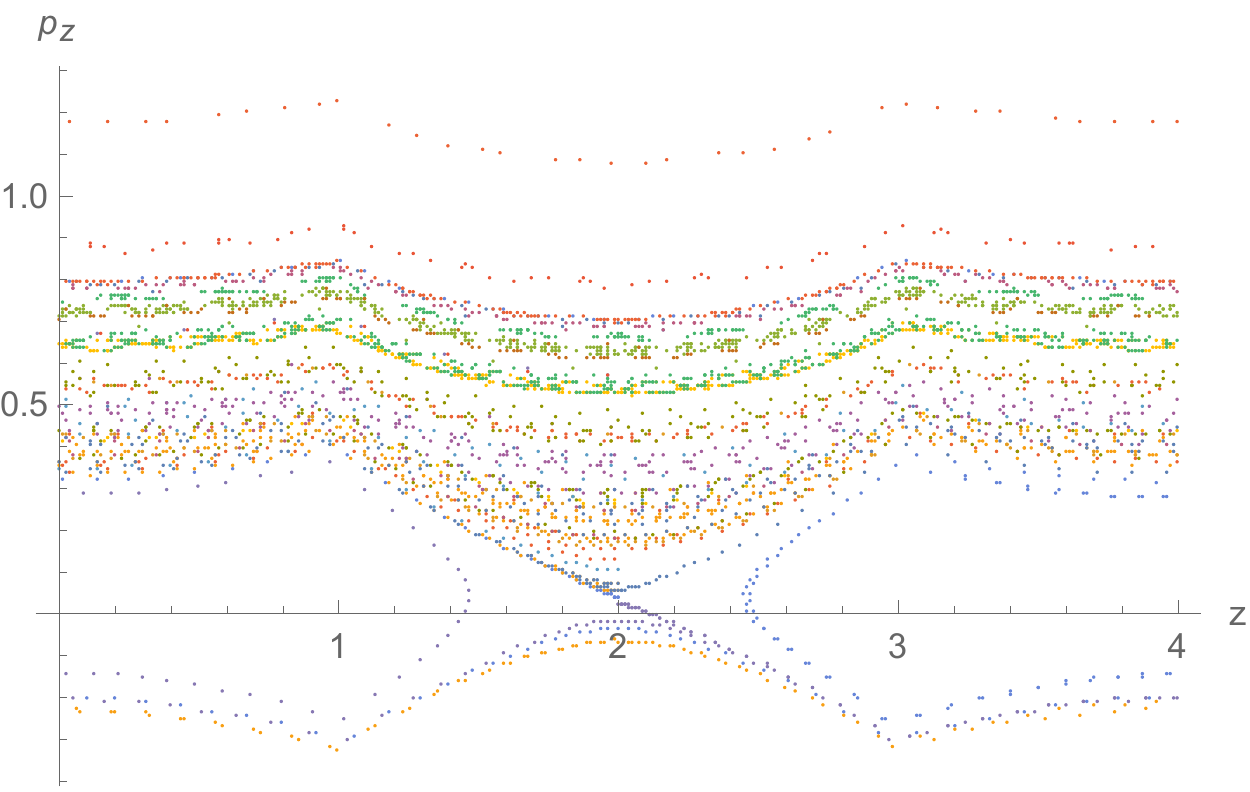}}
 \subfloat[\small $E = 5$ \normalsize]{
   \label{fig_lowE_chi01_Poincare}
    \includegraphics[width=0.5\textwidth]{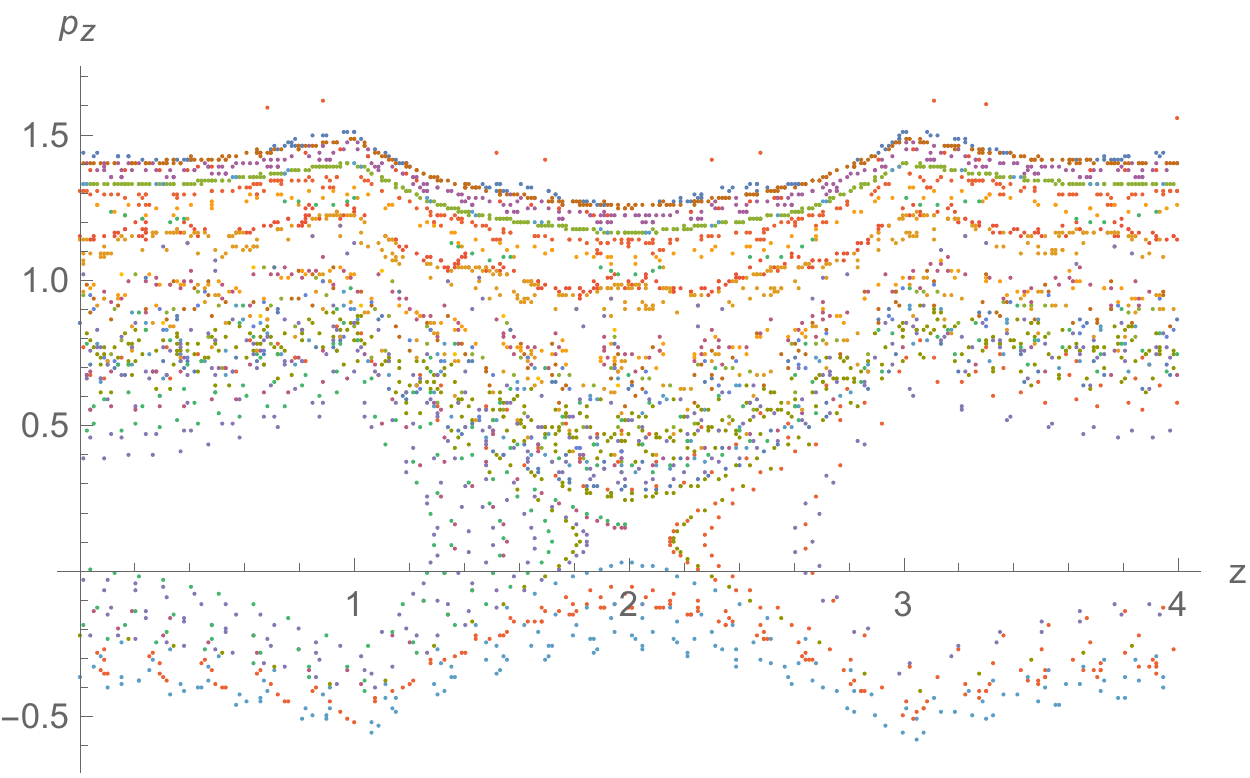}}\\    
 \centering
 \subfloat[\small $E = 10$ \normalsize]{
   \label{fig_lowE_chi025_Poincare}
     \includegraphics[width=0.5\textwidth]{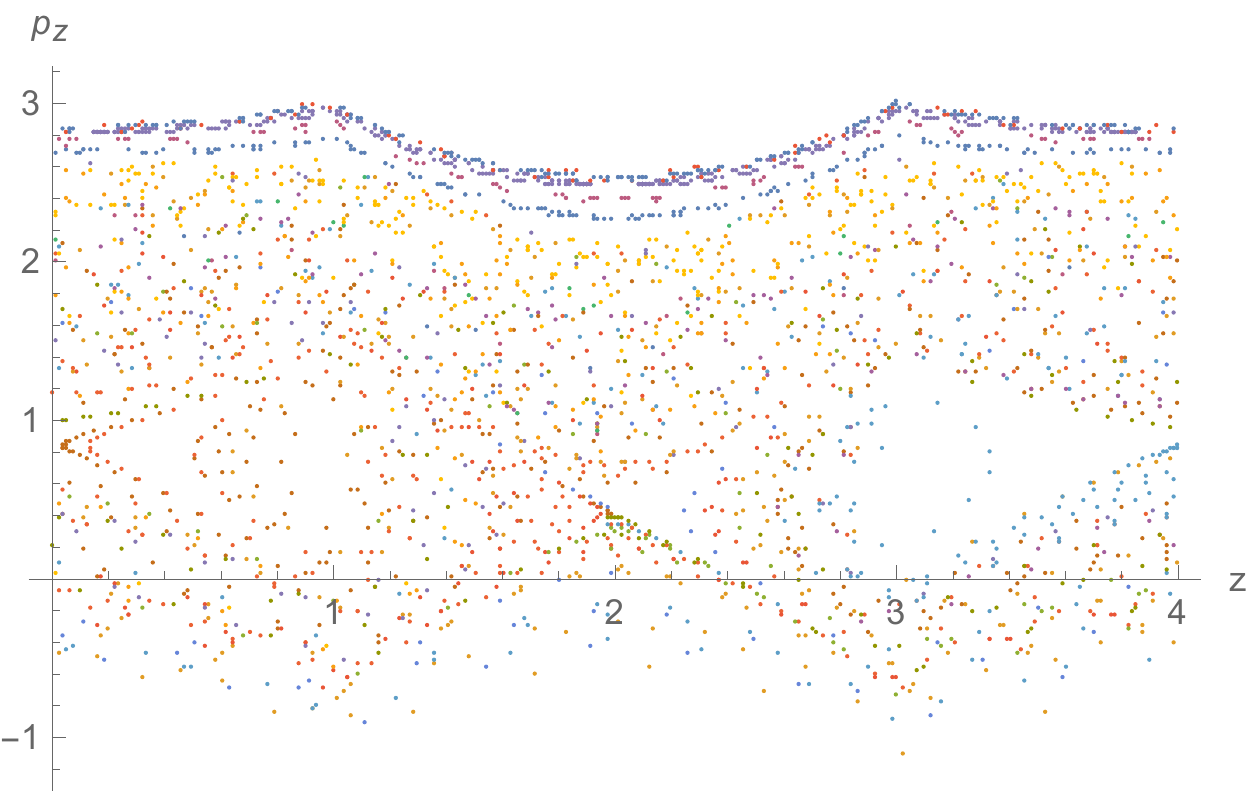}}
 \subfloat[\small $E = 15$ \normalsize]{
   \label{fig_lowE_chi05_Poincare}
    \includegraphics[width=0.5\textwidth]{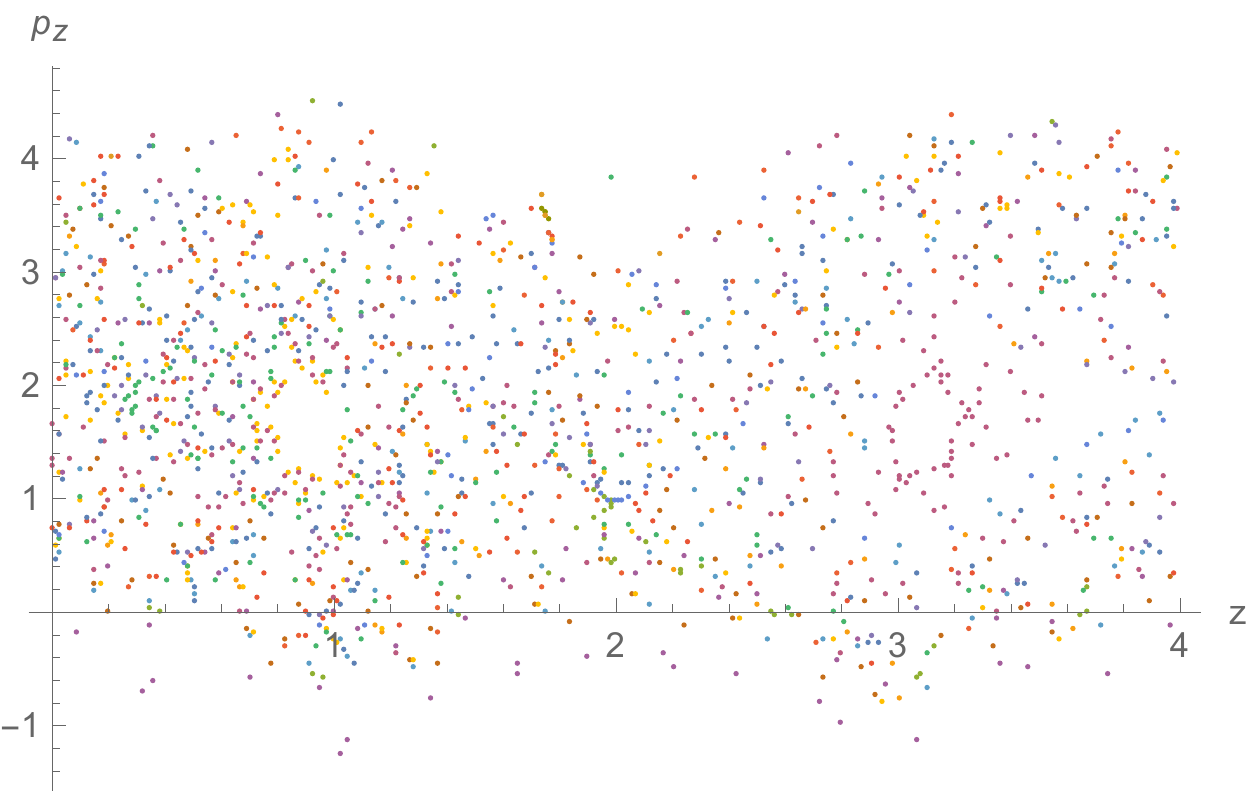}}

\caption{Poincar\'{e} sections for the $(z, p_z)$-plane at $\chi(t)=0$, for the quiver in eq. (\ref{quiver2final}) at different energies.}\label{fig_PoincareQuiver1z}
}
\end{figure}

In order to generate Poincar\'{e} sections for the background solutions in eqs. (\ref{quiver4final}) and (\ref{quiver2final}) we first choose a set of different initial conditions, all having the same energy $E$. We do this by setting, $z(0) = 2$, $p_\chi(0) = 0$, and varying $p_z(0) \in [0, 10]$ and $\chi(0)$ in such a way that the Virasoro constraint in eq. (\ref{virasoro}) is always satisfied for a given value of the energy. We then run the numerical evolution for these initial points, and plot the points $(z, p_z)$ every time $\chi(t) = 0$---see Figure \ref{fig_PoincareQuiver1z}. 

\begin{figure}[h!]
{
 \centering
 \subfloat[\small $E = 15$ \normalsize]{
   \label{Run6_pt15}
     \includegraphics[width=0.5\textwidth]{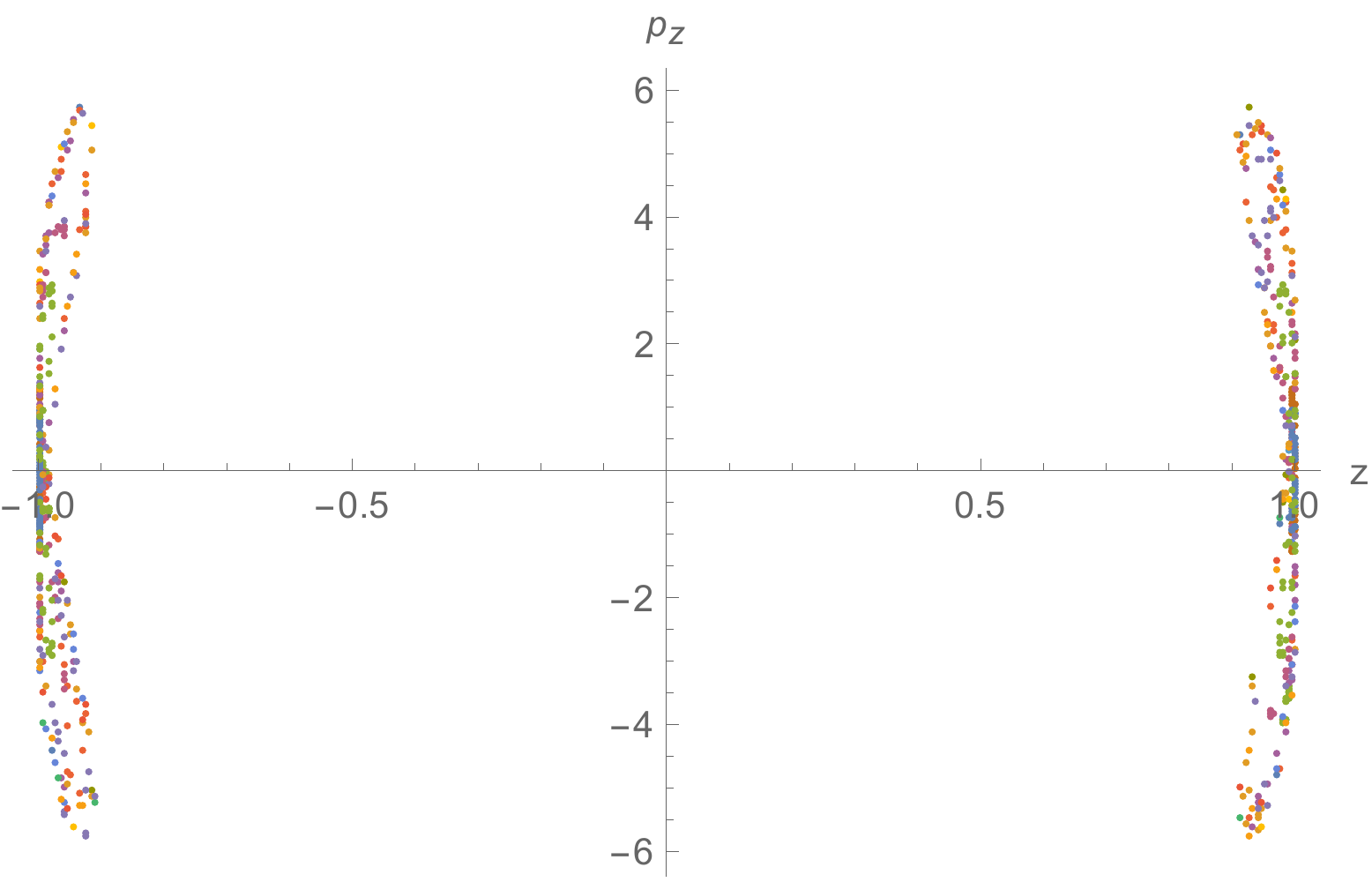}}
 \subfloat[\small $E = 35$ \normalsize]{
   \label{Run6_pt35}
    \includegraphics[width=0.5\textwidth]{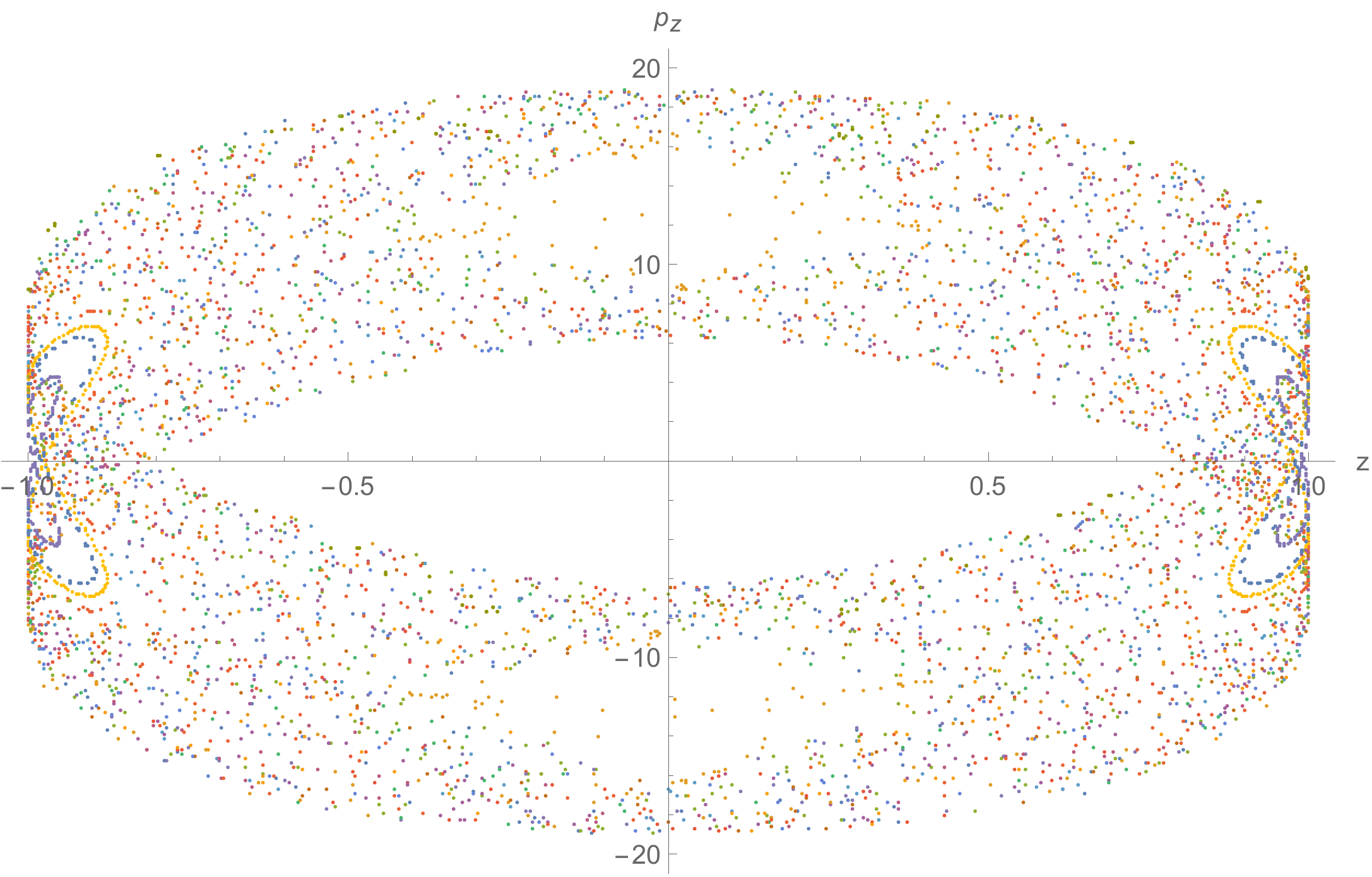}}

\caption{Poincar\'{e} sections for the $(\chi, p_\chi)$-plane at $z(t)=2$, for the quiver in eq. (\ref{quiver2final}) at different energies.}\label{fig_PoincareQuiver1chi}
}
\end{figure}

\begin{figure}[h!]
{
 \centering
 \subfloat[\small $E = 5$ \normalsize]{
   \label{fig_quiver2_pt5_Poincare}
     \includegraphics[width=0.5\textwidth]{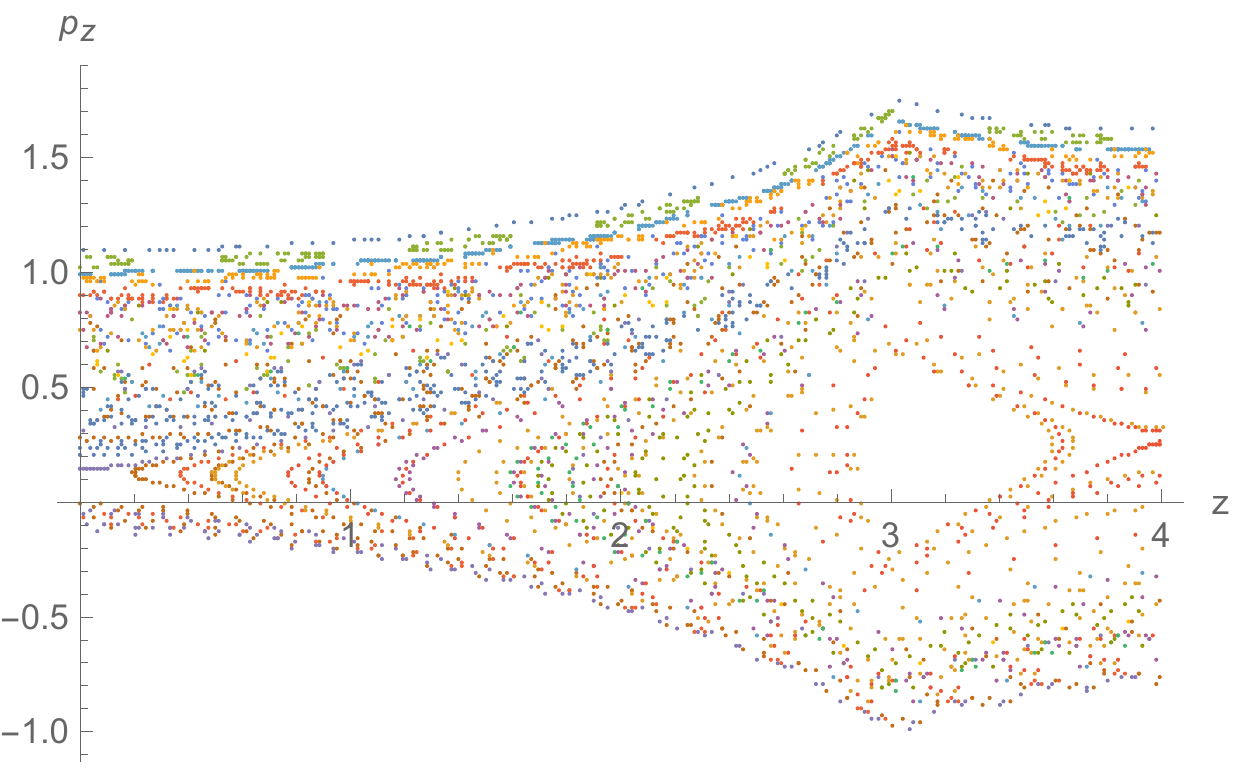}}
 \subfloat[\small $E = 15$ \normalsize]{
   \label{fig_quiver2_pt15_Poincare}
    \includegraphics[width=0.5\textwidth]{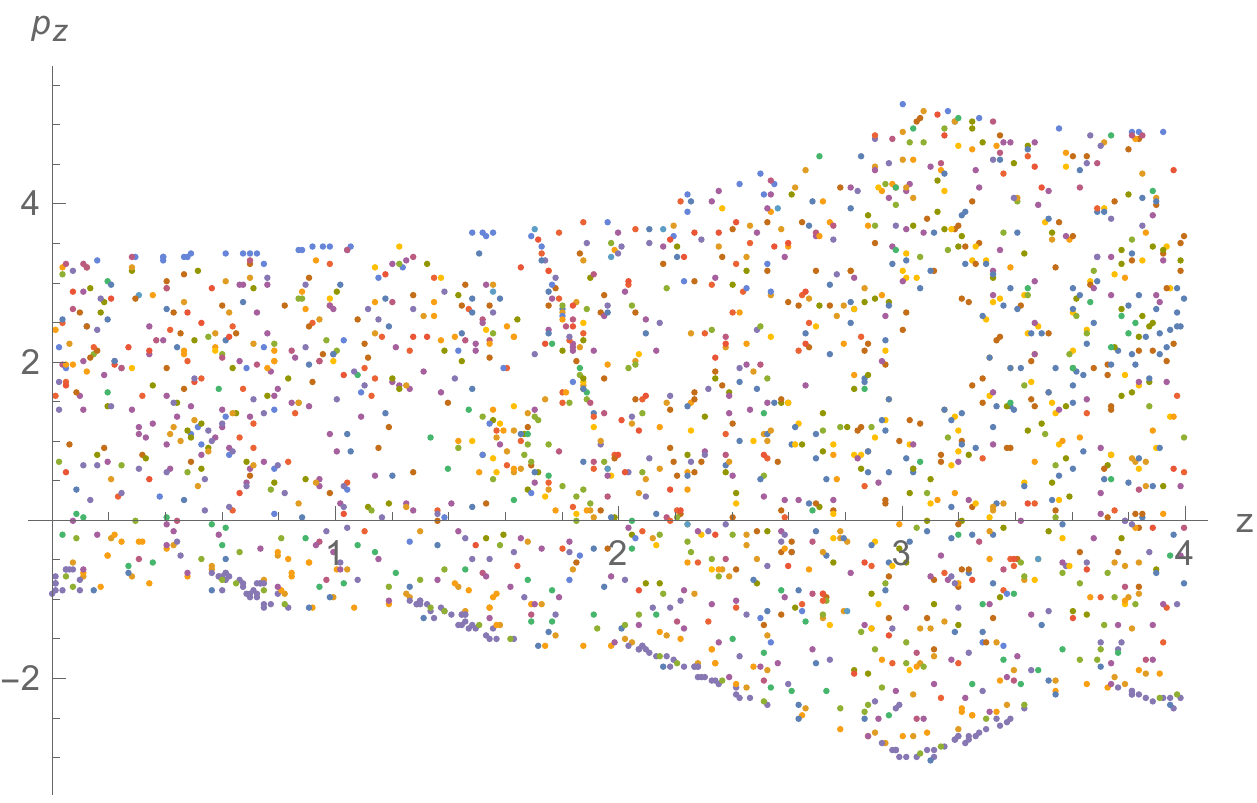}}\\    
 \centering
 \subfloat[\small $E = 35$ \normalsize]{
   \label{fig_quiver2_pt35_Poincare}
     \includegraphics[width=0.5\textwidth]{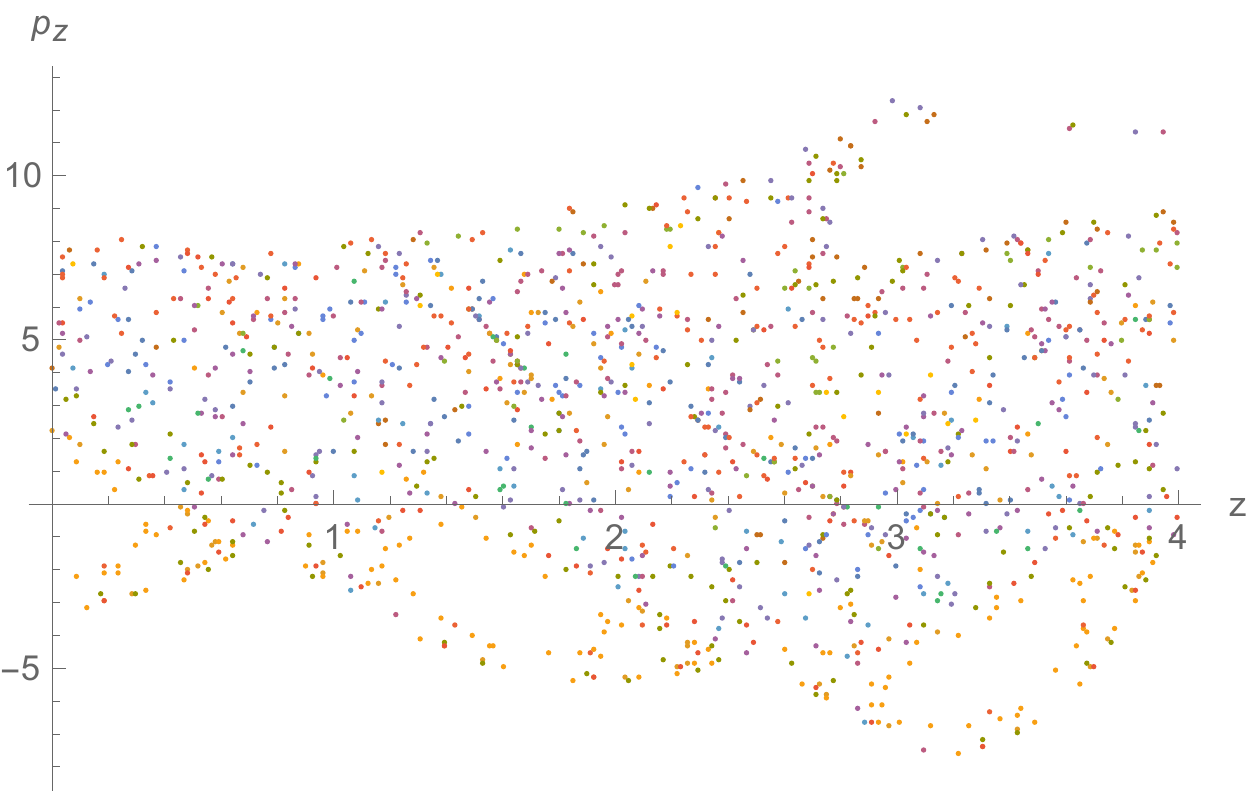}}
 \subfloat[\small $E = 65$ \normalsize]{
   \label{fig_quiver2_pt65_Poincare}
    \includegraphics[width=0.5\textwidth]{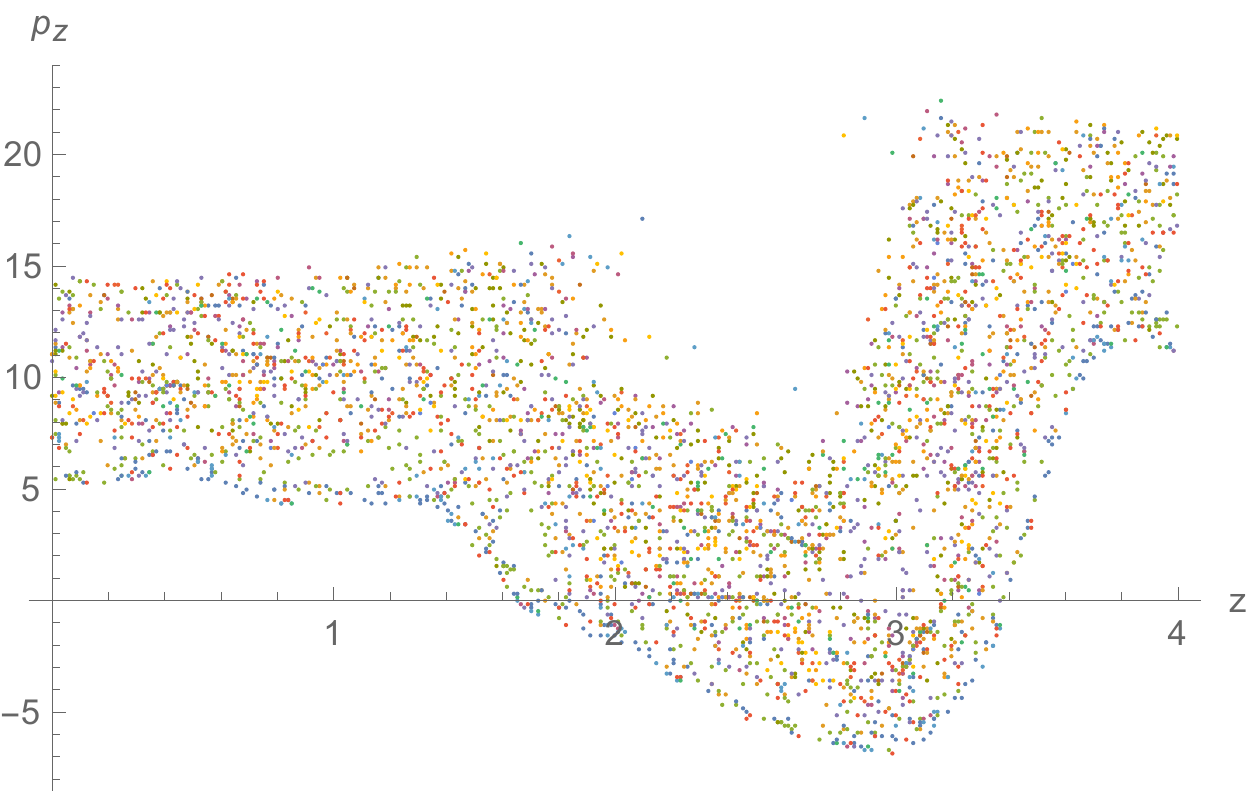}}

\caption{Poincar\'{e} sections for the $(z, p_z)$-plane at $\chi(t)=0$, for the quiver in eq. (\ref{quiver4Pfinal}) at different energies.}\label{fig_PoincareQuiver2z}
}
\end{figure}

If the string motion were integrable, the corresponding trajectories would have been constrained to a 2-dimensional torus in this $(z, p_z, \chi, p_\chi)$ phase-space. The Poincar\'{e} cross-sections of the phase-space would  then show the different resonant tori as embedded circles. The absence of such embedded circular KAM curves--in Figures \ref{fig_PoincareQuiver1z}, \ref{fig_PoincareQuiver1chi} and \ref{fig_PoincareQuiver2z}--- indicates that we are dealing with a non-integrable system, in agreement with the results we found in the earlier sections.

From our earlier study of the numerical evolution in Section \ref{subsec_NumericalEvolution}, we know that for low energies the string oscillates between the different poles of the two-sphere when turning around along the $z$-axis. As we explained, the point $\chi(t) = 0$ corresponds to the string being on the north pole of the two-sphere. We noticed that for low enough energies the string stays localised at this pole while moving along the $z$-axis. This is clearly  seen from the horizontal lines in Figure \ref{fig_lowE_chi0001_Poincare}. Also, notice  that for very low momenta the string does not reach the other side of the $z$-domain and stays localised around one of the endpoints. 

As we increase the energy (and consequently choose a higher value for $\chi$ to satisfy the Virasoro constraint) the string is no longer fixed at the pole but starts to oscillate quasi-periodically around the poles as it transverses the $z$-direction. This is the nature of the circles that we see appearing along the horizontal lines in Figures \ref{fig_lowE_chi01_Poincare} and \ref{fig_lowE_chi025_Poincare}. Finally, as we increase the energy even further we see that the Poincar\'{e} 
section looses all of its structure, since the string seems to move  randomly along the 2-sphere as the string moves along the $z$-direction.

Similar Poincar\'{e} sections for the background in eq. \ref{quiver4final}---quiver 1, can be seen in figure \ref{fig_PoincareQuiver2z}. Though this second quiver solution is not left-right symmetric along the $z$-direction the behaviour of its numerical evolution is in both cases very similar. 

Finally,  a different Poincar\'{e} cross-section for the quiver in eq. (\ref{quiver2final})  
is shown in Figure \ref{fig_PoincareQuiver2z}. 
Here we choose our initial conditions in a similar manner but we now plot the points $(\chi, p_\chi)$ every time $z = 2$. We see again that for low energies the string stays located at the poles where $\cos\chi = 1$ or $\cos\chi=-1$. 
As we go to higher energies the string is  located at random points on the two-sphere every time we cross $z=2$.

\section{Conclusions and future work}\label{concl}
We conclude our paper by first summarising the key ideas  that drive the present analysis and highlighting the main results obtained.  Finally, we list some of the key questions suggested by these results which are worth further investigation. 

The purpose of this work is to explore the dynamics of six-dimensional SCFTs with ${\cal N}=(1,0)$ SUSY. We discussed the holographic representation of the strongly coupled dynamics and  focused on  analytically showing the non-integrability of these theories. 
The fact that the system is strongly coupled at the conformal fixed point, together with  the absence of a local Lagrangian, makes it  a very difficult task to explore anything using the field theory description itself. However,  studying the dual string description is  promising. In the Massive IIA dual background one can  perform semi-classical computations that  unveil some of the issues related to the non-integrability. 

The traditional way of establishing the analytic integrability of  a dynamical system is to find the appropriate Lax pair.
Unfortunately, there is no general prescription to construct Lax pairs for a given field theory. Therefore, we choose  a different procedure, namely to consider a solitonic excitation in the dual string background, and study it as a dynamical system. We used a simple prescription (that goes under the name of Kovacic's algorithm \cite{kovacic}) to show  the presence of  \textit{non}-integrability associated with the  phase space of our classical string embedded in massive type IIA. This implies that the eigenvalues of the dilatation operator of the CFT cannot be determined with the usual techniques.

By probing the dual type IIA background with  a classical string (our soliton) and studying the Hamiltonian dynamics of this excitation,  we gain information on `long' operators of different quivers in the $ \mathcal{N}=(1,0) $ SCFTs. These  operators have large  quantum numbers (scaling and angular momentum). The classical strings correspond to that specific sector within the dual SCFTs. By virtue of the Maldacena duality, the analysis is equivalent to exploring the non-integrability associated to that specific CFT sector. In fact, if the classical dynamics associated to specific string embeddings in the bulk fails to satisfy the Kovacic's criterium, the corresponding phase space dynamics is non-integrable. This, in turn, implies non-integrability in the full dual $ \mathcal{N}=(1,0) $ SCFTs. 

 In order to put our analytic findings on a more solid ground, we  carry out a numerical analysis where we compute various physical quantities like the Lyapunov coefficients, the power spectra and the Poincar\'e sections that eventually display the onset of the  \textit{chaotic} behaviour associated with the corresponding phase space dynamics of the classical string embeddings.

These observations eventually give rise to deeper questions that one might wish to explore further. In the following we mention a few of them. 
 \begin{itemize}
 \item{It would be interesting to explore other correlation functions of different operators using the holographic perspective and examples described here. For example, the study of a BMN sector and associated pp-wave presents similar features as those in \cite{Itsios:2017nou}. Similarly, the study of magnons, spiked-strings, meson spectra and other similar observables are natural  to explore  in the context of the string duals discussed here.}
 \item{On the same line, the generalisation of our  integrability study to the  many $AdS_p\times M_{10-p}$ backgrounds found in recent classifications, might teach generic lessons about CFTs.}
 \item{The examples of non-integrability found in this paper, add to the list of \cite{Basu:2011fw}-\cite{Roychowdhury:2017vdo}. It is then natural to ask: {\it what} is the characteristic of the field theory that  causes  the non-integrability? It is  {\it not}
 the amount  of SUSY, nor the global symmetries. Surely, the stringy character of our solitons is essential, as all these effects vanish when the wrapping $\nu$ vanishes.  What other characteristics of the field theory-background pair should we take into account to diagnose non-integrability?
}
 \item{It would be of interest to study flows away from some of our quiver solutions, to other fixed points. The behaviour of the central charge as we defined it (and other correlators) should depend on the particular quiver we start with.}
 \item{In the same vein, the study of the chaos-indicators (Lyapunov coefficients, Poincar\'e sections, power spectra), is of huge interest. It is natural to wonder if there is some quiver CFT for which the onset of chaos is parametrically delayed. Similarly, the study of our string soliton under a RG-flow to another fix point might  show how these chaos indicators change under RG-flow.}
 \item{It is certainly of interest to relate the chaos observed here to the out-of-order four point correlation function of some hermitian  operators, as studied in \cite{Maldacena:2015waa}. Possibly, string worldsheet  four point correlators on our sigma model display similar behaviour to those in \cite{deBoer:2017xdk}. }
 \end{itemize}
We hope to report on some of these questions in future publications.

\section*{Acknowledgements}

Various colleagues have helped to improve the presentation of this work. We thank: Stefano Cremonesi, S. Prem Kumar, Yolanda Lozano, Leo Pando Zayas, Daniel Thompson, Alessandro Tomasiello and Salom\'on Zacar\'{\i}as.
 
C. N\'u\~nez is a Wolfson Fellow of the Royal Society. D. Roychowdhury is supported through the Newton-Bhahba Fund and he would like to acknowledge the Royal Society UK
and the Science and Engineering Research Board India (SERB) for financial assistance. J. M. Pen\'{\i}n is funded by the Spanish grant FPA2014-52218-P by Xunta de Galicia (GRC2013-024),  by FEDER and by the Maria de Maeztu Unit of Excellence MDM-2016-0692, and is supported by the Spanish FPU fellowship FPU14/06300. J. van Gorsel is supported by an STFC scholarship.


\appendix
\section{Appendix: the function $\alpha(z)$ for the quivers in Figures \ref{fig-quiver2} and \ref{fig-quiver3}}\label{detailsquivers}

In this appendix we plot the function $\frac{\alpha(z)}{81 \pi^2 N}$ and its derivatives for the quivers in Figures \ref{fig-quiver2} and \ref{fig-quiver3}.\\For the quiver in Figure \ref{figure1xx}, we have:

\begin{figure}[h]%
    \centering
    {{\includegraphics[width=4.5cm]{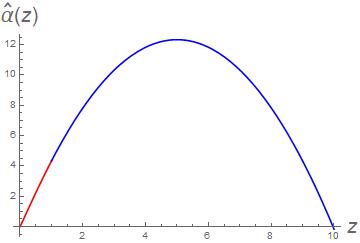} }}%
    \qquad
   {{\includegraphics[width=4.5cm]{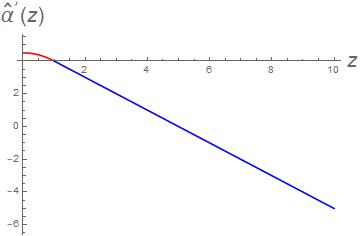} }}%
    \qquad
    {{\includegraphics[width=4.5cm]{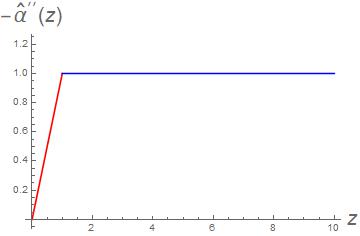} }}

\caption{$\hat{\alpha}(z)$ and its first and second derivatives, for the dual background of the quiver in Figure \ref{fig-quiver2}.}
\label{alphaderivatives2}
\end{figure}

With these values of $\alpha(z)$ and its derivatives, we can easily construct the functions $f_i(z)$ that describe the gravitational background. We plot, in order, $f_i(z)$ where $i \in \{1,2,3,4,5,6\}$.

\begin{figure}[h]%
    \centering
    {{\includegraphics[width=4.5cm]{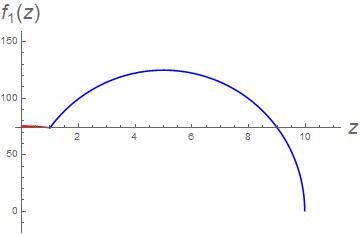} }}%
    \qquad
    {{\includegraphics[width=4.5cm]{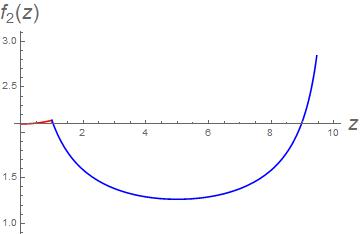} }}%
    \qquad
    {{\includegraphics[width=4.5cm]{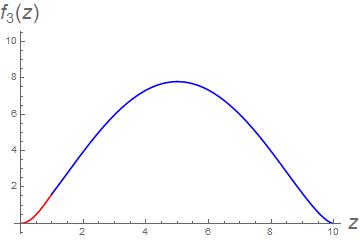} }}
    \qquad
    {{\includegraphics[width=4.5cm]{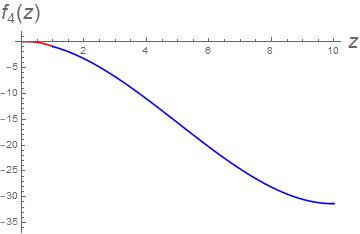} }}
    \qquad
    {{\includegraphics[width=4.5cm]{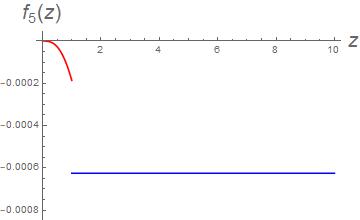} }}
    \qquad
    {{\includegraphics[width=4.5cm]{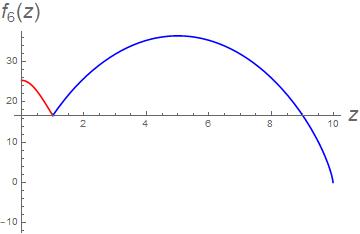} }}

\caption{$f_i(z)$ for the dual background to the quiver in Figure \ref{fig-quiver2}.}
\label{fs2}
\end{figure}
\begin{figure}[h]%
    \centering
    {{\includegraphics[width=4.5cm]{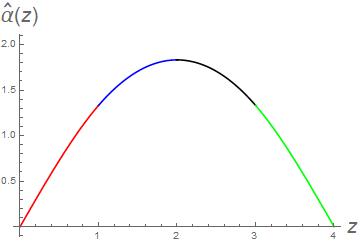} }}%
    \qquad
    {{\includegraphics[width=4.5cm]{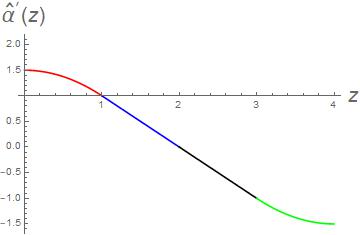} }}%
    \qquad
    {{\includegraphics[width=4.5cm]{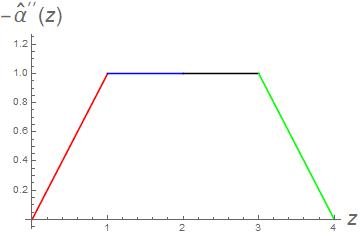} }}

\caption{$\hat{\alpha}(z)$ and its first and second derivatives, for the dual background of the quiver in Figure \ref{fig-quiver3}.}
\label{alphaderivatives3}
\end{figure}
\begin{figure}%
    \centering
    {{\includegraphics[width=4.5cm]{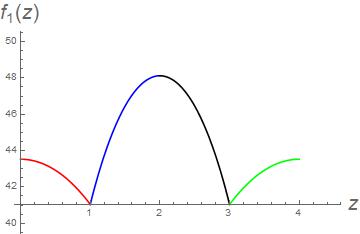} }}%
    \qquad
    {{\includegraphics[width=4.5cm]{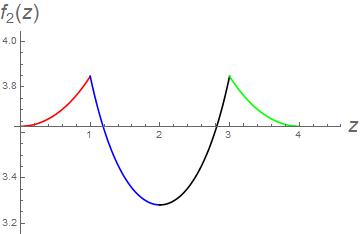} }}%
    \qquad
    {{\includegraphics[width=4.5cm]{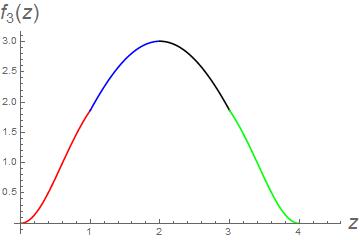} }}
    \qquad
    {{\includegraphics[width=4.5cm]{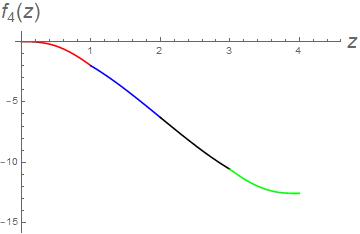} }}
    \qquad
    {{\includegraphics[width=4.5cm]{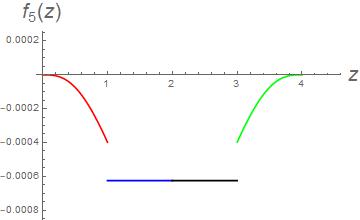} }}
    \qquad
    {{\includegraphics[width=4.5cm]{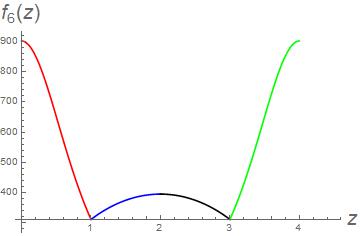} }}

\caption{$f_i(z)$ for the dual background to the quiver in Figure \ref{fig-quiver3}.}
\label{fs3}
\end{figure}

Now we do the same for the quiver in Figure $\ref{fig-quiver3}$. We depict the function $\alpha(z)$ and its derivatives in Figure \ref{alphaderivatives3} and $f_i(z)$ in Figure \ref{fs3}.

\begin{figure}%
    \centering
    {{\includegraphics[width=7.0cm]{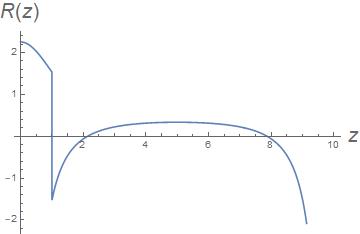} }}%
    \qquad
    {{\includegraphics[width=7.0cm]{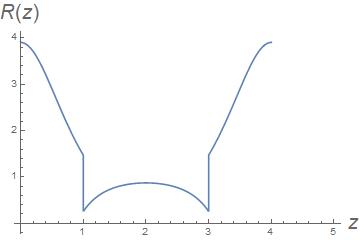} }}

\caption{The Ricci scalars for the backgrounds associated to the quivers in Figures \ref{fig-quiver2} and \ref{fig-quiver3} respectively.}
\label{Ricci23}
\end{figure}

It is easy to construct the Ricci scalar of these geometries. It can be written in terms of $\alpha(z)$ and its derivatives: 
\begin{eqnarray}
& & R=\frac{1}{4 \sqrt{2} \pi \alpha^2 \alpha''^2 (\alpha'^2-2 \alpha \alpha'')^2} \sqrt{-\frac{\alpha}{\alpha''}}  \big[  -21 \alpha'^6 \alpha''^2+42 \alpha \alpha'^5 \alpha'' \alpha^{(3)} \nonumber \\
& & -252 \alpha^2 \alpha'^3 \alpha''^2 \alpha^{(3)} +336 \alpha^3 \alpha' \alpha''^2 \alpha^{(3)}  + 8 \alpha^3 \alpha'^2 \alpha'' (13 (\alpha^{(3)})^2-7 \alpha'' \alpha^{(4)}) \nonumber \\ 
& &+12 \alpha^3 \alpha''^2 (-7 \alpha''^3-15 \alpha (\alpha^{(3) })^2+6 \alpha \alpha'' \alpha^{(4)} ) +\alpha \alpha'^{(4)} (63 (\alpha'')^3-21 \alpha (\alpha^{(3)})^2+10 \alpha \alpha'' \alpha^{(4)}) \big] \nonumber
\end{eqnarray}
In Figure \ref{Ricci23} we plot the Ricci scalar for both geometries.

\section{On Liouvillian integrability and Kovacic's algorithm}\label{kovalg}

In this appendix, we briefly describe Kovacic's algorithm \cite{kovacic}.
Consider a second order differential equation,
\begin{equation}
y''(x)+ B(x) y'(x) +A(x) y(x)=0,
\label{kov1}
\end{equation}
where $A(x),B(x)$ are complex rational functions. We are interested in the existence of closed form solutions, namely solutions that can be expressed
in terms of algebraic, exponential, and trigonometric functions, and integrals of the previous functions. If this is the case, we call the solution Liouvillian. The algorithm
of \cite{kovacic} provides one such solution or shows there is none (in which case we refer to the differential equation (\ref{kov1}) as non-integrable).
We will not describe the algorithm itself (that is efficiently implemented by many different softwares). 
We will limit us to explain the logic behind Kovacic's algorithm and some {\it necessary but not-sufficient} conditions 
that a combination of the functions $A,B, B'$ must satisfy, for the eq. (\ref{kov1}) to be Liouville-integrable.

We start by redefining the function $y(x)$ and rewriting  the differential equation as,
\begin{eqnarray}
& & y(x)= e^{\int w(x)-\frac{B(x)}{2} dx},\nonumber\\
& & w'(x) +w(x)^2=V(x), \;\;\;\;\; V(x)=\frac{2B' + B^2-4 A}{4}.\label{kov2}
\end{eqnarray}
It was shown that if the function $w(x)$ is algebraic of degrees 1,2,4,6, or 12, then the eq. (\ref{kov1}) is Liouville integrable \cite{kovacic}.
This results comes from the application of Galois theory to differential equations (this is called Picard-Vessiot theory).
This formalism studies the most general group of invariances of the differential equation (\ref{kov1}), that is the transformations that act on the solutions of the equation,  that is a subgroup of $SL(2,C)$. Kovacic showed that there are four possible cases of subgroups of $SL(2,C)$ that can arise
\begin{itemize}
\item{{\bf Case 1:} the subgroup is generated by the matrix of the form   
\[
   G=
  \left[ {\begin{array}{cc}
   a & 0 \\
   b & \frac{1}{a} \\
  \end{array} } \right],
\]
with $a,b$ complex numbers. In this case $w(x)$ is a rational function of degree 1.}

\item{{\bf Case 2:} the subgroup of $SL(2,C)$ is generated by matrices of the form (here $c$ is a complex number),
\[
   G=
  \left[ {\begin{array}{cc}
   c & 0 \\
   0 & \frac{1}{c} \\
  \end{array} } \right],\;\;\;
   G=
  \left[ {\begin{array}{cc}
   0 & c \\
   -\frac{1}{c} & 0 \\
  \end{array} } \right],
\]
in this case the function $w(x)$ is rational of degree 2
}
\item{{\bf Case 3:} the situation in which $G$ is a finite group, not included in the two above cases. In this case, the degree of $w(x)$ is either 4,6 or 12.}
\item{{\bf Case 4:} the group is $SL(2,C)$ and the solutions for $w(x)$, if they exist are non-Liouvillian.}
\end{itemize}
Interestingly,  Kovacic provided not only an algorithm to find the solutions in the first three cases above, but also
 a set of {\it necessary but not sufficient} conditions that the function $V(x)$ in eq. (\ref{kov2}) must satisfy to be in any of the first three cases detailed above
 \cite{kovacic}. For each of the cases as ordered above, the conditions are :
\begin{itemize}
\item{{\bf Case 1:} every pole of $V(x)$ has order 1 or  has even order. The order of the function $V(x)$ at infinity is either even or greater than 2.}
\item{{\bf Case 2:} $V(x)$ has either one pole of order 2, or  poles of odd-order greater than 2 . }
\item{{\bf Case 3:} the order of the poles of  $V$ does not exceed 2, and the order of $V$ at infinity is at least 2.}
\end{itemize}
If none of the above is satisfied, the analytic  solution (if it exists), is non-Liouvillian.
\subsection{One example}
Let us work out an example to see the criteria at work. We study the NVE in eqs.  (\ref{NVE})-(\ref{AB}). To simplify matters, we will just study the
NVE in the interval $0\leq z\leq 1$, so that the function $\alpha(z)=- 81\pi^2 N (a_1 z+\frac{z^3}{6})$.
In this case the coefficients are
\begin{eqnarray}
& & {\cal A}=1- \sqrt{3} \frac{(z^4 +20 a_1 z^2- 60 a_1^2)}{\sqrt{-6 a_1- z^2} (z^4+12a_1 z^2-12 a_1^2) }    ,\nonumber\\
& & {\cal B}=\frac{2}{ z}+\frac{3z}{(6 a_1+z^2)} -4\frac{6 a_1 z+z^3}{(z^4+12 a_1 z^2-12 a_1^2)}.\label{calacalb}
\end{eqnarray}
To avoid cluttering the expressions, we have chosen  the coefficients $E=4\pi$ (such that $z=\tau$) and $ \nu=1$.

The coefficients of this NVE  are not rational functions. To amend this,  we change 
from $z$ to a new variable $v$,
\begin{equation}
z=\sqrt{-6a_1- v^2}\label{changevv}
\end{equation}
denoting $x'=\frac{dx}{dv}$, the NVE  reads
\begin{equation}
x''(v) + {\cal C} x'(v)  +{\cal D} x(v)=0,\;\;\; {\cal C}=\frac{\left(  {\cal B}(v)+\frac{d}{dv}\frac{dv}{dz} \right)}{\frac{dv}{dz}}, \;\;\;
 {\cal D}= \frac{{\cal A}(v)}{(\frac{dv}{dz})^2} .
\end{equation}
According to what was explained around eq. (\ref{kov2}), we now 
need to analyse the principal part of the potential $4V=2\frac{d{\cal C}}{dv} +{\cal C}^2 - 4 {\cal D} $.

For the particular case of the ${\cal A}$ and ${\cal B}$, and the change of variables in  eqs. (\ref{calacalb})-(\ref{changevv}) we find,
\begin{eqnarray}
& & {\cal C}=\frac{v^6 -12 a_1 v^4 -240 a_1^2 v^2 -576 a_1^3}{v(v^2+6 a_1)(v^4-48 a_1^2)} ,\nonumber\\
& &  {\cal D}=-1+\frac{6a_1 +5\sqrt{3}v}{ v^2+6a_1}-\frac{4\sqrt{3}v(v^2-4 a_1)}{v^4-48 a_1^2}   ,\nonumber\\
& & 4V=4+ \frac{\gamma_0}{ v^2} +\frac{\gamma_1 }{(v^2+6a_1)^2}
+\frac{\gamma_2 +\gamma_3 v}{(v^2+6a_1)} +\frac{\gamma_4 v^2 }{(v^4-48 a_1^2)^2}+\nonumber\\
& & \frac{\gamma_5 +\gamma_{6} v +\gamma_7 v^2+\gamma_8 v^3}{(v^4-48 a_1^2)}.
\label{potentialquiver1}
\end{eqnarray}
The coefficients $\gamma_i$ are numerical constants, not very relevant for our analysis below.
Notice that the potential has a pole of order one at $v=\pm i \sqrt{6 a_1}$. The order of the potential (leading
power of the degrees of the denominator minus numerator) is one. Hence, $V(x)$ does not fall in any of the three allowed cases.
The solution to the equation should then be non-Liouvillian.

Notice that this analysis covers the cases of our quivers in Figures \ref{figure1xx} and \ref{fig-quiver2}. Indeed,
both these quivers start with the function $\alpha(z)=-81 \pi^2 N( a_1z+\frac{z^3}{6})$.  Hence, both backgrounds and both quiver CFTs are non-integrable.



\section{Integrability of the configuration in Type IIA and M-theory}\label{appendixM2brane}
In this short appendix (the details of which will be fully worked out elsewhere), we will indicate the steps
that lead to the a more detailed study of our string configuration for the case in which
the function $\alpha(z)=(R^2- \mu^2 z^2)$, characterising a background in Type IIA (with mass parameter $m=0$) that lifts to and $AdS_7\times S^4/Z_k$ in eleven dimensional supergravity. This is the case we studied around eq.(\ref{alphamassless}). We shall lift the background to eleven dimensions, and then study the dynamics of a membrane that mimics our type IIA string. The lifted background is,
\begin{eqnarray}
& & ds_{11}^2= f_{6}^{-2/3}\left[  f_1\;ds^2_{AdS_7}+f_2\;dz^2+f_3\;d \Omega^2(\chi, \xi)   \right] + f_6^{4/3}(dy- f_5 \cos\chi\; d\xi)^2,\nonumber\\
& & 6 C_3= f_4 \sin\chi\; d\chi\wedge d\xi\wedge dy.\label{config11d}
\end{eqnarray}
 We define $\gamma_{ij}= G_{\mu\nu}\partial_i X^\mu \partial_j X^\nu$, where $i,j=\tau,\sigma,\rho$ are the world-volume coordinates of the membrane, the functions $f_i$ are all functions of $z$ and are defined in eq.(\ref{functionsf}). We use the action and constraints for a membrane (see for example \cite{Hartnoll:2002th}),
 \begin{eqnarray}
 & & S=\int d\tau d\sigma d\rho \left(\gamma_{\tau\tau} +L^2 \det(\gamma_{\alpha\beta})+ 2 L \epsilon^{ijk}C_{\mu\nu\delta}\partial_i X^\mu \partial_j X^\nu \partial_k X^\delta\right).\nonumber\\
 & & \gamma_{\tau\alpha}=0,\;\;\;\; \gamma_{\tau\tau}+ L^2 \det \gamma_{\alpha\beta}=0.\;\;\;\; (\alpha,\beta=\sigma,\rho)\label{membrane}
 \end{eqnarray}
We propose a membrane configuration that is the natural lift of the string soliton we proposed in the main part of the  paper,
\begin{equation}
t=t(\tau),\;\; z=z(\tau),\;\;\chi=\chi(\tau),\;\; \xi=k\sigma,\;\; y=\lambda \rho.\label{membxxx}
\end{equation}
We find an effective Lagrangian and constraint that can be written as,
\begin{eqnarray}
& & L= f_6^{-2s/3}\left[f_1 \dot{t}^2-f_2 \dot{z}^2 -f_3\dot{\chi}^2+ L^2k^2\lambda^2 f_3 f_6^{4s/3}\sin^2\chi + Lk\lambda f_4 f_6^{2s/3}\dot{\chi}\sin\chi  \right],\nonumber\\
& &0= -f_1\dot{t}^2+f_2 \dot{z}^2 +f_3\dot{\chi}^2+ L^2k^2\lambda^2 f_3 f_6^{4s/3}\sin^2\chi.
\label{lagconstmem}
\end{eqnarray}
One can see that choosing $s=0$ (and identifying $L\lambda k=\nu$), we recover the expressions for strings written below eq.(\ref{solitonxx}).  On the other hand, for $s=1$, we have the Lagrangian and constraint for the membrane. 
 
 We now study the equations of motion derived from eq.(\ref{lagconstmem}). We find,
 \begin{eqnarray}
 & & \dot{t}=\frac{E}{f_1}f_6^{2s/3},\label{eqsmemb}\\
 & & 2 f_3 \ddot{\chi}=-2L^2k^2\lambda^2 f_3 \cos\chi\sin\chi f_6^{4s/3}-2 \dot{z}\dot{\chi} f_3' +2Lk\lambda \sin\chi \dot{z} f_4' f_6^{2s/3}+\frac{4s}{3}\frac{f_3 f_6'}{f_6}\dot{z}\dot{\chi}.\nonumber\\
 & & \ddot{z}+ E^2 \frac{f_6^{4s/3}}{2f_1 f_2}\Big(\frac{f_1'}{f_1} -\frac{2s}{3}\frac{f_6'}{f_6}\Big) +
 \dot{z}^2 \Big(\frac{f_2'}{2 f_2} -\frac{s}{3}\frac{f_6'}{f_6} \Big)+\dot{\chi}^2 \frac{f_3}{2f_2}\Big(-\frac{f_3'}{f_3} +\frac{2s}{3}\frac{f_6'}{f_6} \Big)+\nonumber\\
 &  & + L k\lambda \sin\chi\;\dot{\chi} \frac{f_4'}{f_2}f_6^{2s/3} +L^2k^2\lambda^2 \frac{f_3 f_6^{4s/3} \sin^2\chi}{2f_2}\Big(\frac{f_3'}{f_3} +\frac{2s}{3}\frac{f_6'}{f_6}\Big)=0.\nonumber
 \end{eqnarray}
The reader can check that for $s=0$ the equation of motion of the string are recovered.

We now apply the same algorithmic procedure as in the main body of the paper. The configuration with $\chi(\tau)=\dot{\chi}(\tau)=\ddot{\chi}(\tau)=0$ solves the $\chi-$equation of motion and leaves the $z-$equation as,
\begin{equation}
\ddot{z}+ E^2 \frac{f_6^{4s/3}}{2f_1 f_2}\Big(\frac{f_1'}{f_1} -\frac{2s}{3}\frac{f_6'}{f_6}\Big) +
 \dot{z}^2\Big(\frac{f_2'}{2 f_2} -\frac{s}{3}\frac{f_6'}{f_6}\Big)=0.
\end{equation}
Calculating explicitly for the function $\alpha(z)=\mu (1-z^2)$ (after   choosing constants appropriately), using the explicit expression for $f_i(z)$, we find that $z-$equation is solved by
\begin{equation}
z_s(\tau)=\cosh\tau.\label{solmemb}
\end{equation}
Fluctuating the $\chi-$equation as $\chi(\tau)=0+\epsilon f(\tau)$, we find the NVE,
\begin{eqnarray}
& & \ddot{f}+ {\cal B}\dot{f} +{\cal A}f=0,\label{nvememb}\\
& & {\cal B}= \dot{z}(\tau)\left(\frac{f_3'}{f_3} -\frac{2s}{3}\frac{f_6'}{f_6}   \right)|_{z_s}=2\coth\tau  ,\nonumber\\
& & {\cal A}=L^2k^2\lambda^2 f_6^{4s/3} -L\lambda kf_6^{2s/3}\frac{f_4'}{f_3}\dot{z}(\tau)|_{z=z_s} = n_1 \sinh\tau+n_2\sinh^2\tau .\nonumber
\end{eqnarray}
With $n_1,n_2$ two numbers.  In what follows we take $s=1$ to discuss the case of the membrane only. It is convenient to change the variable $v=e^{-\tau}$, to have an NVE that reads,
\begin{equation}
f''+\frac{3v^2+1}{v(v^2-1)}f' + \left(\frac{n_1}{2v^3}(1-v^2) + \frac{n_2}{4v^4}(1-v^2)^2\right)f=0.
\end{equation}
We denoted $f'=\frac{df}{dv}$. We can construct the effective potential of the
associated Schr\"odinger problem, as indicated in eq.(\ref{kov2}),
\begin{eqnarray}
& & V(v)=\frac{2{\cal B}' + {\cal B}^2-4 {\cal A}}{4}=\frac{3}{4v^2}-\frac{n_1}{2v^3}+\frac{n_1}{2v}-\frac{n_2}{4v^4} -\frac{n_2}{4} +\frac{n_2}{2v^2}.
\end{eqnarray}
We observe that the first of the necessary conditions discussed in Appendix \ref{kovalg}
is satisfied. The Kovacic algorith should produce a Liouvillian solution for the membrane, making the membrane configuration in eq.(\ref{membxxx}) Liouville-integrable.

We see that the problem with the string is that it `misses' the effects of the dilaton, represented above by the various powers of $f_6^{2/3}$. It is the presence of the dilaton (that the `classical limit' of the Polyakov action misses), what changes the equation to introduce integrability. Note that the dilaton goes very large at the ends of the interval $z=\pm 1$ (in these units), hence it cannot be neglected.

\section{Computation of the Lyapunov spectrum}\label{appendixlyapunov}


In the following we discuss the algorithm used to compute the Lyapunov characteristic exponents (LCEs) for a generic system, in particular for a system of canonical equations such as the ones we have previously studied. We will make use of the prescription described in \cite{Lyapunov}.\\
Let us consider a generic n-dimensional smooth dynamical system, which can generically be written as:
\begin{equation}
\dot{q}=V(q)
\label{dynsys}
\end{equation}
where $q(\tau)$ is the $n$-dimensional state vector $q=\left(\vec{X}(\tau), \vec{P}(\tau) \right)$ at time $\tau$, $\dot{q}=\frac{dq}{d\tau}$ and $V$ is a vector field on an open set $U$ of the phase space manifold, which generates a flow $f$:
\begin{equation}
\dot{f}^\tau(q)=V(f^\tau(q)) \ \ \  \mathrm{for \ all} \ q \in U,\tau \in \mathrm{R}
\end{equation} 
where $f^\tau(q)=f(q,\tau)$.

Consider the evolution under the flow of two nearby points in the phase space, $q_0$ and $q_0+\delta_0$, being $\delta_0$ a small perturbation of the initial point $q_0$.  After a time $\tau$, the perturbation $\delta_\tau$ will become:
\begin{equation}
\delta_\tau \equiv f^\tau(q_0+\delta_0)-f^\tau(q_0) \approx D_{q_0} f^\tau(q_0) \cdot \delta_0
\label{tangent}
\end{equation}
The average exponential rate of divergence (or convergence) of two trajectories is defined by:
\begin{equation}
\lambda(q_0,\delta_0)=\lim_{\tau \rightarrow \infty} \frac{1}{\tau} \log \frac{||\delta_\tau||}{||\delta_0||}=\lim_{\tau \rightarrow \infty} \frac{1}{\tau} \log ||D_{q_0} f^\tau(q_0) \cdot \delta_0||
\label{LCE}
\end{equation} 
being $||\delta||$ the length of the vector $\delta$. If $\lambda(x,u)>0$, we have exponential divergence of nearby orbits. Under weak conditions on the nature of the dynamical system, the limit \ref{LCE} exists, it is finite and it is equal to the largest LCE $\lambda_1$, see \cite{Oseledec1968} for reference.

The LCEs of order $p$, $1 \leq p \leq n$, are introduced to describe the mean rate of growth of a p-dimensional volume in the tangent space. Considering a parallelepiped $U_0$ in the tangent space whose edges are the $p$ vectors $\delta_1,...,\delta_p$, the LCEs of order $p$ are defined by:
\begin{equation}
\lambda^p(q_0,U_0)=\lim_{\tau \rightarrow \infty} \frac{1}{\tau} \log [\mathrm{Vol}^p (D_{q_0} f^\tau(U_0))]
\label{volp}
\end{equation}
being $\mathrm{Vol}^p$ the p-dimensional volume defined in the tangent space. It can be seen \cite{Oseledec1968} that there exist $p$ linearly independent vectors $u_1,...,u_p$ satisfying:
\begin{equation}
\lambda^p(q_0,U_0)= \lambda_1+ ... + \lambda_p
\label{LCEp}
\end{equation}
The tangent vector $\delta_\tau$ defined in \ref{tangent} evolves in time satisfying:
\begin{equation}
\dot{\Phi}_\tau(q_0)=D_q V(f^\tau(q_0)) \cdot \Phi_\tau(q_0), \ \ \ \Phi_0(q_0)=I
\label{fullsys}
\end{equation}
where $\Phi_\tau(q_0)=D_{q_0} f^\tau(q_0)$. To calculate the trajectory, we have to integrate the system:
\begin{align}
 \Bigg \{\begin{array}{ccr}
\dot{q}  \\
\dot{\Phi}  
\end{array} \Bigg \}= \Bigg \{\begin{array}{ccr}
V(q) \\
D_q V(q) \cdot \Phi
\end{array} \Bigg \},  \ \ \  \Bigg \{\begin{array}{ccr}
q(\tau_0)  \\
\Phi(\tau_0)  
\end{array} \Bigg \}= \Bigg \{\begin{array}{ccr}
q_0 \\
I
\end{array} \Bigg \}
\label{fullsysarray}
\end{align}



To compute the spectrum of LCEs, we will use the algorithm discussed in \cite{Benettin1980}, based on the calculation of the order-$p$ LCEs defined in equation \ref{LCEp} and on a repeated application of a Gram-Schmidt orthonormalization procedure (which avoids technical difficulties that arise in the implementation of the recipe described in \cite{Oseledec1968}) that we briefly summarize here. 




Recall that if we compute an orthonormal set of vectors $\{\hat{\delta}_i \}$ out of the original set of vectors $\{\delta_i \}$, 
by using the Gram-Schmidt orthogonalisation procedure, the volume of the parallelepiped spanned by ${\delta_1, ..., \delta_p}$ is 
\begin{equation}
\mathrm{Vol} \{\delta_1, ..., \delta_p \}= ||\hat{\delta}_1||...||\hat{\delta}_p||
\end{equation}


Then the algorithm starts by choosing an initial condition $q_0$ and a $n \times n$ matrix $\Delta_0=[\delta^0_1,...,\delta^0_n]$. Using the Gram-Schmidt procedure, we calculate the corresponding matrix of orthonormal vectors  $\hat{\Delta}_0=[\hat{\delta}^0_1,...,\hat{\delta}^0_n]$ and integrate the equation \ref{fullsysarray} from $\{q_0,
\Delta_0 \}$ for a short interval $T$, to obtain $q_1=f^T(q_0)$ and 
\begin{equation}
\Delta_1 \equiv [\delta^1_1,...,\delta^1_n]=D_{q_0} f^T(\Delta_0)= \Phi_T(q_0) \cdot [\delta^0_1,...,\delta^0_n]
\end{equation}
The algorithm proceeds by repeating this integration-orthonormalization procedure $K$ times. During the $k$-th step, the $p$-dimensional volume $\mathrm{Vol}^p$ defined in \ref{volp} increases by a factor of $||w^k_1||...||w^k_p||$, where $\{ w^k_1,...,w^k_p \}$ is the set of orthogonal vectors calculated from $U_k$ using the Gram-Schmidt technique. Then:
\begin{equation}
\lambda^p(q_0, \Delta_0)=\lim_{k \rightarrow \infty} \frac{1}{kT} \sum^{k}_{i=1} \log(||\hat{\delta}^i_1||...||\hat{\delta}^i_p||)
\end{equation} 
From which we can derive
\begin{equation}
\lambda_p=\lim_{k \rightarrow \infty} \frac{1}{k T} \sum^{k}_{i=1} \log ||\hat{\delta}^i_p||
\end{equation}
To obtain the Lyapunov spectrum, we continue calculating the quantities:
\begin{equation}
\frac{1}{KT} \sum^{K}_{i=1} \log ||\hat{\delta}^i_1|| \approx \lambda_1,...,\frac{1}{KT} \sum^{K}_{i=1} \log ||\hat{\delta}^n_1|| \approx \lambda_n 
\end{equation}
for a suitable value of $T$, until they show convergence.

\section{A relation with non-Abelian T-duality?}\label{natdsection}
Let us briefly discuss a possible (and quite weak at the moment of this writing!) relation between the metrics in the Cremonesi-Tomasiello backgrounds (see Section \ref{cremotoma}) and 
non-Abelian T-duality. 

Since the work of Sfetsos and Thompson \cite{Sfetsos:2010uq}, the non-Abelian version of the usual T-duality has regained interest and played a role as a solution generating technique.
Various papers taking a perspective inspired by holography, have made clear that when applied to symmetric enough backgrounds (characteristically $AdS_{p+1}\times M_{9-p}$ backgrounds) the generated solutions correspond to CFTs in $p$ dimensions, realised on a $D_p-NS_5-D_{p+2}$ system---for a sample of results, see the papers \cite{natd} for early attempts and \cite{Lozano:2016kum}, \cite{Lozano:2016wrs} for more recent and precise connections between non-Abelian T-duality and brane set-ups. 

It is natural to ask if the Massive IIA $AdS_7$ backgrounds studied in this paper can be thought of as the non-Abelian T-dual of some background, conjecturally in Type IIB, with dilaton and $F_3$ flux.

Let us give some comments in the direction of realising the previous idea. We consider a solution in Massive IIA that is the simplest possible. Consider, for example, the case in which the function $\alpha(z)$ in Massive IIA is
\begin{equation}
\alpha(z)= A \sin\omega z.\nonumber
\end{equation}
This is a solution to the equation of motion if $\alpha'''=-162\pi^2 F_0=-A\omega^3 \cos\omega z$, which implies that the mass-parameter is actually position dependent. A possible way to understand this, suggests a position dependent smearing of D8-branes. Since $\alpha''=-\omega^2\alpha$, we have that the background and the Ricci scalar read,
\begin{eqnarray}
& & ds^2=8\pi\frac{\sqrt{2}}{\omega} AdS_7 +\sqrt{2}\pi \omega dz^2+ \frac{\sqrt{2}\pi}{\omega}\left(\frac{\sin^2\omega z}{1+\sin^2\omega z}\right)d\Omega_2,\nonumber\\
& & e^{-2\phi}=e^{-2\phi_0}(1+\sin^2\omega z),\;\;\; B_2=\pi\left(-z +\frac{\sin\omega z\cos\omega z}{\omega(1+\sin^2 \omega z)}\right)d\Omega_2,\;\;\; F_0= -\frac{A\omega^3 \cos\omega z}{162\pi^2}\nonumber\\
& & F_2=- \frac{A \omega^2}{81 \pi^2}  \left(\frac{\sin^3\omega z}{1+\sin^2 \omega z}\right) d\Omega_2,\;\;\;
R=\frac{\omega \sin^4\omega z}{4\sqrt{2}\pi}\left(\frac{12 +100 \cot^2 \omega z + 75 \cot^4 \omega z}{1+\sin^2 \omega z}\right)
\label{cremotomazz}
\end{eqnarray}
Notice that the background is non-singular. We expand this Massive IIA solution close to $z\to 0$ and we find,
\begin{eqnarray}
& & ds^2\sim8\pi\frac{\sqrt{2}}{\omega} AdS_7 +\sqrt{2}\pi \omega \left(dz^2+ z^2 d\Omega_2\right),\nonumber\\
& & e^{-2\phi}\sim e^{-2\phi_0}(1+\omega^2 z^2),\; B_2\sim -\frac{5\pi \omega^2}{3} z^3 d\Omega_2,\;
F_2\sim- \frac{A \omega^5}{81 \pi^2}  z^3 d\Omega_2,\;F_0\sim -\frac{A\omega^3 }{162\pi^2} .
\label{cremotomazzexp}
\end{eqnarray}

We want to think about this background as obtained by non-Abelian T-duality applied on some seed-solution. We can consider a space-time of the form $AdS_7\times S^3$ in Type IIB. Notice that this is {\it not} a solution to the equations of motion. Hence we need to consider a more complicated background, depending on the coordinates of $S^3$. Importantly, notice that this background needs not be SUSY, it may be the case that the duality creates the supersymmetry. This putative background should have warp factors that can be decomposed in harmonics of $S^3$. Consider the s-wave and perform non-Abelian T-duality on it. We insist, this s-wave should not be a solution of the equations of motion. We shall obtain, 
\begin{eqnarray}
& & ds^2= L_{AdS}^2 AdS_7 + L^2 \left(dr^2+ \frac{r^2}{r^2+1} d\Omega_2\right),\nonumber\\
& & e^{-2\phi}= e^{-2\phi_0}(1+ r^2),\;\;\; B_2=\mu_0 \frac{r^3}{1+r^2} d\Omega_2,\;\;\;
F_2=\nu_0 \frac{r^3}{1+r^2}  z^3 d\Omega_2,\;\;\;\; F_0= f_0.
\label{natd}
\end{eqnarray}
Where $\mu_0,\nu_0,L_{AdS}, L, f_0, \phi_0$ are some constants.
Consider an expansion of this background near $r\to0$. We find that after appropriately choosing the constants,
 it has the form in eq. (\ref{cremotomazzexp}).  

Notice that the same result would be obtained by starting with the background  
described around eq. (\ref{quiver4Pfinal}) and expanding  for $z\to 0$. 
This is obvious as close to $z=0$ the function $\alpha= a_1 z+\frac{z^3}{6}$ and $\alpha= A \sin \omega z$
coincide for some choice of $A,\omega,a_1$. 

This is reminiscent of the observations in the paper 
\cite{Lozano:2016kum}. In fact, there the non-Abelian T-dual of $AdS_5\times S^5$  (the Sfetsos-Thompson solution \cite{Sfetsos:2010uq}) was considered. A solution in type IIA with  
a linear charge density $\lambda=N z$ is found and a completion of the geometry is proposed.  Analogously, for the background defined around eq. (\ref{quiver4Pfinal}), we have linear charge density, and for small values of the $z$-coordinate, the charge density (that is the rank function $R(z)=-\alpha''$ ) is also linear for the background in eq. (\ref{cremotomazz}).

In other words, the backgrounds in eq. (\ref{quiver4Pfinal}) or that in eq. (\ref{cremotomazz}) provide a completion to a background like that in eq. (\ref{natd}), obtained via non-Abelian duality on a putative Type IIB background. This structure, observed in examples with $AdS_4, AdS_5, AdS_6$
could be repeated for the case of $AdS_7$ if a true-solution in Type IIB could be found, such that its non-Abelian T dual has the form given in eq. (\ref{natd}).

\end{document}